\pdfoutput=1
\documentclass[acmtog]{acmart}
\acmSubmissionID{514}

\AtBeginDocument{%
  \providecommand\BibTeX{{%
    \normalfont B\kern-0.5em{\scshape i\kern-0.25em b}\kern-0.8em\TeX}}}

\setcopyright{acmcopyright}\acmJournal{TOG}
\acmYear{2021}\acmVolume{40}\acmNumber{4}\acmArticle{110}\acmMonth{8} \acmDOI{10.1145/3450626.3459864}

\author{Mengdi Wang}
\affiliation{
\institution{Dartmouth College}
}
\email{mengdi.wang.gr@dartmouth.edu}

\author{Yitong Deng}
\affiliation{
\institution{Dartmouth College}
}
\email{yitong.deng.gr@dartmouth.edu}

\author{Xiangxin Kong}
\affiliation{
\institution{Dartmouth College}
}
\email{xiangxin.kong.gr@dartmouth.edu}

\author{Aditya H. Prasad}
\affiliation{
\institution{Dartmouth College}
}
\email{aditya.h.prasad.21@dartmouth.edu}

\author{Shiying Xiong}
\affiliation{
\institution{Dartmouth College}
}
\email{shiying.xiong@dartmouth.edu}

\author{Bo Zhu}
\affiliation{
\institution{Dartmouth College}
}
\email{bo.zhu@dartmouth.edu}

\usepackage{wrapfig}
\usepackage{diagbox}
\usepackage{verbatim}
\usepackage{amssymb}
\usepackage{latexsym}
\usepackage{lineno}
\usepackage{color}
\graphicspath{{figures/}}
\usepackage{amsmath,bm}
\usepackage{psfrag}
\usepackage{mathtools}

\theoremstyle{definition}

\usepackage{array}

\usepackage{pifont}

\usepackage{psfrag}
\usepackage{makecell}
\usepackage{algorithm}
\usepackage{threeparttable,booktabs}
\usepackage{graphicx}
\usepackage{subcaption}
\usepackage{flushend}

\usepackage{algpseudocode}

\usepackage{xcolor}
\usepackage{graphicx}

\newcommand{\FourTeaserFig}[5]
{
\begin{teaserfigure}
    \centering
    \includegraphics[width=.24\textwidth]{#1}
    \includegraphics[width=.24\textwidth]{#2}
    \includegraphics[width=.24\textwidth]{#3}
    \includegraphics[width=.24\textwidth]{#4}
  \caption{#5}
  \label{fig:teaser}
\end{teaserfigure}
}

\newcommand{\supermagicfigure}[9]{
{
    \hfill
    \hbox{
        \hspace{#3}
        \vbox to #2{
            \hbox{
                \includegraphics[width=#1]{#4}\hspace{#3}
                \includegraphics[width=#1]{#5}\hspace{#3}
                \includegraphics[width=#1]{#6}
            }
            \vfill
            \hbox{
                \includegraphics[width=#1]{#7}\hspace{#3}
                \includegraphics[width=#1]{#8}\hspace{#3}
                \includegraphics[width=#1]{#9}
            }
        }
    }
    \hfill
    }
}

\newif\ifverbose
\verbosetrue

\ifverbose

\newcommand{\vb}[1]{\textcolor{red}{#1}}
\else

\newcommand{\vb}[1]{}
\fi

\usepackage{url}
\usepackage{xcolor}
\definecolor{newcolor}{rgb}{.8,.349,.1}

\begin{document}

\title{Thin-Film Smoothed Particle Hydrodynamics Fluid}

\renewcommand{\shortauthors}{Mengdi Wang, Yitong Deng, Xiangxin Kong, Aditya H. Prasad, Shiying Xiong, Bo Zhu}
\newcommand{\revise}[1]{{#1}}
\newcommand{\revisea}[1]{{#1}}

\begin{abstract}
    We propose a particle-based method to simulate thin-film fluid that jointly facilitates aggressive surface deformation and vigorous tangential flows.
    We build our dynamics model from the surface tension driven Navier-Stokes equation with the dimensionality reduced using the asymptotic lubrication theory and customize a set of differential operators based on the weakly compressible Smoothed Particle Hydrodynamics (SPH) for evolving point-set surfaces.
    The key insight is that the compressible nature of SPH, which \revise{is} unfavorable in its typical usage, is helpful in our application to co-evolve the thickness, calculate the surface tension, and enforce the fluid incompressibility on a thin film.
    In this way, we are able to two-way couple the surface deformation with the in-plane flows in a physically based manner.
    We can simulate complex vortical swirls, fingering effects due to Rayleigh-Taylor instability, capillary waves, Newton's interference fringes, and the Marangoni effect on liberally deforming surfaces by presenting both realistic visual results and numerical validations.
    The particle-based nature of our system also enables it to conveniently handle topology changes and codimension transitions, allowing us to marry the thin-film simulation with a wide gamut of 3D phenomena, such as pinch-off of unstable catenoids, dripping under gravity, merging of droplets, as well as bubble rupture.
\end{abstract}


\FourTeaserFig{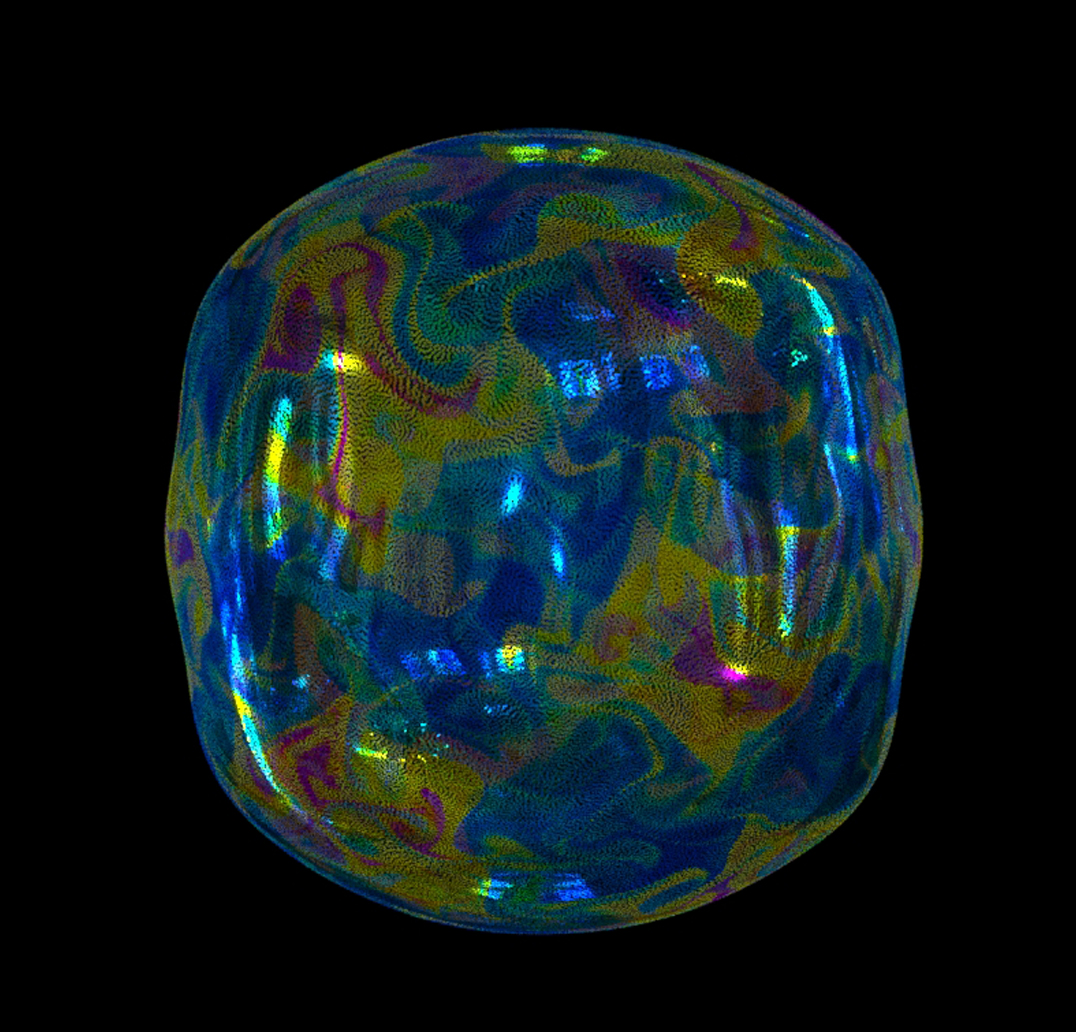}{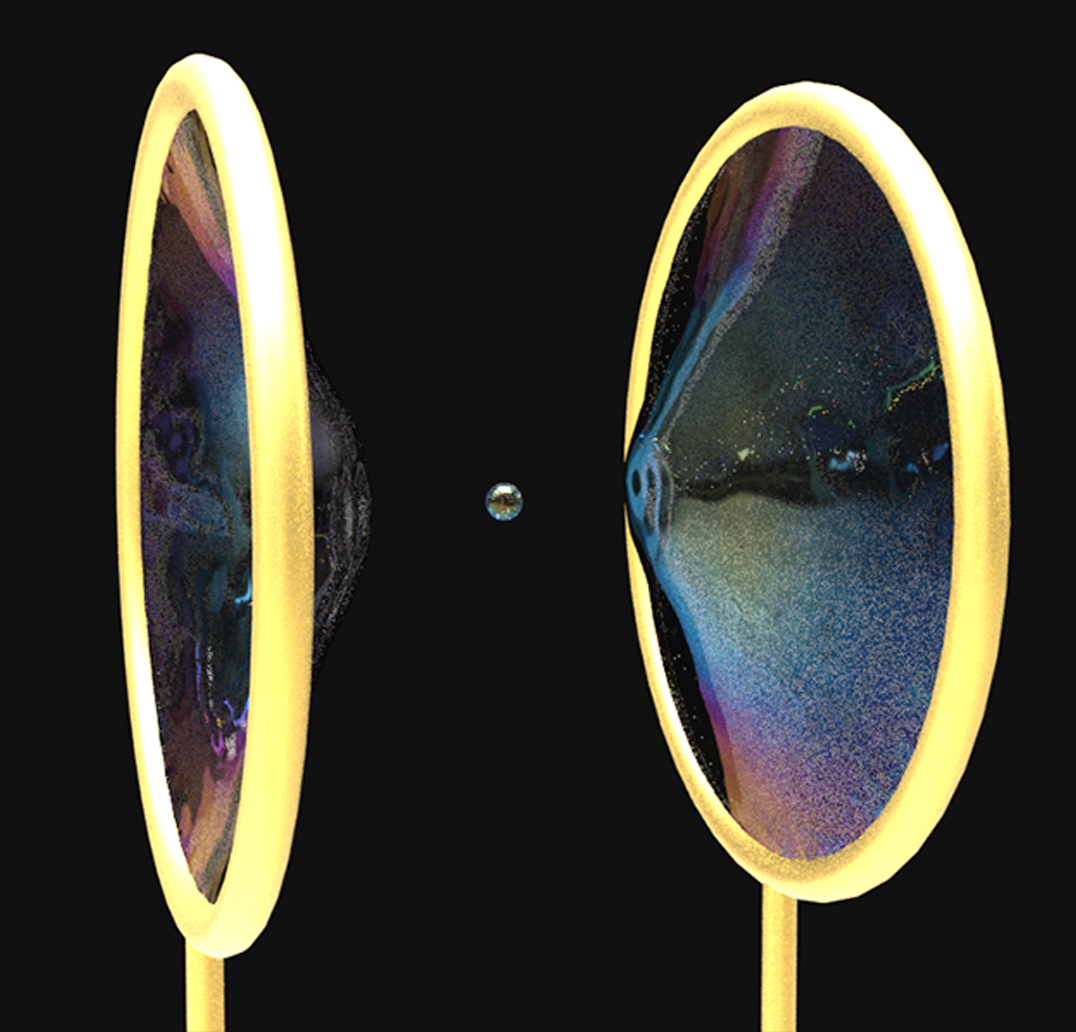}{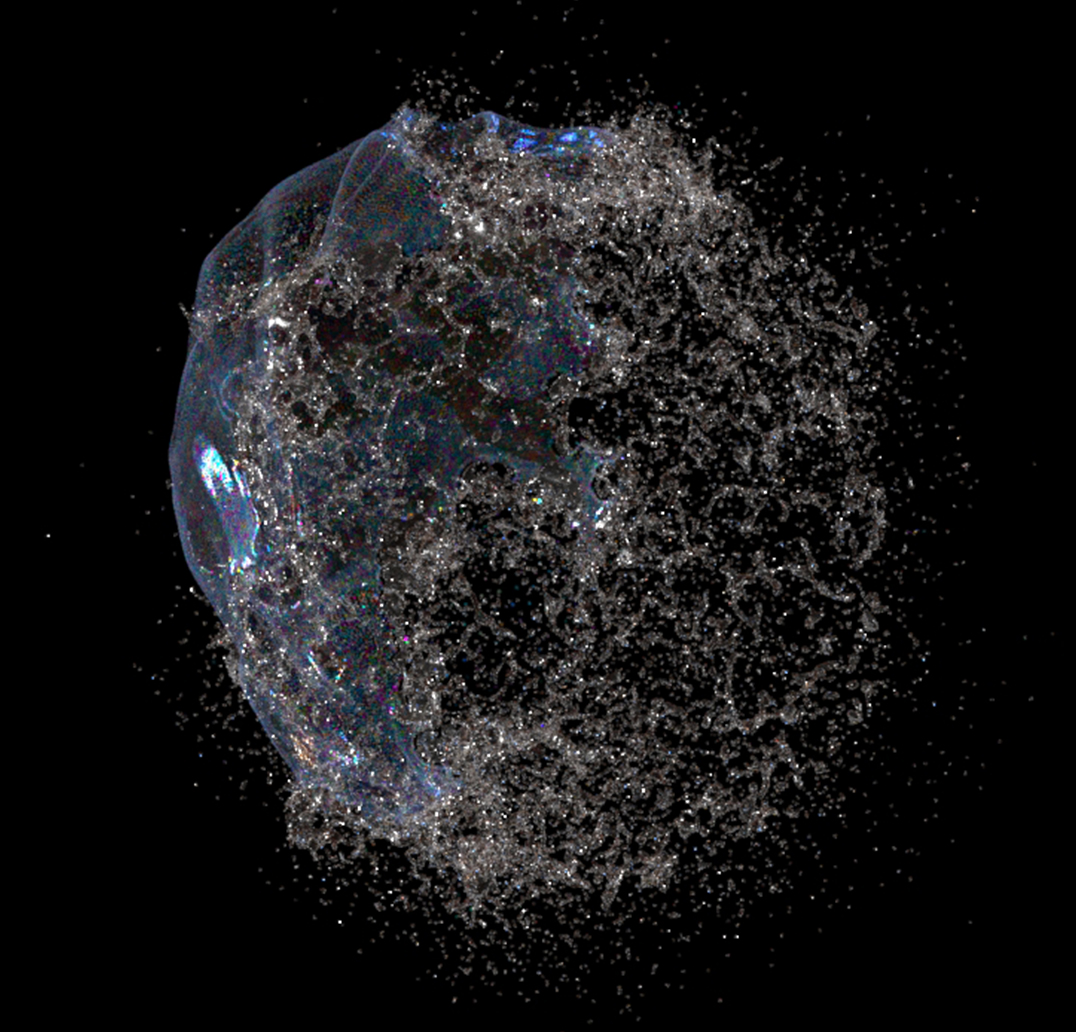}{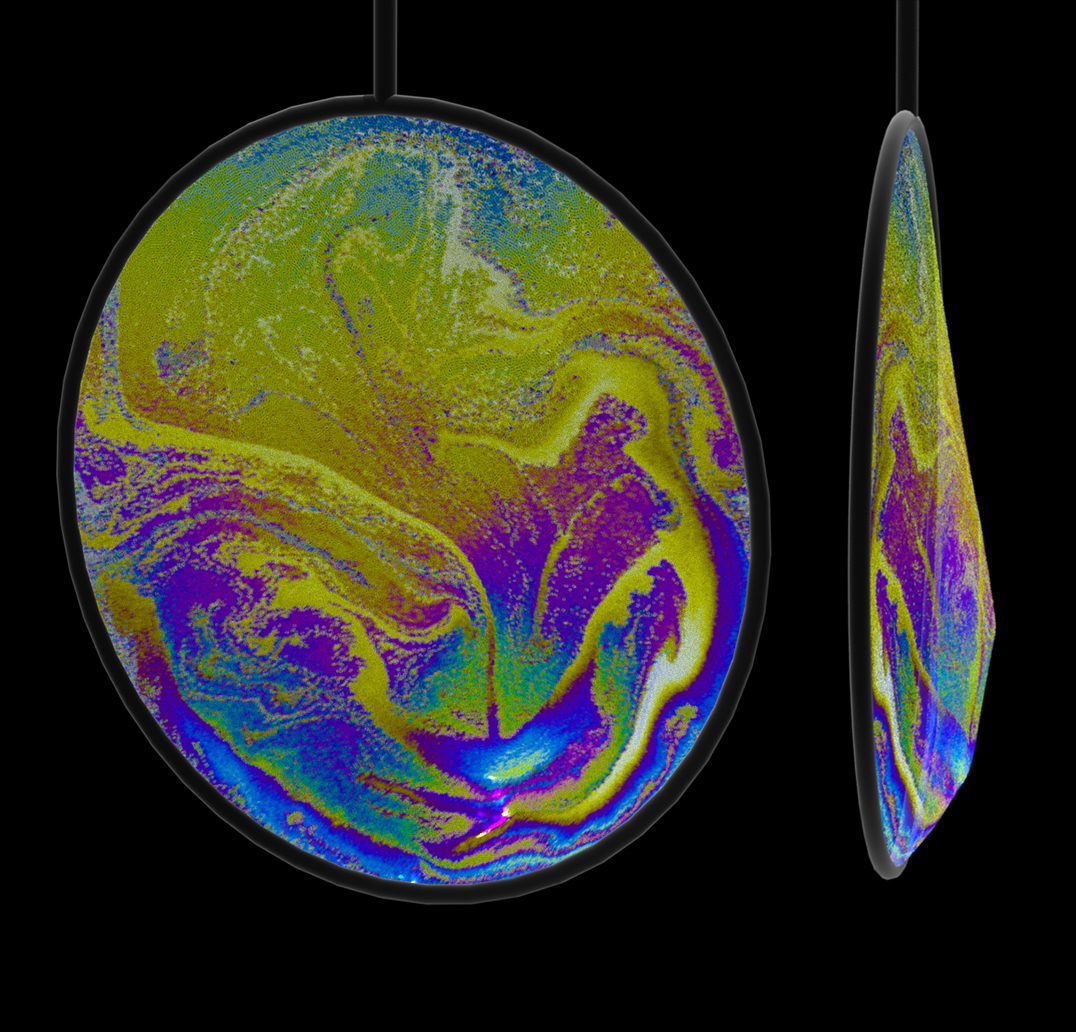}{Several thin-film phenomena as simulated using our proposed method. Counting from left to right: (1) surface flow on a oscillating soap bubble, (2) pinched-off droplets between two circular rims pulling away from one another, (3) a soap bubble bursting into tiny droplets and filaments after being poked from the right, (4) the vibrant, opal-like color pattern caused by Rayleigh-Taylor instability on a large-deforming, wet thin-film surface.}

\maketitle
\section{Introduction}

Thin films are fascinating fluid phenomena on two levels. On the macroscopic scale, their surface-tension-driven dynamics morph them into elegant, minimal-surface geometries such as catenoids in an energy-optimized manner, which are captivating feats \revise{both} artistically and mathematically. When coupled with external forces of air pressure, wind, or gravity, \revise{these tendencies} create the unique bounciness we see in soap bubbles that is satisfying to watch. On the microscopic scale, thin films carry vibrant and delicate color patterns that arise from the interference of light bouncing between the varying film thicknesses, while the turbulent flows precipitate the surface to constantly evolve and contort, thereby creating a smoothly flowing color palette as rich as that of oil paint.

Research in computer graphics has not ceased to strive and capture the beauty that thin films bring about. Seminal previous works
\cite{foam-zheng-2009,related-zhu-2014,foam-saye-2013,related-ishida-2017,related-da-2015,kim2015giant} devise methods to obtain visually appealing simulations of surface tension driven thin-film phenomena, focusing on the metamorphosis of thin-film surfaces. On the other end, many works \cite{ physics-huang-2020,azencot2015functional, vantzos2018real,azencot2014functional,related-hill-2016,stam2003flows} push towards the simulation of the surface tension-driven thickness evolution on the deformed surface, which, when combined with the optical insight in thin-film interference \cite{glassner2000soap,belcour2017practical,smits1992newton,iwasaki2004real,jaszkowski2003interference} can be used to recreate the distinct beauty of iridescent bubbles. Recently, \revise{Ishida et al. \shortcite{physics-ishida-2020}} highlight the importance of integrating the two aspects in achieving enhanced richness and plausibility; and proposes a successful method to jointly simulate surface deformation and thickness evolution. Nevertheless, the endeavor is far from being a closed case, and it yet calls for the realization of surface flow with more liveliness and sophistication, surface deformation with higher frequency details, and the integration with more 3D surface tension phenomena, which together bring a significant challenge for a simulation's robustness, efficiency, and versatility.

In this paper, we attempt to tackle this challenge with a particle-based method. We represent the thin film with a set of codimension-1 particles and devise a set of differential operators based on the Smoothed Particle Hydrodynamics on the evolving surface. In the full 3D simulation, the inevitable compressibility of SPH causes volume loss that results in visual implausibilities. However, because our simulation is principally performed on a codimension-1 surface, the in-plane compression can be compensated by an expansion in its \revise{codimension-direction}, which grants each particle a varying thickness that together outlines a curved surface in the three-dimension space. The surface such prescribed allows for the computation of surface-tension-related forces that the codimension-1 SPH particles will carry out. In this way, a simulation cycle is completed with the SPH playing a dual role: both as the \textit{initiator} who forms the curved surface with the compressible nature and the \textit{executor} who carries out the dynamics resulting from such a curved surface. Using this design, we can simulate thin-film-specific behaviors while inheriting the SPH's virtue of simplicity, adding little additional cost. With the incorporation of a physically-based surface tension model derived from the lubrication assumption, our simulation algorithm effectively reproduces a wide range of phenomena such as the Newton interference patterns, the Marangoni effect, capillary waves, and the Rayleigh-Taylor Instability. 

Our particle-based, codimensional simulation method for thin-film surfaces bridges the mature research literature of SPH with the point-set surface representation, a promising avenue that is not yet thoroughly studied.
Compared to mesh-based paradigms \cite{related-ishida-2017,physics-ishida-2020,related-da-2015,related-zhu-2014}, our particle-based system is flexible for dealing with the topology changes and codimension transitions, which are particularly pertinent to this application, since thin films are delicate constructions that are marked for their tendency to disintegrate, yielding some of the most exciting spectacles such as the pinching-off of the film surface into countless filaments and droplets. To this end, our method conveniently implements particle sharing with an auxiliary, standard 3D SPH solver following previous works \cite{sph-muller-2003,rim-akinci-2013}. Our particles can be directly copied from codimension-1 to codimension-0 and vise versa without any extra processing, enabling the realization of a variety of thin-film phenomena with 3D components.

In summary, our contribution includes:
\begin{itemize}
  \item A thin-film SPH simulation framework that jointly simulates large and high-frequency surface deformation and lively in-surface flows
  \item A meshless simulation framework with a set of SPH-based differential operators to discretize the thin-film physics on a curved, point surface
  \item The leverage of the SPH compressibility to couple the in-surface thickness evolution and the surface metamorphosis via physically-based surface tension
  \item A versatile particle surface representation method that conveniently handles topology and codimension 0-1 transition to integrate surface and volumetric SPH simulation.
\end{itemize}

\section{Related works}

\paragraph{\revise{Mesh-Based} Dynamic Thin-Film} To precisely capture the \revise{dynamic flow and irregular geometry} of a thin film, a myriad of previous work adopts the idea of representing a thin film using a triangle mesh. With the thickness of a thin film coupled into the fluid simulation, modifications can be applied to the projection method of incompressible fluid simulation to fit the requirement of simulating flow dynamics on thin films  \cite{related-zhu-2014,physics-ishida-2020}. 
The vortex sheet model, which uses circulation as a primary variable, is also a feasible option \cite{related-da-2015}. 
Another branch of work \cite{related-ishida-2017} focuses on the surface area-minimizing effect of surface tension. 
Apart from fluid simulation based on \revise{Navier-Stokes} equations, the continuum-based model can also be used for simulating highly-viscous fluid \cite{related-bergou-2010,related-batty-2012}, where surface tension is modeled as an area-minimizing term within the energy-optimization elastic model. \revise{These works not only propose some computational algorithms but also studied the physical model of dynamical thin films, thus set up a remarkable baseline for this topic. However, they have to carefully trade-off between the ability to drastically change the topology of thin-film and solve non-trivial tangential flow on it.}

\paragraph{\revise{Mesh-Based Static} Thin-Film} When constraining the thin film to a fixed shape, the complexity and challenges emerging from an arbitrary-shaped triangle mesh can be alleviated. When the fixed shape is a sphere, special treatments are required to \revise{perform discretization in spherical coordinates} to solve \revisea{the Navier-Stokes} equations \cite{related-bridson-2015,related-hill-2016}, or governing equations coupled with the evolution of thin-film thickness and surface tension \cite{physics-huang-2020}. Beyond bubble, Azencot et al. \shortcite{related-mirela-2015} further explores the flow dynamics of a thin layer of fluid sticking to an arbitrarily shaped object. \revisea{These works show drastic flow convection on the surface of a fixed geometry. We seek to liberate the fixed-surface restriction with our particle method.}

\paragraph{Level-Set Bubble and Foam}
\revise{Triangle mesh is not the only data structure for solving thin-film dynamics. When taking the air within bubbles into consideration, the whole physical system can be easily viewed as a two-phase fluid system. In this case, the level set on grids is a widely adopted tool.} It can be used to model surface tension in different fluids such as ferrofluids \cite{related-ni-2020}. \revise{The multi-phase fluid system, where bubbles are surrounded by large bulks of water, \textit{i.e.}, foams \cite{foam-aanjaneya-2013,foam-albadawi-2013,foam-hong-2003,foam-kang-2008,foam-patkar-2013}, can also be conveniently simulated.} Further, with a relatively symmetric treatment to fluid and water, simulate both foams and water drops \cite{foam-hong-2005}, or standalone bubbles in the air \cite{foam-zheng-2009} after taking delicate care respecting to the thin of soap film. \revise{Unfortunately, level-set immediately means that large bulks of water and air away from the interface have to be included in the computation, limiting the algorithm's complexity and performance.}

\paragraph{SPH Methods} 
It can be seen that all Eulerian methods will face the dilemma between surficial flow and topology changes. 
Therefore, we turn to the \revisea{Lagrangian formalism, where the fluid within a thin film} is represented by particles. The most popular \revisea{particle-based fluid simulation method, Smoothed Particle Hydrodynamics} (SPH) \cite{sph-gingold-1977,sph-brookshaw-1985,sph-monaghan-1985,sph-monaghan-1992,sph-morris-1997,sph-solenthaler-2009,sph-liu-2010,sph-koschier-2019}, where discretization formulation based on radial smoothing kernels is used to approximate differential operators. The simplicity and high parallelizability make SPH useful for interactive fluid simulation \cite{sph-muller-2003}, \revisea{and can readily operate in conjunction with other fluid simulation algorithms \cite{extra-ihmsen-2013,sph-cornelis-2014,extra-band-2018}. It can also be extended to different applications such as simulating multiphase flow, computing magnetohydrodynamics, and blue noise sampling} \cite{sph-tartakovsky-2005,sph-price-2012,sph-jiang-2015}. Recently, SPH has been used for snow \cite{extra-gissler-2020}, elastic solids \cite{extra-peer-2018}, ferrofluids \cite{extra-huang-2019} and viscous fluids \cite{extra-peer-2015,extra-bender-2016}.
On different geometries, simulation algorithms based on SPH are also developed \cite{sph-tavakkol-2017,sph-omang-2006}. \revisea{The} SPH method can also be coupled with a grid-based simulation \revisea{\cite{related-losasso-2008}} or rigid-body simulation \cite{extra-gissler-2019}. As a particle-based method, the treatment of multi-phases, including thin films and foams \cite{rim-yang-2017,sph-yang-2017-unified}, is straightforward in SPH. Some works modeled surface tension with the SPH method \cite{rim-yang-2016,rim-akinci-2013,rim-becker-2007}. The work of Ando and Tsuruno \shortcite{related-ando-2011} captures and preserves thin fluid films with particles. However, the main body of the algorithm is grid-based, and surface tension is not shown in the physical model.

\paragraph{Point-Based Thin Film} \revise{The modeling of thin-film needs geometry information, which can be implicitly given by the shape of the point cloud. A shallow water simulation facilitated by a particle-based height field \cite{extra-solenthaler-2011} inspired us to represent the film thickness similarly. On this basis, we still need some geometrical operators on the surface.} Laplacian operator can be generalized to codimension manifolds \cite{schmidt2014laplace} and discretized in an SPH way \cite{petronetto2013mesh}. Closest point method (CPM)  \cite{ruuth2008simple, cheung2015localized} provides a way to approximate functions on the Cartesian space defined on manifolds, and it thus can be used to solve PDEs on the surface, which includes the governing equations of surface flow \cite{auer2012real, kim2013closest, auer2013semi,related-morgenroth-2020}. Other methods are also available, including graph \revise{Laplacian} \cite{belkin2008towards}, local triangular mesh \cite{belkin2009constructing, lai2013local} and moving least squares \cite{lancaster1981surfaces, levin1998approximation, nealen2004short,saye2014high}. The last one is used by Wang et al. \shortcite{wang-2020} to solve surface tension flows, and \revisea{is} also adopted by our framework.

\section{Method Overview}
\revisea{Overall, our thin-film SPH method is an enhanced version of a conventional SPH solver with geometric information}. We represent the thin film by a set of codimension-1 \revisea{(surface)} particles with local coordinates carried by each of them. The thickness of the film is concurrently estimated both by a surface convolution method resembling the computation of density in classic SPH, and also by equations derived from the mass-conservation law. This point-cloud representation not only allows a Lagrangian in-surface flow simulation but also gives us the exact shape of the manifold thin film. Simulating the thin-film deformation then becomes possible as we can compute the mean curvature of the particle-surface using our codimension-1 differential operators. Finally, our particle-based data structure provides a way to naturally \revisea{transit} a codimension-1 thin film into a codimension-0 fluid bulk, with the latter handled by a traditional SPH fluid solver. 
In the following sections, we will provide details of this algorithm.

\section{Thin-Film Geometry}

\begin{table}
  \label{tab-symbols}
  \begin{tabular}{ccl}
    \toprule
    Symbol & Meaning\\
    \midrule
    $\bm{S}_C
    $ & center surface of soap film\\
    $\bm{S}_{I}^{+}$ & upper interface of soap film\\
    $\bm{S}_{I}^{-}$ & lower interface of soap film\\
    $h$ & distance between interfaces and center surface\\
    $\rho$ & density of fluid in soap film\\
    $\xi^1,\xi^2,z$ & local coordinate variables\\
    $\bm{e_1}, \bm{e_2}, \bm{n}$ & basis vectors of local frame\\
    $\bm{u}$ & velocity of fluid\\
    $\gamma$ & surface tension coefficient\\
    $\Gamma$ & surfactant concentration\\
    $u,v$ & tangential components of velocities on soap film\\
    $w$ & normal component of velocities on soap film\\
    $\kappa_{c}$ & the curvature of center surface\\
    $\kappa_{h}$ & the local curvature of interfaces\\
    $\alpha_c$ & diffusion coefficient of surfactant\\
    $\alpha_h,\alpha_k,\alpha_d$ & pressure coefficients\\
  \bottomrule
\end{tabular}
\caption{A list of symbols \revisea{used in our thin-film SPH model}}
\end{table}

\begin{figure}
\centering
 \includegraphics[width=.45\textwidth]{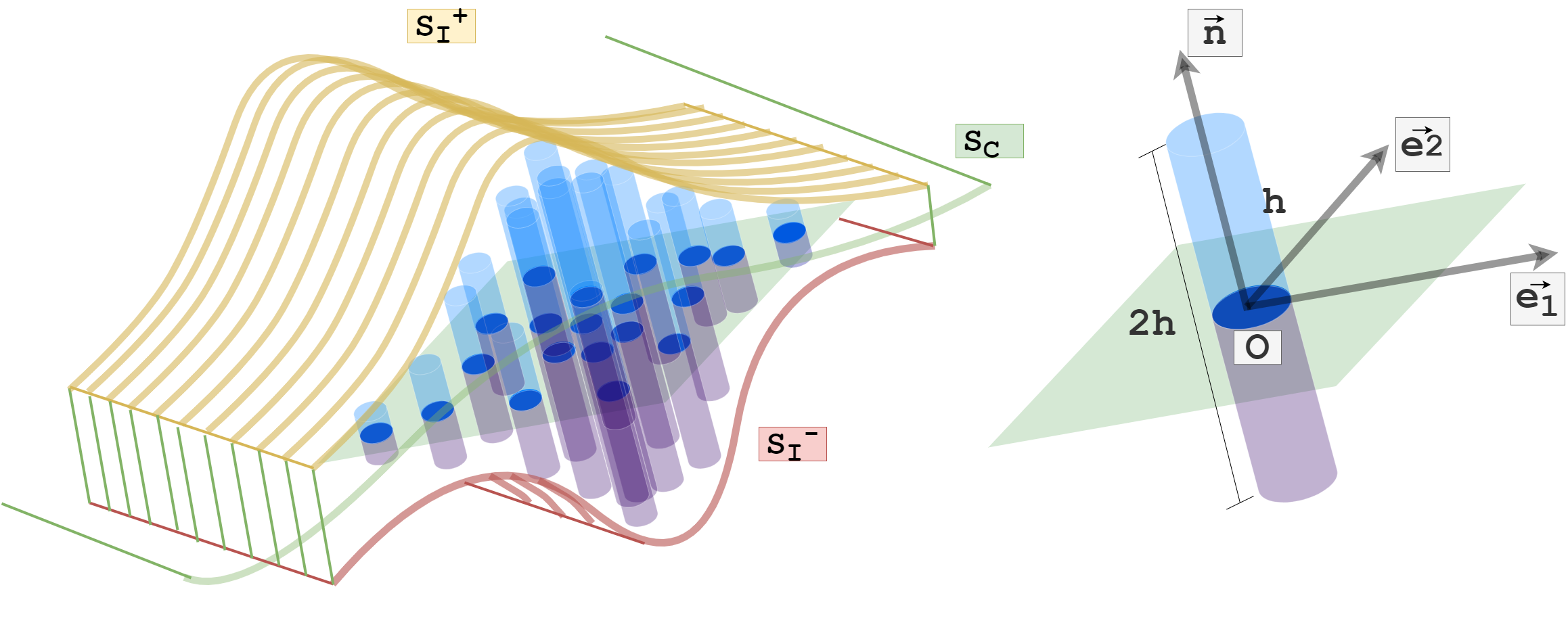}
 \caption{The schematic diagram of \revisea{the soap film's} internal structure. \revise{Central surface $S_C$ is enveloped by two interfaces $S_I^{\pm}$. For point $O$ on $S_C$, the coordinate system is established with normal axis $\bm{n}$ and tangential axes $(\bm e_1, \bm e_2)$. The thickness of film is $2h$ along $\bm{n}$.}}
 \label{fig:film-structure}
\end{figure}

We assume that a soap film is a three-layered, volumetric structure whose characteristic thickness is far less than its characteristic length. As shown in Figure \ref{fig:film-structure}, we denote the upper and lower interfaces as $S_I^{\pm}$, and central surface $S_C$. We build a surface coordinate system $(\bm e_1, \bm e_2)$ near point $O$ on the central surface $S_C$, with a corresponding normal vector $\bm n$, making the coordinates $(\xi^1,\xi^2,z)$ near point $O$. Since we assume that the interfaces $S_I^{\pm}$ are symmetric about the central surface, we take the heights of the upper(yellow) and lower(red) surfaces in the local coordinate system as $z=\pm h(\xi^1,\xi^2)$ respectively. Thus the thickness of \revise{the} thin film is $2h$.

For a codimension-1 surface $S_C$, we follow the work of Wang et al. \shortcite{wang-2020} to define the gradient of a scalar field $s$,
the divergence of a velocity field $\bm v = v^1\bm e_1+v^2\bm e_2$ and the Laplacian of a scalar field $s$ as
\begin{equation}
\begin{dcases}
\bm{\nabla}_s s= \sum_{k=1}^{2} \sum_{l=1}^{2}g^{kl}\frac{\partial s}{\partial \xi^{l}} \bm e_k,\\
\bm{\nabla}_s \cdot \bm v = \frac{1}{\sqrt{g}}\sum_{k=1}^2\frac{\partial }{\partial \xi^k}\left( \sqrt{g} v^k \right),\\
\nabla_s^2 s = \frac{1}{\sqrt{g}}\sum_{k=1}^{2}\sum_{l=1}^{2} \frac{\partial}{\partial \xi^k}\left(\sqrt{g}g^{kl}\frac{\partial s}{\partial \xi^{l}}\right),
\end{dcases}
\end{equation}
where
\begin{equation}
    g^{kl} = \frac{\partial \bm x}{\partial \xi^k} \frac{\partial \bm x}{\partial \xi^l}
\end{equation}
is the metric tensor on $S_C$, $\bm x$ is the 3D Euclidean coordinates, and $g=\det(g^{ij})$.

For \revisea{a} soap film, the scale of the normal direction, which \revisea{is} typically in the order of $10^{-7}m$, is very small in comparison to that of the tangential direction. Hence, we will not consider small quantities of higher order with respect to $h$ in the numerical simulation. The normal direction, tangential direction, and curvature on the upper and lower surfaces $S_I^{\pm}$ can be approximated as
\begin{equation}
\begin{dcases}
 \bm n^{\pm} =  \frac{ \pm\bm n - \bm \nabla_s h}{|\pm\bm n -\bm \nabla_s  h|} =\pm \bm n  - \bm \nabla_s h + O(h^2),\\
 \bm t_1^{\pm} = \frac{\bm e_1 \mp\bm e_2 \times \bm \nabla_sh}{|\bm e_1 \mp\bm e_2 \times \bm \nabla_sh|}= \bm e_1 \mp\bm e_2 \times \bm \nabla_sh + (h^2),\\
 \bm t_2^{\pm} = \frac{\bm e_2 \pm\bm e_1 \times \bm \nabla_sh}{|\bm e_2 \pm\bm e_1 \times \bm \nabla_sh|}= \bm e_2 \pm\bm e_1 \times \bm \nabla_sh + O(h^2),\\
 \kappa^{\pm} = -\frac{1}{2}\bm\nabla \cdot \bm n^{\pm} =  \pm \kappa_c+\kappa_h+ O(h^2),
 \end{dcases}
 \label{eq:geometry}
\end{equation}
where $\bm \nabla$ is the three dimensional gradient operator that satisfies $\bm \nabla_s= \bm \nabla  - \bm n (\bm n \cdot \bm \nabla)$. \revise{The curvature of the center surface is given by $\kappa_c = 0.5\bm{\nabla}_s^2 S_C$, and the local curvature caused by the surface thickness is defined as $\kappa_h=0.5 \nabla^2_s h$.}
In addition, the gradient operator on the upper and lower surfaces $S_I^{\pm}$ can be approximated as
\begin{equation}
    \bm \nabla_{s^\pm}= \bm \nabla  - \bm{n}^\pm (\bm{n}^\pm \cdot \bm \nabla) = \bm \nabla_s \pm  \bm \nabla_s h(\bm n \cdot \bm \nabla ) \pm  \bm n( \bm \nabla_s h \cdot \bm \nabla ) + O(h^2).
    \label{eq:npm}
\end{equation}

\section{Governing equations}
In this section, we derive the evolution equation of the central surface $S_C$.
Considering the flow field near point $O$, we assume that the velocity field $\bm u $ within $-h<z<h$ satisfies 3D incompressible Navier--Stokes equations:
\begin{equation}
    \begin{dcases}
       \rho \left(\frac{\partial\bm{u}}{\partial t}+\bm{u}\cdot \bm \nabla\bm{u}\right)=-\bm \nabla p+\mu\nabla^2\bm{u}+\bm{f},\\
        \bm \nabla\cdot\bm{u}=0,
    \end{dcases}~~~~-h<z<h.
    \label{eq:ns_buck}
\end{equation}
The term $\bm{f}$ here \revisea{denotes the} external force. Following Chomaz \shortcite{physics-chomaz-2001}, we assume a force-balancing boundary condition at \revisea{the} interfaces $S_I^{\pm}$:
\begin{equation}
    (p-p_a+2\kappa^{\pm} \gamma)\bm{n}^\pm = -\bm \nabla_{s^{\pm}}\gamma+\mu[\bm \nabla\bm{u}+(\bm \nabla\bm{u})^T]\cdot\bm{n}^\pm,~~~~z=\pm h,
\label{eq:ns_boundary}
\end{equation}
where $p_a$ is the air pressure. 


\revisea{By substituting \eqref{eq:geometry} and \eqref{eq:npm} into  \eqref{eq:ns_boundary} and employing Taylor expansion}, we can obtain the dominant governing equations at the center surface $S_C$ where $z=0$ as:

\begin{equation}
\begin{aligned}
    &\rho \frac{D \bm{u}}{D t}+O(h^2)=\\
    &\quad2\bm \nabla_s (\kappa_h\gamma+\bm \nabla_s \cdot \bm u_s)+\frac{2\gamma}{h}\kappa_c \bm n+\frac{1}{h}\bm \nabla_s \gamma+ \mu\nabla_{s}^2\bm{u}+\bm{f}.
\end{aligned}
\label{eq:ns_u}
\end{equation}

\revise{Here $\bm u_s=\bm u-(\bm n\cdot \bm u)\bm n$ is the tangential velocity.} 
\revisea{We refer the readers to Appendix \ref{appendix:expansion} for details.}
The mass conservation equation in \eqref{eq:ns_buck} immediately yields the equation \revise{for the} conservation of \revisea{the} height field \cite{extra-solenthaler-2011} as

\begin{equation}
    \frac{D h}{D t} = -(\bm \nabla_s \cdot \bm u_s ) h.
\label{eq:ns_h}
\end{equation}

The strength of surface tension of soap water is under \revise{the} influence of multiple factors, including the water temperature and the concentration level of surfactant. We will only consider the latter under the assumption of always keeping a constant room temperature. \revisea{Surfactants} refer to chemical substances that \revisea{can} significantly reduce water’s surface tension when dissolved by \revisea{them}, like soap, oil, and kitchen detergents. The relationship between the concentration of \revisea{a} surfactant $\Gamma$ and \revisea{the thin film's} surface tension coefficient $\gamma$ can be described by the following  \textit{Langmuir equation of state}:
  $\gamma(\Gamma) = \gamma_0 + RT\Gamma_\infty\log{\left(1-\frac{\Gamma}{\Gamma_\infty}\right)}$ \cite{physics-muradoglu-2007},
where $\gamma_0$ is the surface tension coefficient of \revise{an} untainted interface ($\Gamma=0$), $\Gamma_\infty$ \revisea{is} the maximum packing concentration, $R$ and $T$ \revisea{are} ideal gas constant and temperature respectively. All these four quantities are considered as constants here.
In our case, where $\Gamma\ll\Gamma_0$,
\revisea{The \textit{Langmuir equation of state} can be approximated by $\gamma(\Gamma)=\gamma_0-\gamma_a\Gamma$ with $\gamma_a=RT$ \cite{physics-xu-2005}.}
We further assume that the surfactant concentration $\Gamma$ is invariant with the normal coordinate \revisea{and it} satisfies the classical convection--diffusion equation with \revisea{the} diffusion coefficient $\alpha_c$ as:

\begin{equation}
\frac{\mathrm{D} \Gamma}{\mathrm{D} t} = \alpha_c \nabla_s^2 \Gamma.
\label{eq:ns_gamma}
\end{equation}

Discarding the quantities of the magnitude $O(h^2)$ and combining \eqref{eq:ns_u}, \eqref{eq:ns_h}, and \eqref{eq:ns_gamma}, we arrive at the our functional dynamics model:

\begin{equation}
    \begin{dcases}
    \rho\frac{\mathrm{D} \bm{u}}{\mathrm{D} t}=2\bm \nabla_s (\kappa_h\gamma+\bm \nabla_s \cdot \bm u_s)+\frac{2\gamma}{h}\kappa_c \bm n+\frac{1}{h}\bm \nabla_s \gamma+ \mu\nabla_{s}^2\bm{u}+\bm{f},\\
    \frac{D h}{D t} = -(\bm \nabla_s \cdot \bm u_s ) h,\\
    \frac{\mathrm{D} \Gamma}{\mathrm{D} t} = \alpha_c \nabla_s^2 \Gamma.
    \end{dcases}
    \label{eq:ns_uhgamma}
\end{equation}

\section{SPH Discretization}

\subsection{\revisea{Particle Height}}

\begin{figure}[t]
\begin{center}
\resizebox{0.45\textwidth}{!}{%
\includegraphics[height=3cm]{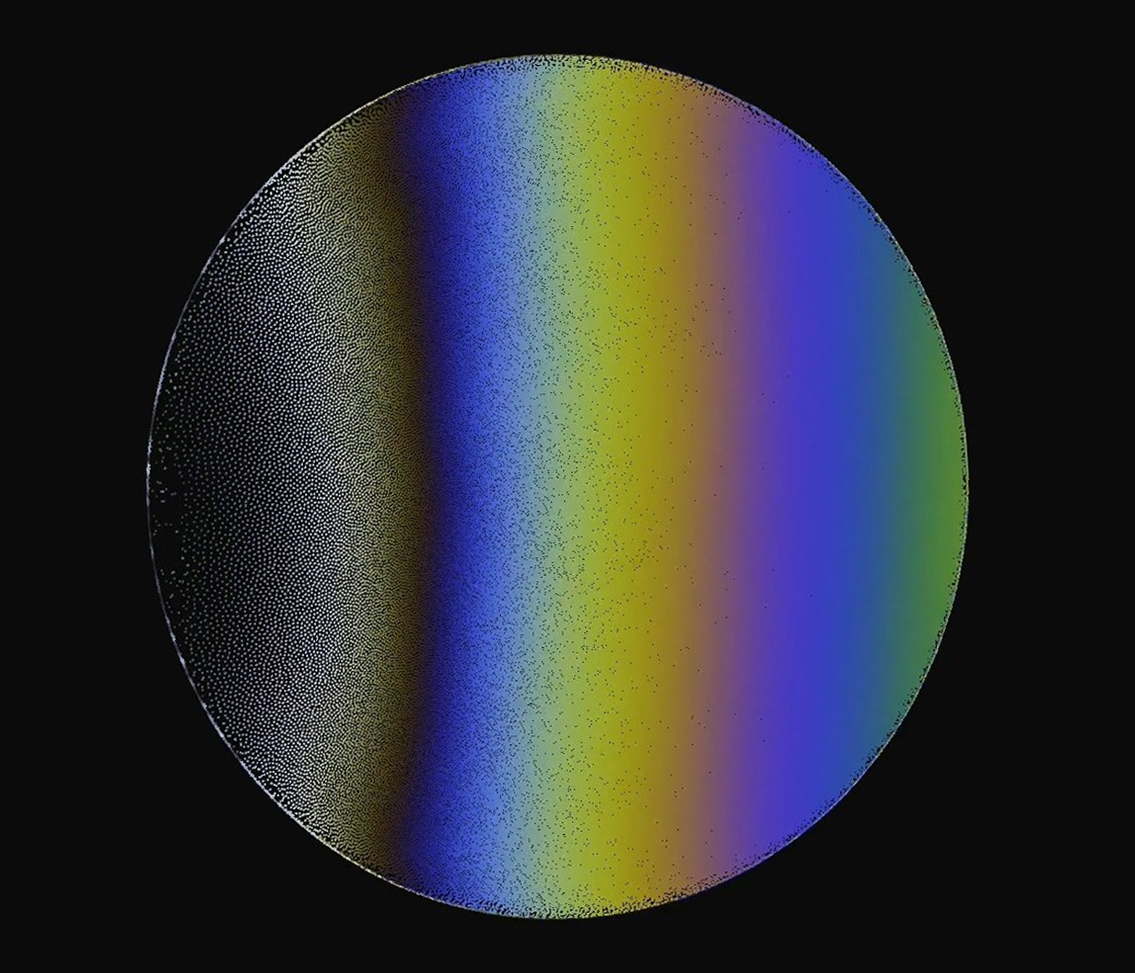}%
\quad
\includegraphics[height=3cm]{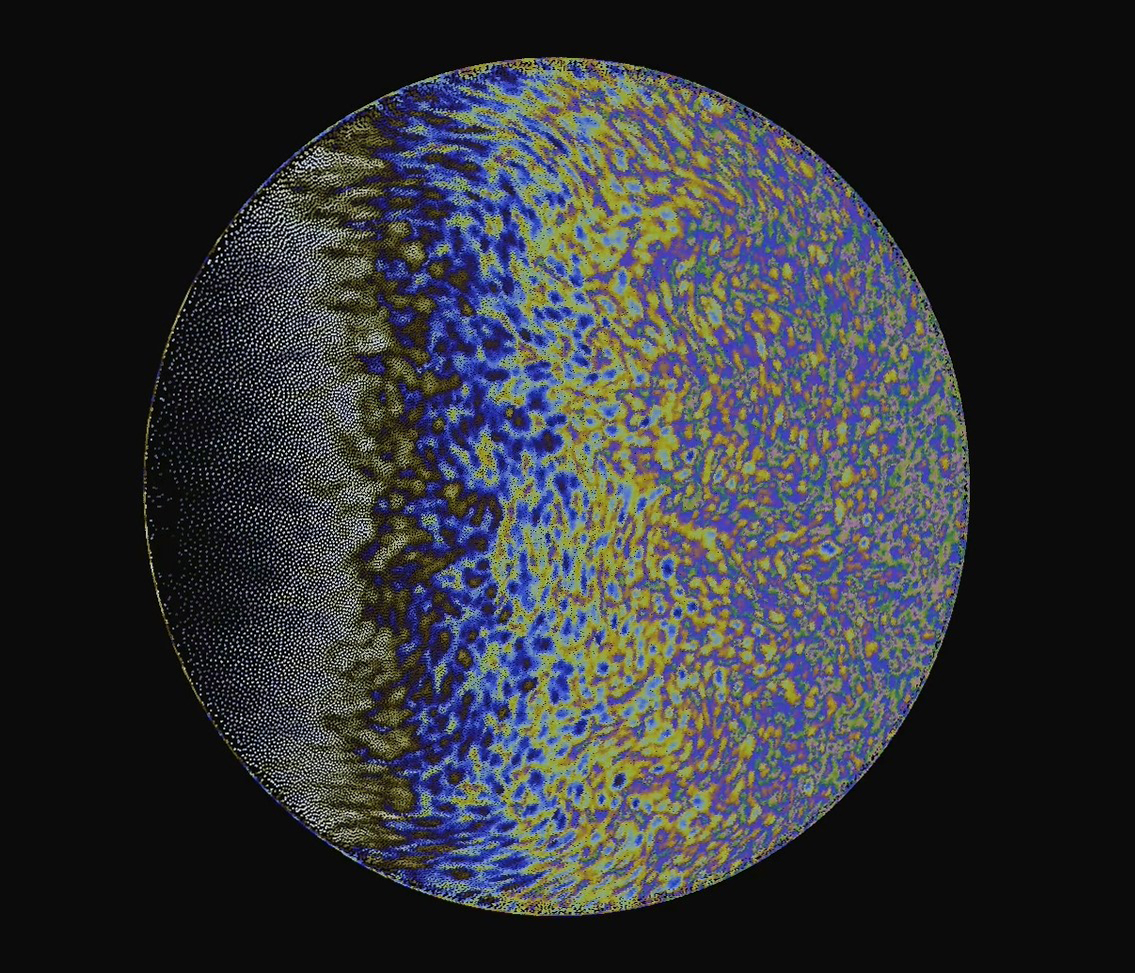}%
}
\end{center}
\caption{\revisea{The comparison between numerical height (left) and advected height (right).}}
\label{fig:h_comparison}
\end{figure}

\revisea{
The key intuition underpinning our SPH particle height model in codimension-1 space is that each thin-film particle carries along with a certain mass of fluid. In the regions where particles are denser, the thin film is deemed thicker because more fluid is contained per unit area on the tangential plane. This intuition can be mathematically captured by the codimension-1 SPH framework.
Given the SPH integration of a physical quantity $A$ on the thin film $A_i = \sum_{j} s_j A_j W_{ij} =\sum_{j} \frac{V_j}{h_j} A_j W_{ij}$, where $s_j$ represents the control area of particle $j$. If we set the function $A$ to equal $h$, then we obtain the SPH integrated expression of the (half-)height field $h$ on a thin film as:
\begin{equation}
\label{eq:sph-h}
    h_i = \sum_{j}V_jW_{ij}.
\end{equation}
In this way, the height of a particle is decided by (i) the particle distribution near its neighborhood and (ii) the volume that each neighboring particle carries. To view it another way, this translates the typical SPH mass density (how much mass per unit area) \cite{sph-tartakovsky-2005} to the volume density (how much volume per unit area), which naturally equates the concept of height on a thin film, an correlation previously explored by Solenthaler et al. \shortcite{extra-solenthaler-2011}. 
Note that $h$ and $V$ are one-sided quantities because we assume symmetry about $S_C$, and the half volume $V_i=m_i/\rho_i$ is assigned with every particle at the beginning and is conserved throughout the simulation. 
}
%

The height $h_i$ that is so computed will be termed the \textit{numerical} height, and we use it for all computation \revise{relevant} to the dynamics of soap bubbles, \textit{e.g.}, the discretization of all differential operators, and the first equation of \eqref{eq:ns_uhgamma} describing the momentum evolution of fluid. It offers a smooth surface that prevents numerical instabilities, however "blurs" the height field with a mollifying kernel. To preserve the vibrant and turbulent flow patterns originating from the advection-diffusion evolution of \revisea{the} height field in \eqref{eq:ns_uhgamma}, we introduce an additional \textit{advected} height $\hat{h}$, that actually follows the Lagrangian evolution equation \eqref{eq:ns_h},
\begin{equation}
    \frac{\mathrm{D}\hat{h}}{\mathrm{D}t}=-\left(\bm\nabla_s\cdot\bm{u}_s\right)\hat{h},
    \label{eq:rh_evolution}
\end{equation}
which will be used to compute the interference color for rendering.

\revisea{Now we compare the two forms of height computation.} The advected height is obtained by temporally integrating the flux about a particle from the velocity field. The numerical height is obtained by first temporally integrating the particles’ positions using the same velocities and then estimating the volume density. The two forms are equivalent if there are no numerical errors. In practice, they deviate in their behavior, each with its advantages and drawbacks. The numerical height is free from error accumulation, as it is recomputed at each step, but it is inevitably smoothed due to the SPH kernel convolution. The advected height can suffer from numerical drift, \revisea{but} it can preserve higher-frequency details. In our system, we use both to their advantages. We use the numerical height \revisea{to compute the dynamics}, where smoothness is beneficial for robustness, and \revisea{we} use the advected height in the color computation for more appealing visual results. \revisea{To reduce the} numerical drift, we normalize the advected height at each timestep to conserve the total volume. 
\revise{As shown in Figure \ref{fig:h_comparison}, the numerical height is much smoother and does not preserve the flow pattern introduced by convection. We provide further comparisons in the appendix.}

\begin{figure}[t]
\centering
    \begin{subfigure}[b]{0.24\linewidth}
    \centering
        \includegraphics[width=\linewidth]{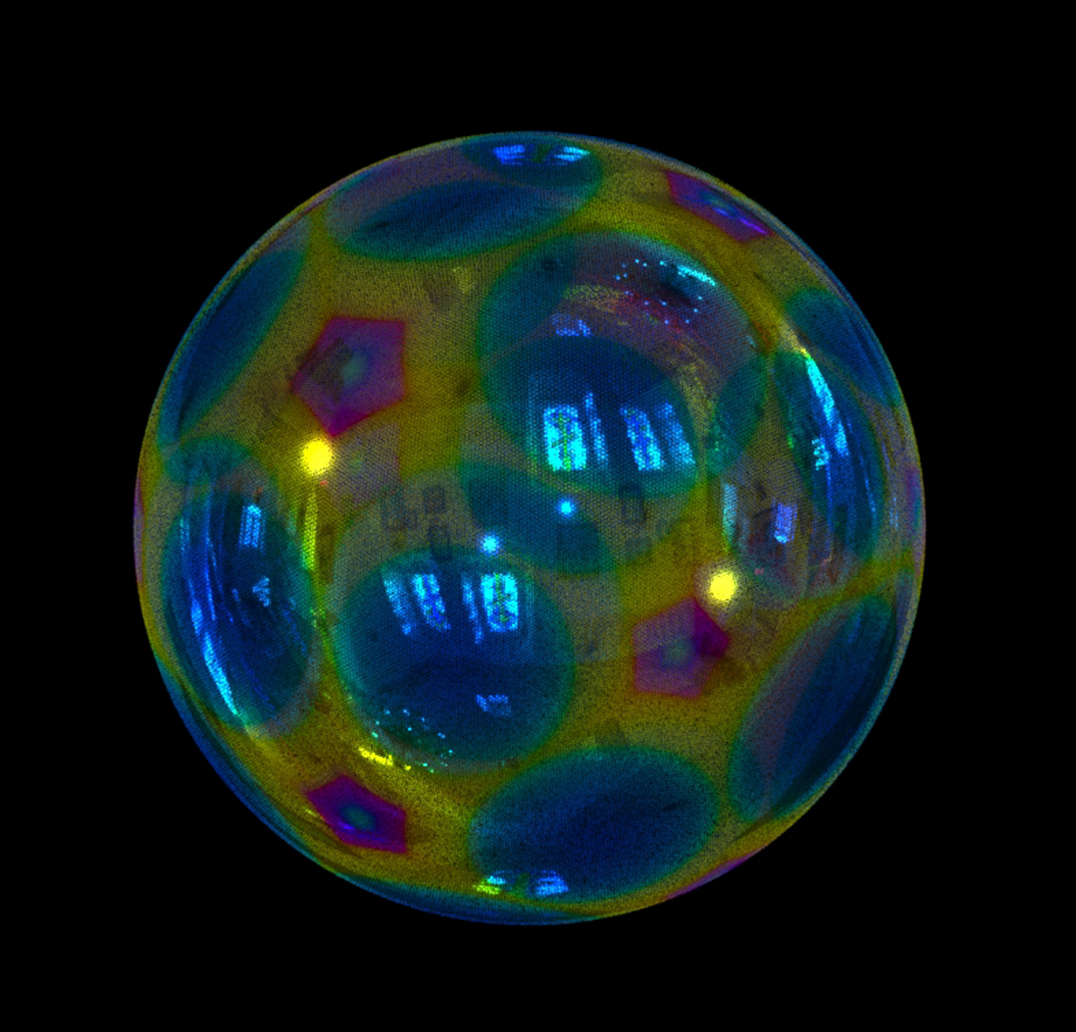}
    \end{subfigure}
    \hfill
    \begin{subfigure}[b]{0.24\linewidth}
    \centering
        \includegraphics[width=\linewidth]{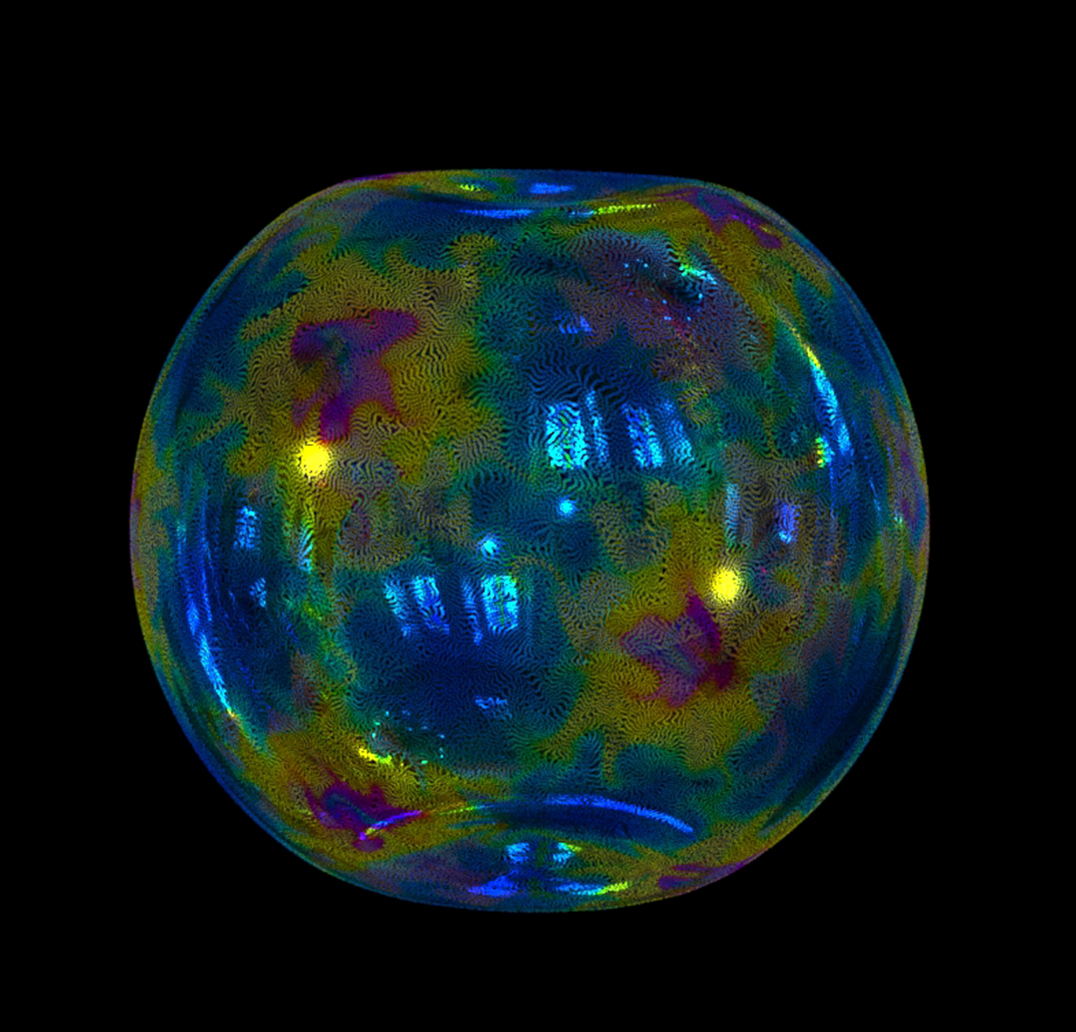}
    \end{subfigure}
    \hfill
    \begin{subfigure}[b]{0.24\linewidth}
    \centering
        \includegraphics[width=\linewidth]{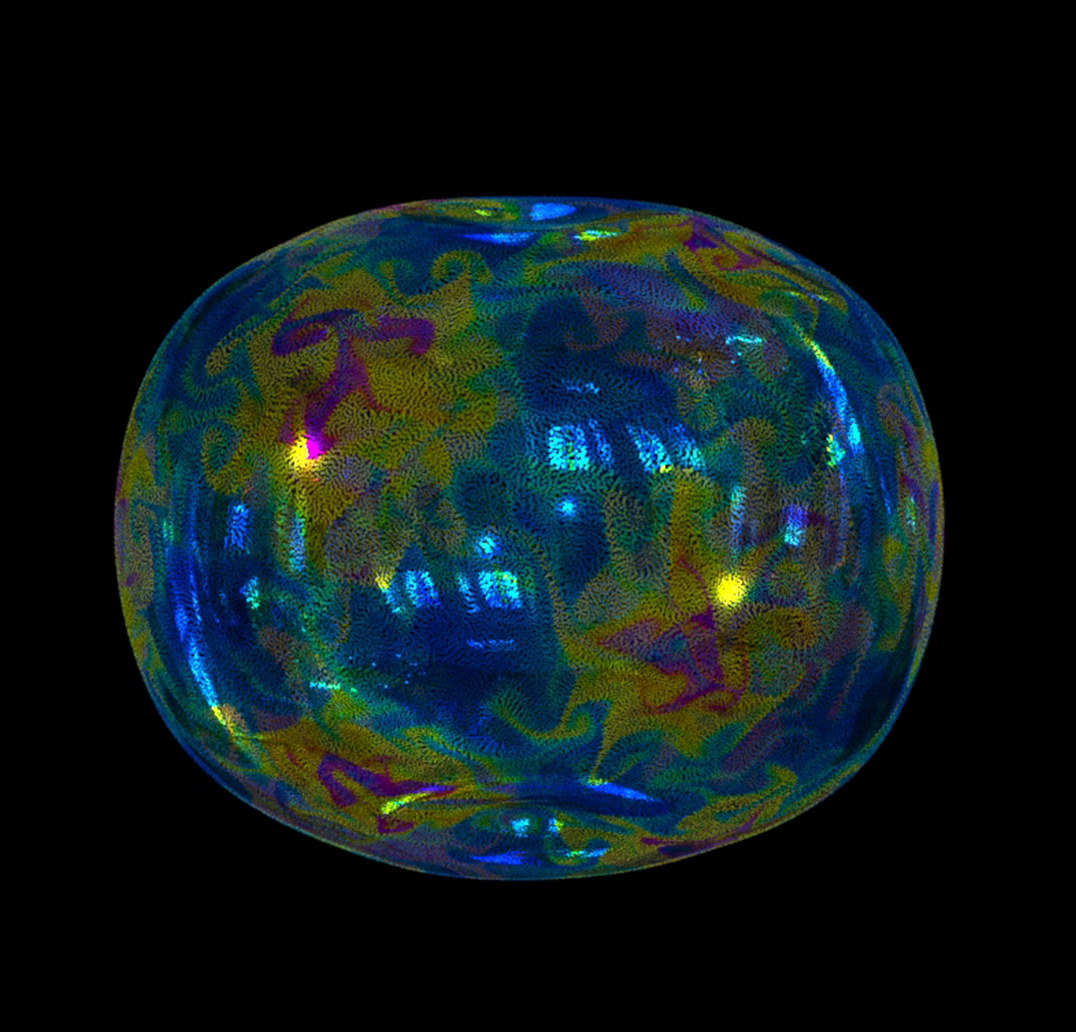}
    \end{subfigure}
    \hfill
    \begin{subfigure}[b]{0.24\linewidth}
    \centering
        \includegraphics[width=\linewidth]{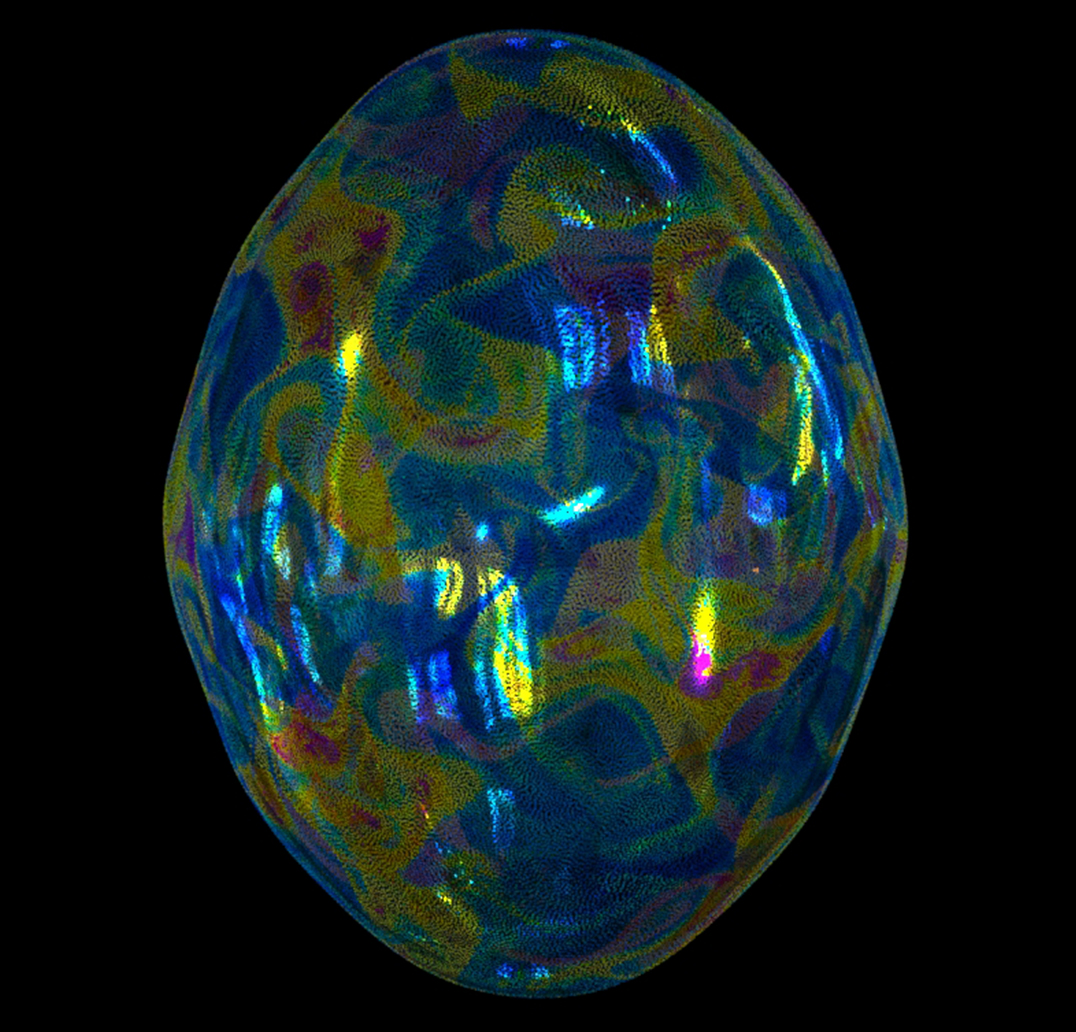}
    \end{subfigure}
    \vskip 0.1\baselineskip
    \begin{subfigure}[b]{0.24\linewidth}
    \centering
        \includegraphics[width=\linewidth]{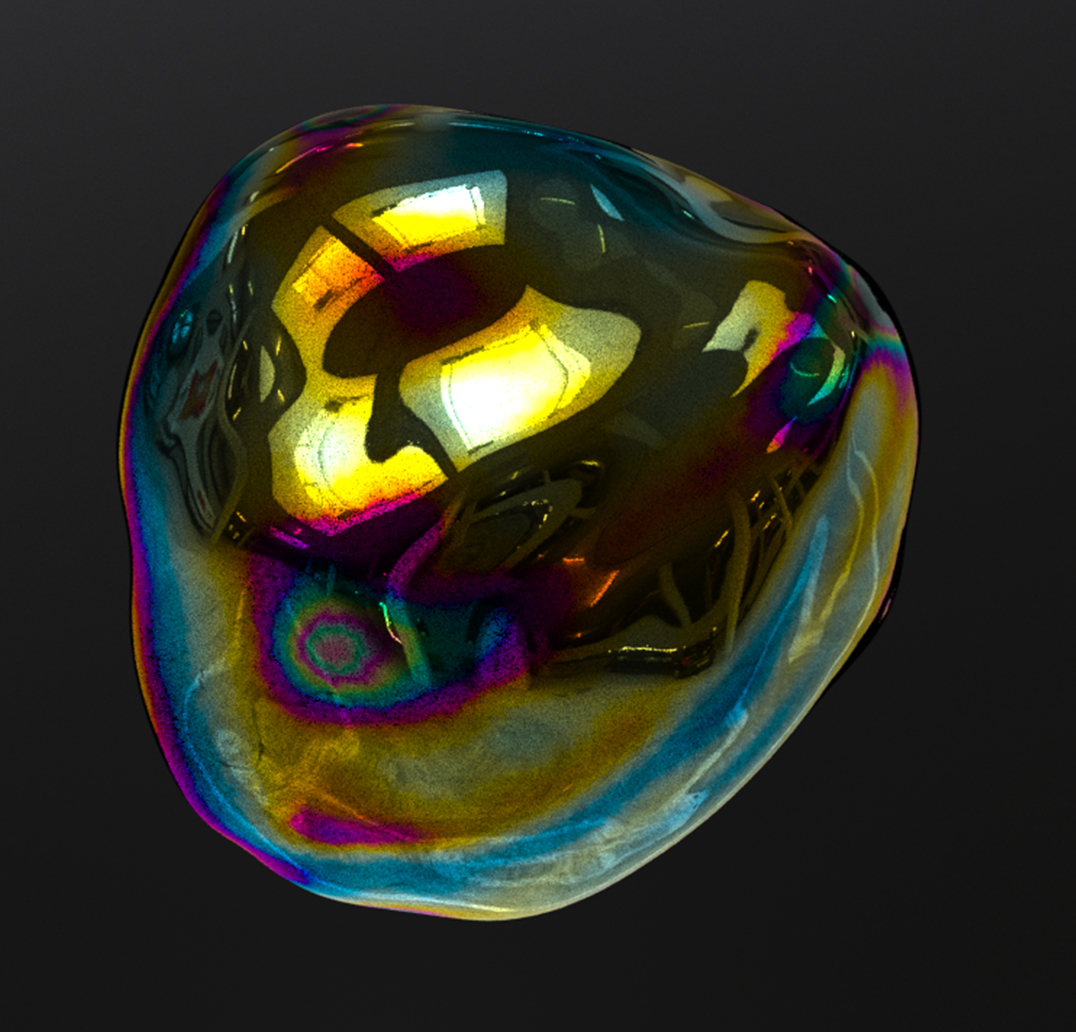}
    \end{subfigure}
    \hfill
    \begin{subfigure}[b]{0.24\linewidth}
    \centering
        \includegraphics[width=\linewidth]{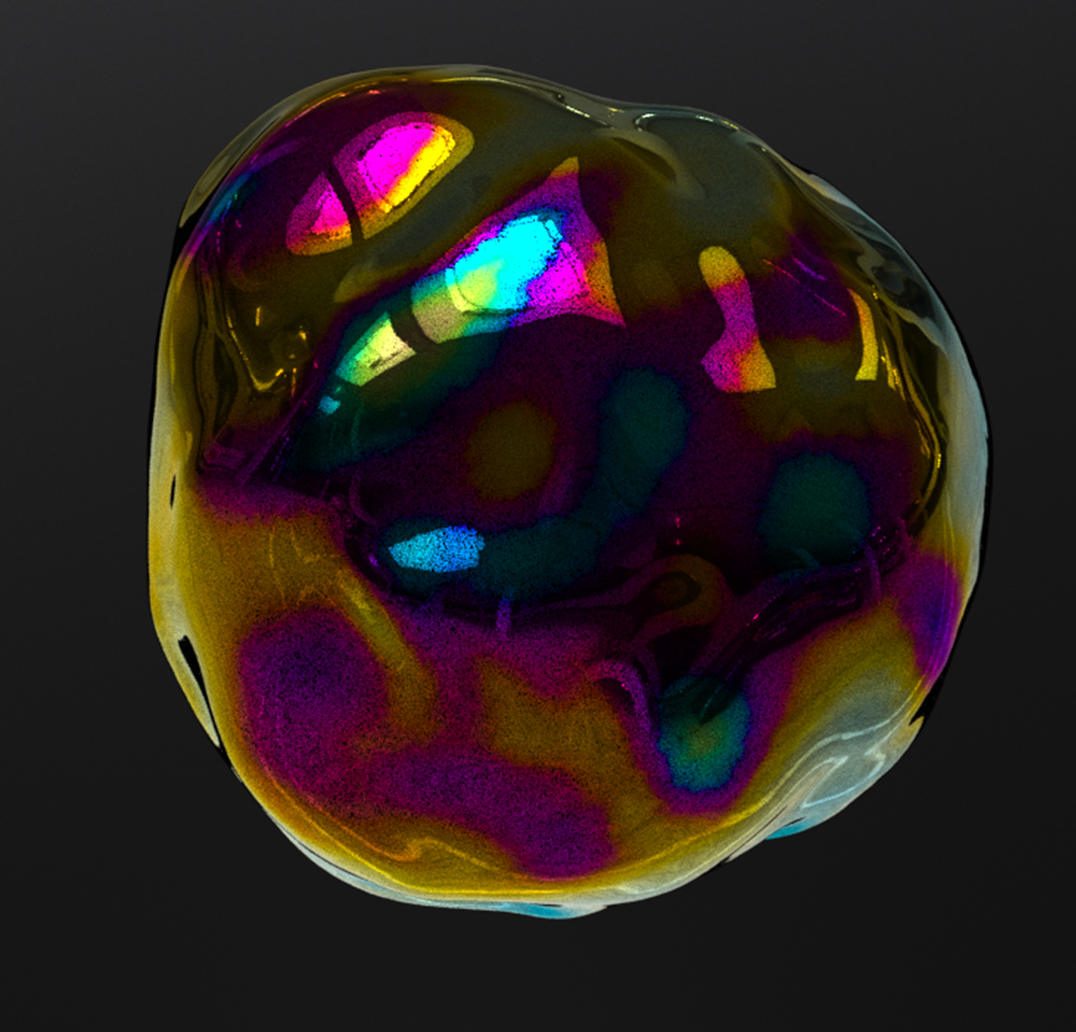}
    \end{subfigure}
    \hfill
    \begin{subfigure}[b]{0.24\linewidth}
    \centering
        \includegraphics[width=\linewidth]{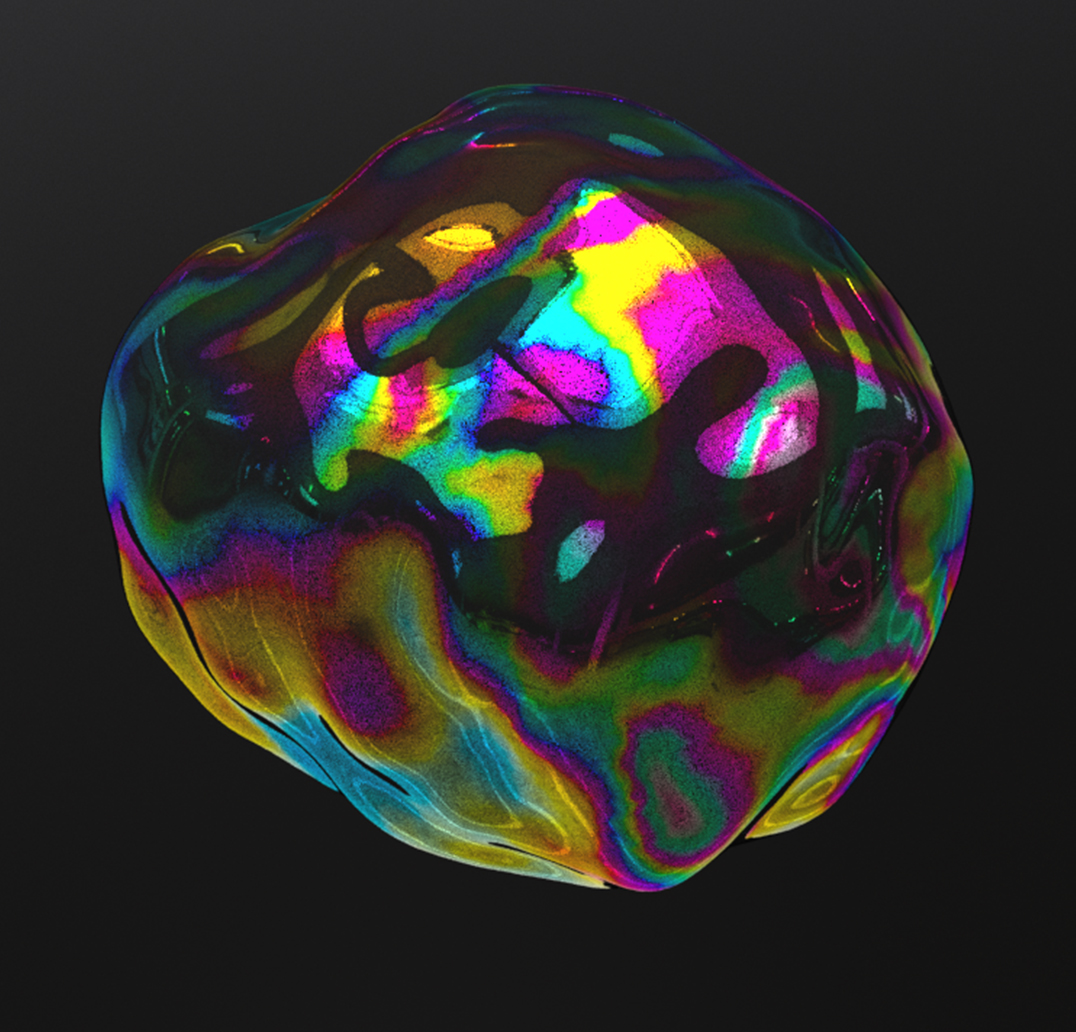}
    \end{subfigure}
    \hfill
    \begin{subfigure}[b]{0.24\linewidth}
    \centering
        \includegraphics[width=\linewidth]{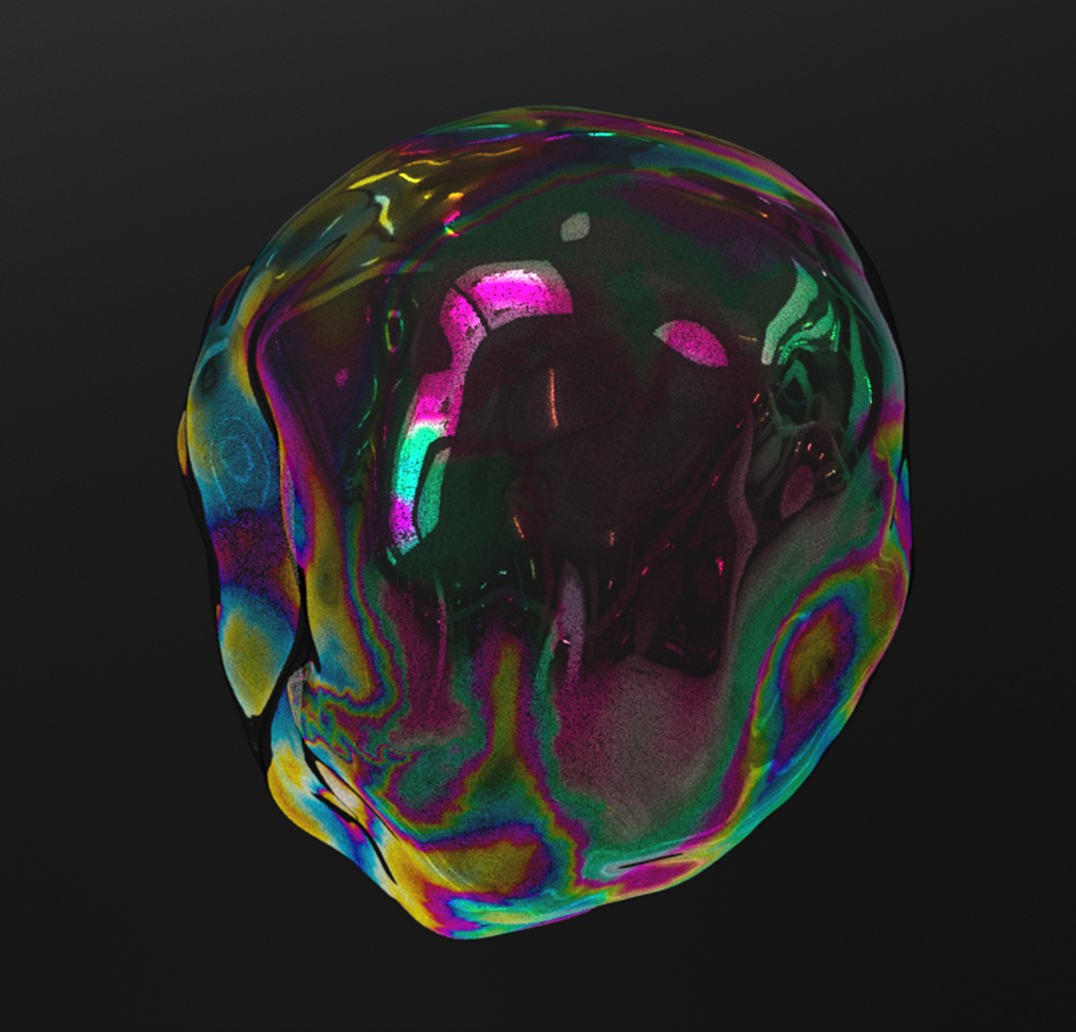}
    \end{subfigure}
\caption{Bubble oscillation: A perfect sphere ocillates under perturbations and vortices (upper part), and an irregular bubble (lower part).}
\label{fig:oscillation}
\end{figure}

\begin{figure*}
    \centering
    \includegraphics[width=\textwidth]{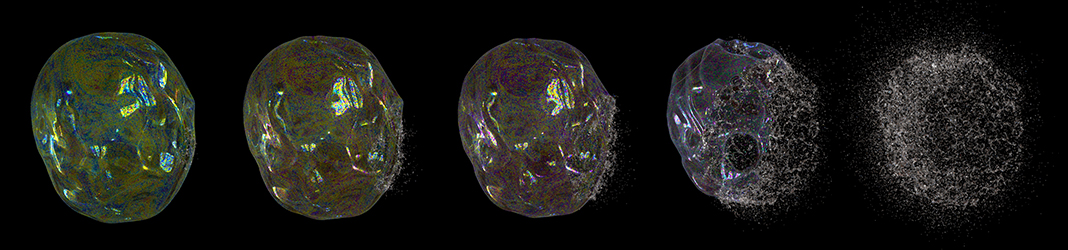}
    \caption{Bubble rupture: The bubble bursts into filaments and droplets after being "punctured" in a small region on the right. The color change on the bubble comes as a result of the surface contraction, which \revise{increases} the thickness of the thin film.}
    \label{fig:bubble-rupture}
\end{figure*}

\subsection{\revise{Differential Operators}}
In our scheme, \revise{the} discretization of surface differential operators, \textit{i.e.}, $\bm{\nabla}_s$,$\left(\bm{\nabla}_s\cdot\right)$ and $\nabla_s^2$ at particle $i$ are obtained by projecting all \revise{of} its neighbor particles to the plane defined by its local frame $\bm{e}^i$, and performing traditional SPH operators on that codimension-1 plane, as shown in \eqref{eq:sph-operators}:

\begin{equation}
\label{eq:sph-operators}
\begin{dcases}
    \left(\bm{\nabla}_s f\right)_i=\sum_{j}h_i V_j\left(\frac{f_i}{h_i^2}+\frac{f_j}{h_j^2}\right)\bm{\nabla}_s W_{ij}, (\textit{symmetric form})\\
    \left(\bm{\nabla}_s f\right)_i=\sum_j\frac{V_j}{h_j}\left(f_j-f_i\right)\bm{\nabla}_s W_{ij},(\textit{difference form})\\
    \left(\bm{\nabla}_s\cdot\bm{u}_s\right)_i=\sum_j\frac{V_j}{h_j}\widetilde{\bm{u}}_{ij}\cdot\bm{\nabla}_s W_{ij},\\
    \left(\nabla_s^2 f\right)_i =\sum_j\frac{V_j}{h_j}\left(f_j-f_i\right)\frac{2|\bm{\nabla}_s W_{ij}|}{|\bm{\widetilde{r}}_{ij}|}.
\end{dcases}
\end{equation}

Distinguishing between the \textit{symmetric} and \textit{difference} forms of the surface gradient operator is a common \revisea{numerical} technique \revisea{employed} in SPH fluid simulations \revise{\cite{sph-muller-2003,sph-koschier-2019}}. \revisea{In our discretization}, we primarily use the symmetric form to take advantage of its momentum conserving nature, with two exceptions. One is the Marangoni force, where we want to highlight the directional quality of the motion. The other is when treating particles near the boundary to negate their tendencies to be pulled towards the center due to the lack of particles on the other side.

The surface vector $\bm{\widetilde{r}}_{ij}$ in \eqref{eq:sph-operators} is obtained by subtracting the local normal component $\bm{n}_i$ from the vector $\bm{r}_{ij}=\bm{r}_i-\bm{r}_j$ \revisea{pointing} from particle $j$ to $i$ as
$
    \bm{\widetilde{r}}_{ij}=\bm{r}_{ij}-\left(\bm{r}_{ij}\cdot\bm{n}_i\right)\bm{n}_i.
$

\revisea{We further define the surface gradient $\bm{\nabla}_s W_{ij}$ of a codimension-1 kernel scalar kernel function $W=W(\widetilde{r})$ as}

\begin{equation}
\revise{
    \bm{\nabla}_s W_{ij}=g_i^{-1}\frac{\mathrm{d}W}{\mathrm{d}\widetilde{r}}\frac{\bm{\widetilde{r}}_{ij}}{|\bm{\widetilde{r}}_{ij}|}.
}
\end{equation}
Here $g_i$ is the metric tensor at particle $i$ and $\widetilde{r}=|\bm{\widetilde{r}}_{ij}|$.
We use the fourth-order spline function \cite{sph-tartakovsky-2005} to compute \revisea{the} numerical height $h_i$ \revisea{and} the Spiky kernel \cite{sph-muller-2003} to compute all other terms to avoid clustering caused by inappropriate repulsive forces. \revisea{Mathematically, these kernel functions are all in 2D forms because they're operating on a codimension-1 plane.}

\begin{figure}[h]
 \centering
 \includegraphics[width=.35\textwidth]{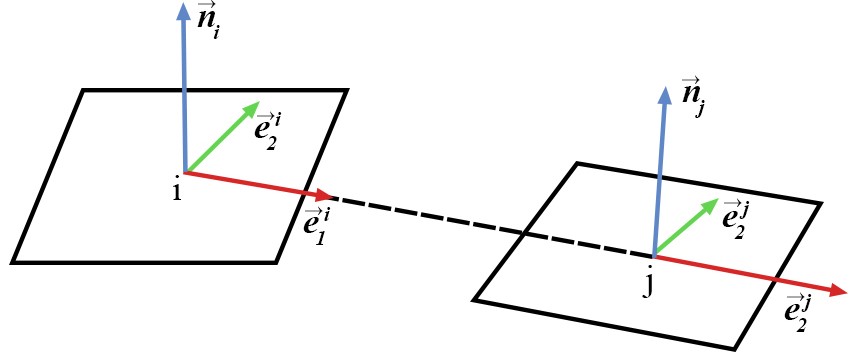}
 \caption{Definition of $\widetilde{\bm{u}}_{ij}$ in \eqref{eq:sph-operators}. Here $\vec{n}$ represents normal axes, $\vec{e}_1$ is a \revisea{a unit vector} pointing from $i$ to $j$, and $\vec{e}_2$ is the cross product of $\vec{n}$ and $\vec{e}_1$.}
 \label{fig:nominal-vector}
\end{figure}

The velocity difference $\widetilde{\bm{u}}_{ij}$ we used for divergence operator is different from the projection of $\bm{u}_j-\bm{u}_i$ to codimension-1 plane $\bm{e}^i$. The reason arises from the insight that a streamline on the surface, with a constant velocity rate $\alpha$ (\textit{i.e.}, $|\bm{u}_i|=|\bm{u}_j|=\alpha,\quad\bm{u}_i\cdot\bm{n}_i=\bm{u}_j\cdot\bm{n}_j=0$ ), should yield a zero divergence. However, if we directly project $\bm{u}_j$ to the local plane at particle $i$, the deviation of normal directions $\bm{n}_i-\bm{n}_j\neq 0$ will \revisea{incorrectly distort the project velocity rate as $|\bm{u}_j-\left(\bm{u}_j\cdot\bm{n}_i\right)\bm{n}_i|\neq \alpha$.}
%
%
To avoid this numerical artifact, we imitate the idea of the  great-circle-advection algorithm used by Huang et al.  \shortcite{physics-huang-2020}; in particular, we pass the same coordinate values from the local frame at $j$ to that at $i$ instead of performing a projection, as shown in Figure \ref{fig:nominal-vector}.

First, we build temporary coordinate systems $\left(\widetilde{\bm{e}}^i,\widetilde{\bm{n}}_i\right)$ and $\left(\widetilde{\bm{e}}^j,\widetilde{\bm{n}}_j\right)$ at local frames of $i$ and $j$, with their normal directions unchanged, \revise{and one tangential axis, say, $\bm{e}_1$,} along the direction of $\bm{r}_{ij}$:

\begin{equation}
    \begin{aligned}
        &\widetilde{\bm{n}}_i=\bm{n}_j,\quad\widetilde{\bm{n}}_j=\bm{n}_j,\\
        &\widetilde{\bm{e}}^i_1=\widetilde{\bm{e}}^j_1=\frac{\bm{r}_{ij}}{|\bm{r}_{ij}|},\\
        &\widetilde{\bm{e}}^i_2=\widetilde{\bm{n}}_i\times\widetilde{\bm{e}}^i_1,\quad\widetilde{\bm{e}}^j_2=\widetilde{\bm{n}}_j\times\widetilde{\bm{e}}^j_1.
    \end{aligned}
\end{equation}

Then, with the the decomposition of $\bm{u}$ at temporary coordinates $\left(\widetilde{\bm{e}},\widetilde{\bm{n}}\right)$ \revisea{we have}

\begin{equation}
\begin{aligned}
    \bm{u}_i=u_i\widetilde{\bm{e}}^i_1+v_i\widetilde{\bm{e}}^i_2+w_i\widetilde{\bm{n}}_i,\\
    \bm{u}_j=u_j\widetilde{\bm{e}}^j_1+v_j\widetilde{\bm{e}}^j_2+w_j\widetilde{\bm{n}}_j.
\end{aligned}
\end{equation}
Finally, as shown in Figure \ref{fig:nominal-vector}, we compute the actual velocity difference $\widetilde{\bm{u}}_{ij}$ used for divergence as:
\begin{equation}
    \widetilde{\bm{u}}_{ij}=\left(u_j-u_i\right)\widetilde{\bm{e}}^i_1+\left(v_j-v_i\right)\widetilde{\bm{e}}^i_2.
\end{equation}

\subsection{State Equation}
With a linear formulation, the state equation which relates the compression of particles and \revisea{the} pressure \revisea{of fluid} in \revisea{a} classical SPH scheme can be written as

\begin{equation}
    p_i=\alpha_p\left(\frac{\rho_i}{\rho_0}-1\right),
\end{equation}
with $\alpha_p$ as a constant parameter \revisea{specifying the stiffness of the feedback}. In our numerical algorithm, we combine this equation of state \revisea{with} the \revisea{thin-film} pressure we derived in the previous section \revisea{to obtain}
\begin{equation}
    p=\kappa_h\gamma+\bm{\nabla}_s\cdot\bm{u}_s,
\end{equation}
which suggests that the fluid pressure inside the thin film is a function of \revisea{the} local curvature $\kappa_h$ and \revisea{velocity's divergence on surface}. With the analogy of density $\rho$ in \revise{the} classical SPH scheme and \revisea{the} numerical height $h$ in ours, we acquire a final formulation of \revisea{the} particle pressure as
\begin{equation}
    p_i=\alpha_h\left( \frac{h_i}{h_0}-1\right)+\alpha_k\gamma_i\left(\kappa_h\right)_i+\alpha_d\left(\bm{\nabla}_s\cdot\bm u_s\right)_i.
\end{equation}
Here $h_0$ is the rest thickness of thin films, and $\alpha_h,\alpha_k,\alpha_d$ are three constant parameters controlling the tangential incompressibility of thin-film system. The first term administrates \revisea{the} weakly compressibility to ensure that the particles are relatively evenly populated in the simulation domain to ensure numerical robustness, while the latter two terms \revise{achieve} the surface-tension-driven behaviors of the thickness evolution. \revisea{Specifically, the stiffness of our system is much lower than that of a standard SPH algorithm used for volumetric incompressible flow simulation. Therefore, the dynamics of particles are dominated by physical evolution instead of by the tendency to be evenly distributed.}

\begin{figure}[t]
\centering
    \begin{subfigure}[b]{0.495\linewidth}
    \centering
        \includegraphics[width=\linewidth]{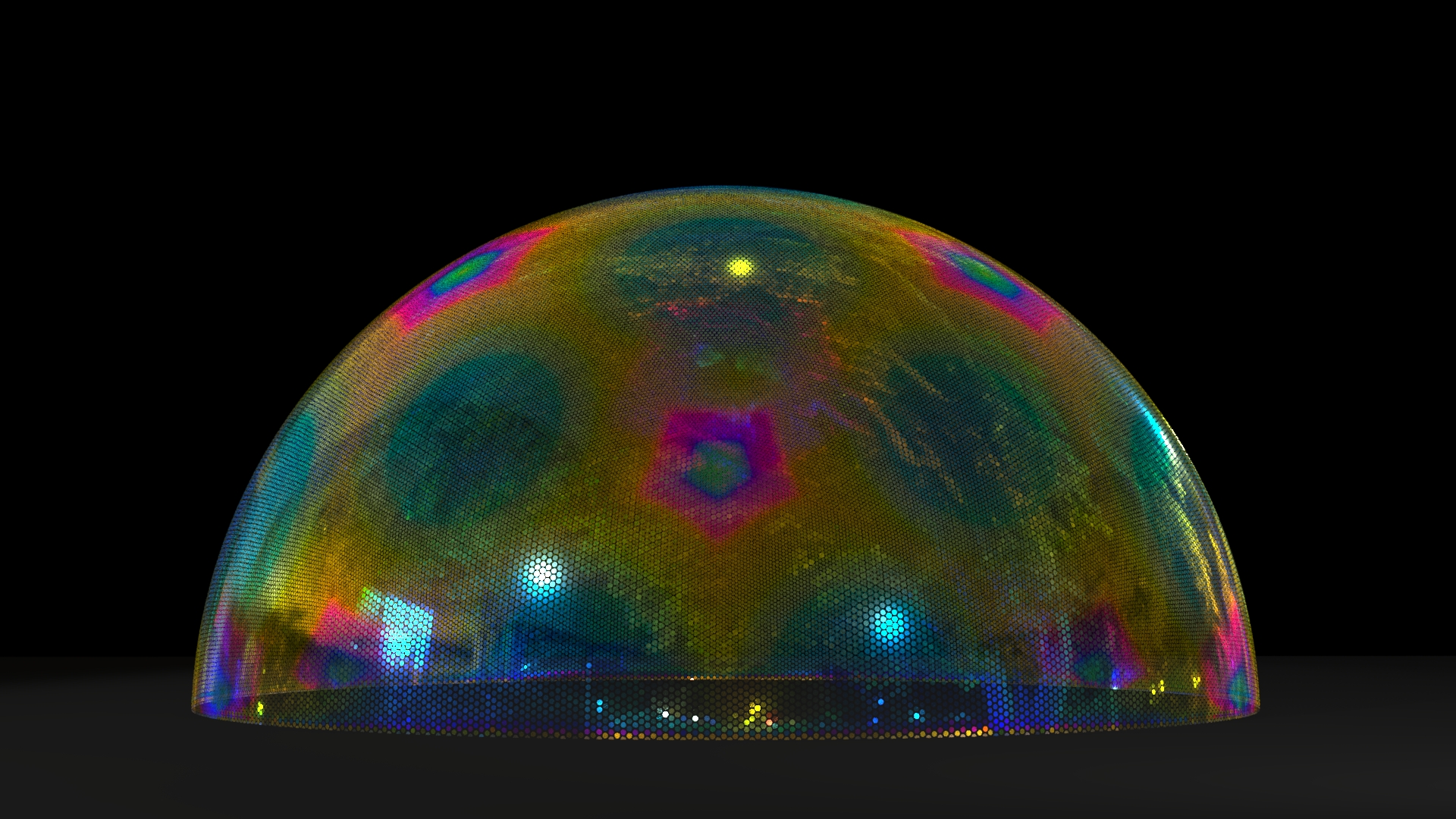}
    \end{subfigure}
    \hfill
    \begin{subfigure}[b]{0.495\linewidth}
    \centering
        \includegraphics[width=\linewidth]{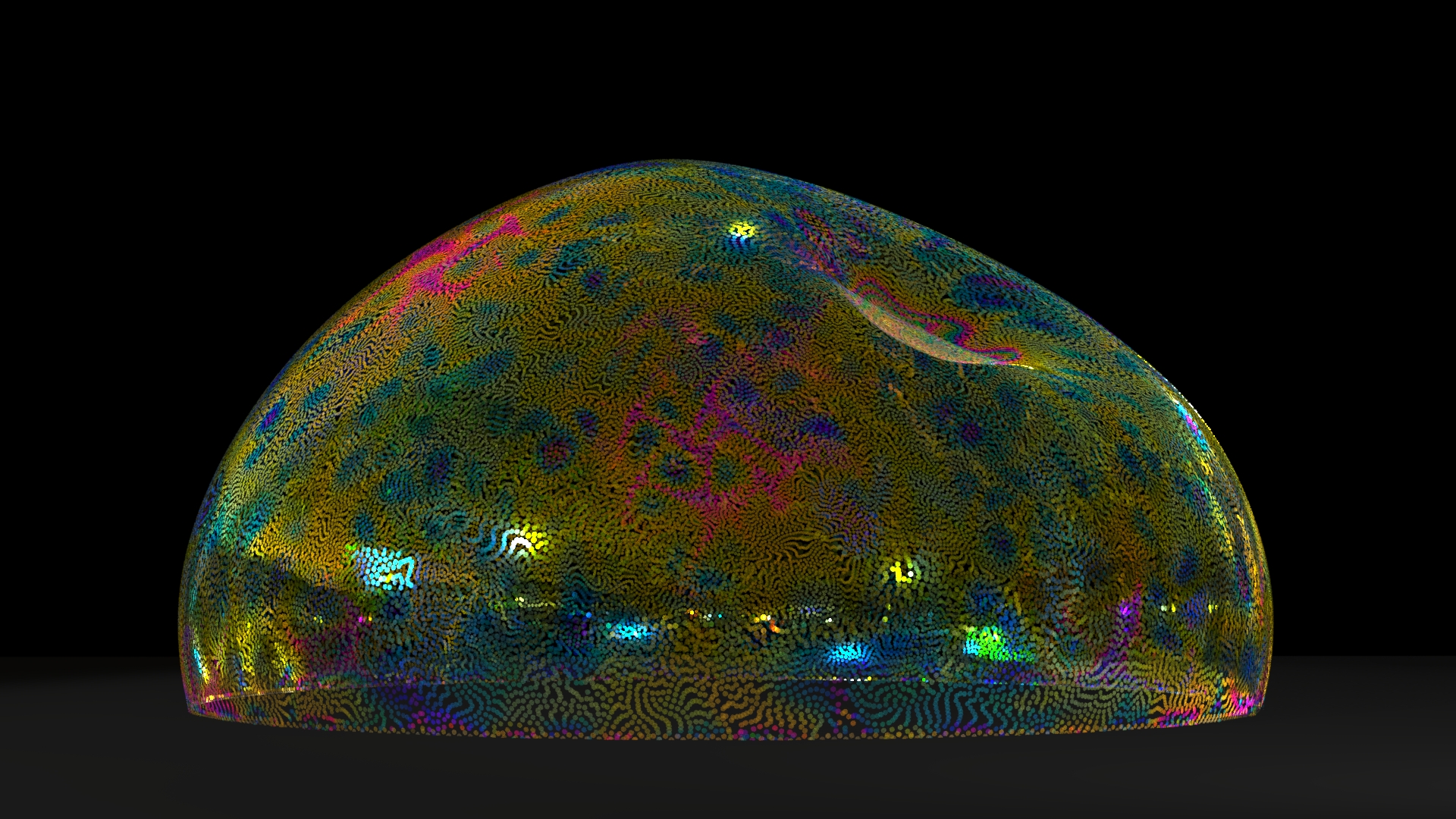}
    \end{subfigure}
    \vskip 0.1\baselineskip
    \begin{subfigure}[b]{0.495\linewidth}
    \centering
        \includegraphics[width=\linewidth]{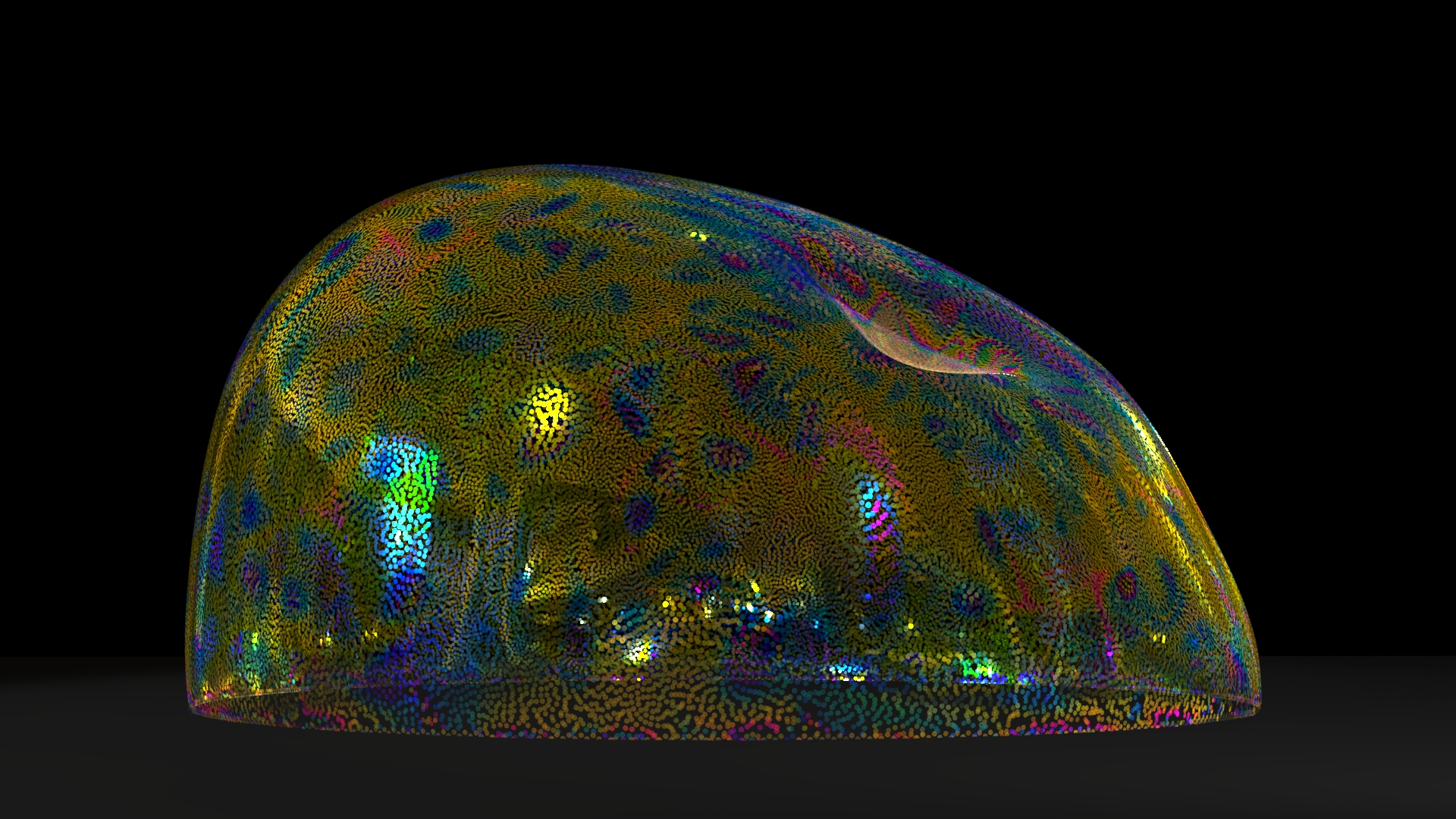}
    \end{subfigure}
    \hfill
    \begin{subfigure}[b]{0.495\linewidth}
    \centering
        \includegraphics[width=\linewidth]{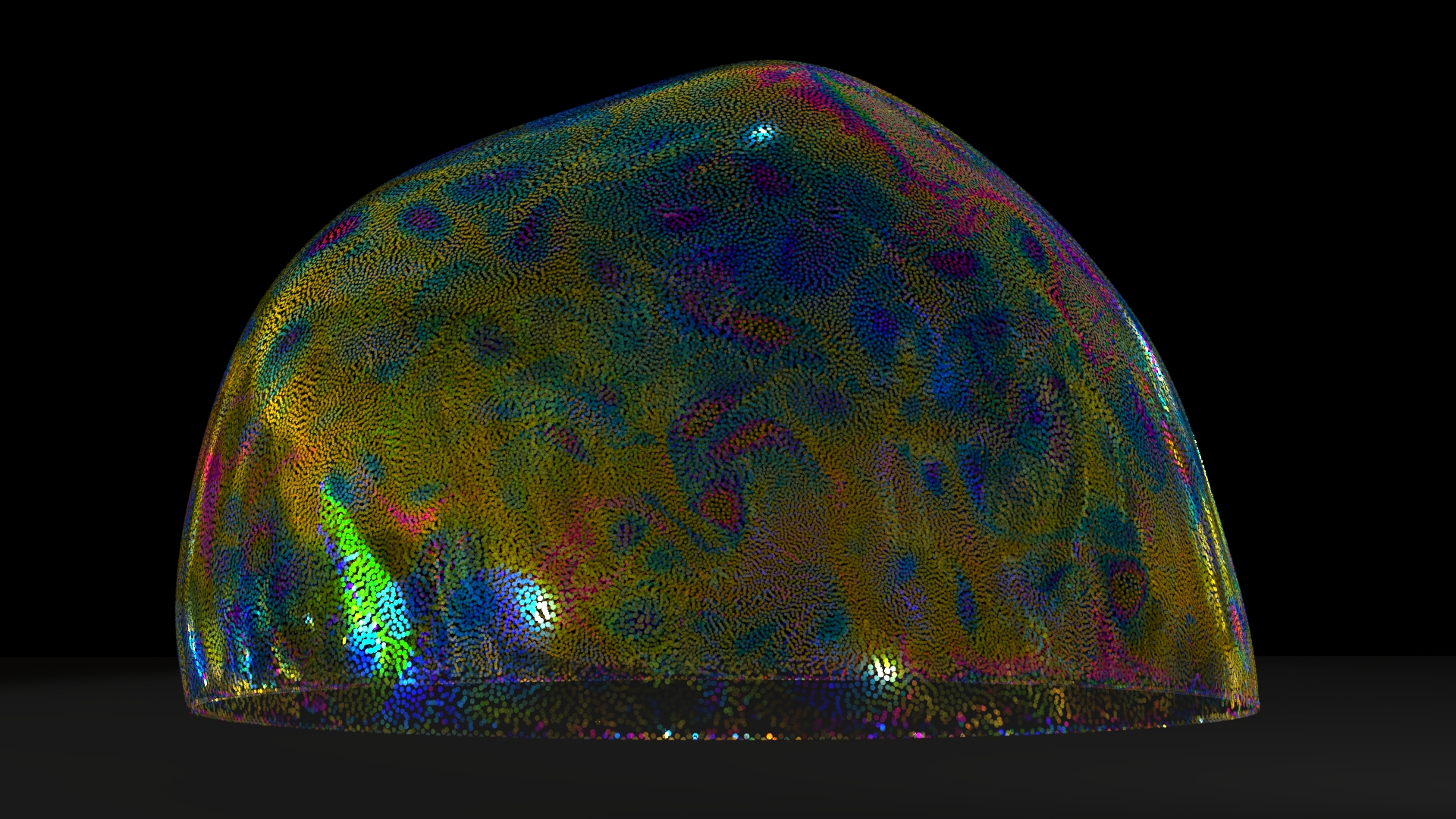}
    \end{subfigure}
\caption{Half bubble: The \revise{surface-tension-driven} oscillation for a half bubble standing on a plane. \revise{In} the beginning, a push toward the center is applied on the top right region of \revise{the} bubble. A vorticity confinement force is employed to enhance the rotational motion on the bubble surface.}
\label{fig:half-bubble}
\end{figure}

\begin{figure}[t]
    \centering
    \includegraphics[width=\linewidth]{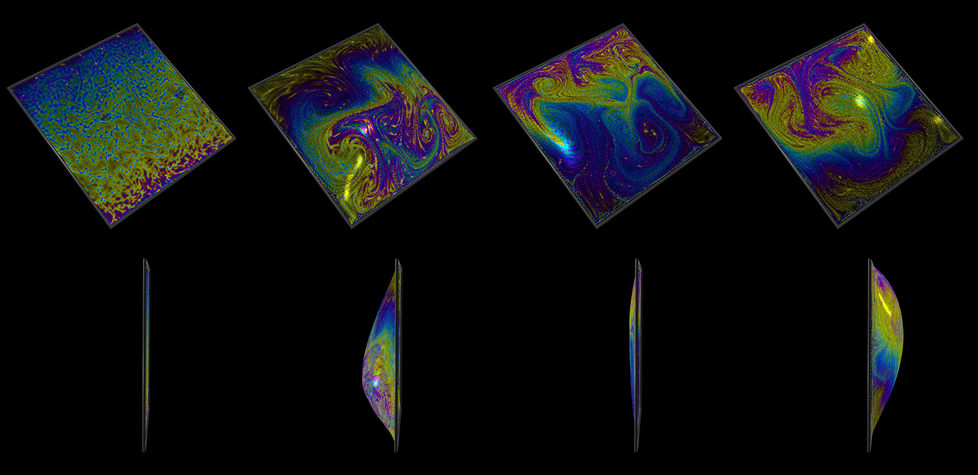}
    \caption{\revise{Thin-film} confined in a square: A gravitational pull is being rotated periodically around the surface, expediting the formation of appealing color patterns.}
   \label{fig:square-RT}
\end{figure}

\begin{figure*}[t]
    \centering
    \includegraphics[width=\textwidth]{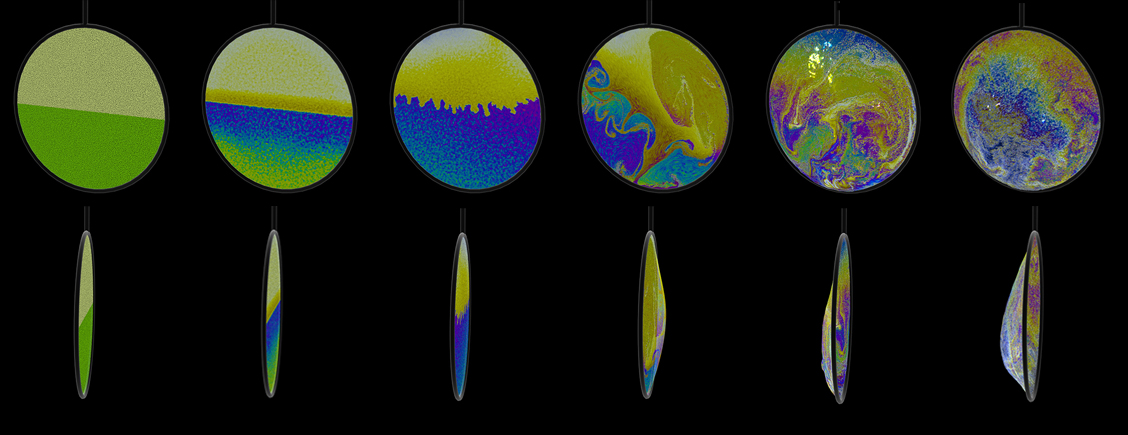}
    \caption{Rayleigh-Taylor instability simulated on a piece of dynamically deforming thin film. Upon initialization, the upper half of the thin film carries a larger density, and the lower half carries a larger volume. The interface \revise{evolves} into finger-like patterns, which further disintegrates into numerous delicate, colorful threads, making up the rich and realistic palette as seen in the last two subfigures.}
    \label{fig:circle-rt}
\end{figure*}

\subsection{\revisea{Force Discretization}}
\revisea{We discretize the fluid forces in \eqref{eq:ns_uhgamma} using our surface SPH operators \eqref{eq:sph-operators}.}
The mean curvature $\kappa_c = 0.5\bm{\nabla}_s^2 S_C$, where $S_C: z=z(x,y)$, is calculated \revisea{using} the surface \revise{Laplacian} operator

\begin{equation}
    2\kappa_c=\sum_j\frac{V_j}{h_j}\left(-\bm{r}_{ij}\cdot\bm{n}_i\right)\frac{2|\bm{\nabla}_s W_{ij}|}{|\bm{\widetilde{r}}_{ij}|}.
\end{equation}

We further introduce revisea{a} vorticity confinement force \cite{yoon-2009,selle-2005,zhu-2010} \revisea{to enhance the vortical motion on the bubble surface}.
We carry the vorticity $\zeta_i$ \revisea{on} every particle $i$, which is subject to a similar convection-diffusion equation of \revisea{the} surfactant concentration \eqref{eq:ns_gamma}. The vorticity confinement force $\bm{f}_i^{\zeta}$ takes the form
$
    \bm{f}_i^{\zeta}=-\sum_j\widetilde{\bm{r}}_{ij}\times\left(\zeta_j\bm{e}_z^j\right)
$.


\section{Time Integration}
Before computing the forces from the SPH discretization of \eqref{eq:ns_uhgamma}, we first calculate the surface tension coefficient $\gamma_i$ and the fluid pressure $p_i$ for each particle as shown in Algorithm \ref{alg:update_status}. After that, we compute all forces in Algorithm \ref{alg:compute_force}.
%
Following the position update, the local frames and metric tensors are reconstructed at every particle with the PCA-based method \cite{wang-2020}. Finally, the numerical height $h$ is updated with \eqref{eq:sph-h}.
\revisea{We summarize our time integration scheme in Algorithm \ref{alg:advance_step}.}

\begin{algorithm}
\caption{Compute Preliminary Variables for particle $i$}
\label{alg:update_status}
\begin{algorithmic}[1]
\State Update surface tension: $\gamma_i=\gamma_0-\gamma_a\Gamma_i$.
\For{each particle $j$ in the neighborhood of $i$}
\State Update divergence: $\left(\bm{\nabla}_s\cdot\bm u_s\right)_i=\sum_j\frac{V_j}{h_j}\widetilde{\bm{u}}_{ij}\cdot\bm{\nabla}_s W_{ij}$.
\State Update local curvature: $\left(\kappa_h\right)_i=\sum_j\frac{V_j}{h_j}\left(h_j-h_i\right)\frac{2|\bm{\nabla}_s W_{ij}|}{|\bm{\widetilde{r}}_{ij}|}$.
\EndFor{}
\State Update pressure: \Statex\qquad $p_i=\alpha_h\left( \frac{h_i}{h_0}-1\right)+\alpha_k\gamma_i\left(\kappa_h\right)_i+\alpha_d\left(\bm{\nabla}_s\cdot\bm u_s\right)_i$.
\end{algorithmic}
\end{algorithm}

\begin{algorithm}
\caption{Compute Forces for Particle $i$}
\label{alg:compute_force}
\begin{algorithmic}[1]
\State Compute external force: $\bm{f}_i^e=m_i\bm{g}+\bm{f}_{\textit{ext}}$.
\For{each particle $j$ in the neighborhood of $i$}
\State Compute vorticity confinement force: $\bm{f}_i^{\zeta}=-\sum_j\widetilde{\bm{r}}_{ij}\times\left(\zeta_j\bm{n}_j\right)$.
\State Compute pressure force with symmetric form: \Statex \qquad$\bm{f}_i^p=2V_i\sum_{j}h_i  V_j\left(\frac{p_i}{h_i^2}+\frac{p_j}{h_j^2}\right)\bm{\nabla}_s W_{ij}$.
\State Compute Marangoni force with difference form:
\Statex \qquad$\bm{f}_i^m=\frac{V_i}{h_i}\sum_j\frac{V_j}{h_j}\left(\gamma_j-\gamma_i\right)\bm{\nabla}_s W_{ij}$.
\State Compute capillary force:
\Statex \qquad$\bm{f}_i^c=\frac{\gamma_i V_i}{h_i}\bm{n}_i\sum_j\frac{V_j}{h_j}\left(-\bm{r}_{ij}\cdot\bm{n}_i\right)\frac{2|\bm{\nabla}_s W_{ij}|}{|\bm{\widetilde{r}}_{ij}|}$.
\State Compute viscosity force:
\Statex \qquad $\bm{u}_{ij}=\bm{u}_j-\bm{u}_i$,
\Statex \qquad$\bm{f}_i^v=V_i\mu\sum_j\frac{V_j}{h_j}\left(\bm{u}_{ij}-\left(\bm{u}_{ij}\cdot\bm{n}_i\right)\bm{n}_i\right)\frac{2|\bm{\nabla}_s W_{ij}|}{|\bm{\widetilde{r}}_{ij}|}$.
\EndFor{}
\end{algorithmic}
\end{algorithm}

\begin{algorithm}
\caption{Advance a Time Step}
\label{alg:advance_step}
\begin{algorithmic}[1]
\For{each particle $i$ }
\State Compute preliminary variables with Algorithm \ref{alg:update_status}.
\State Update advected height: $\hat{h}_i+=-\hat{h}_i\left(\bm{\nabla}_s\cdot\bm u_s\right)_i\Delta t$.
\State Diffuse surfactant concentration:
\Statex \qquad $\Gamma_i+=\alpha_c\Delta t\sum_j\frac{V_j}{h_j}\left(\Gamma_j-\Gamma_i\right)\frac{2|\bm{\nabla}_s W_{ij}|}{|\bm{\widetilde{r}}_{ij}|}$.
\State Compute forces with Algorithm \ref{alg:compute_force}.
\State Update velocity: $\bm{u}_i+=\frac{\Delta t}{m_i}\left(\bm{f}_i^e+\bm{f}_i^\zeta+\bm{f}_i^p+\bm{f}_i^m+\bm{f}_i^c+\bm{f}_i^v\right)$.
\State Update position: $\bm{r}_i+=\bm{u}_i\Delta t$.
\State Update local frame $\bm{e}^i$, normal $\bm{n}_i$ and metric tensor $g_i$.
\State Update numerical height: $h_i = \sum_{j}V_jWij$.
\EndFor{}
\end{algorithmic}
\end{algorithm}

\subsection{Codimension Transition}

\begin{figure}[ht]
\begin{center}
\resizebox{0.45\textwidth}{!}{%
\includegraphics[height=3cm]{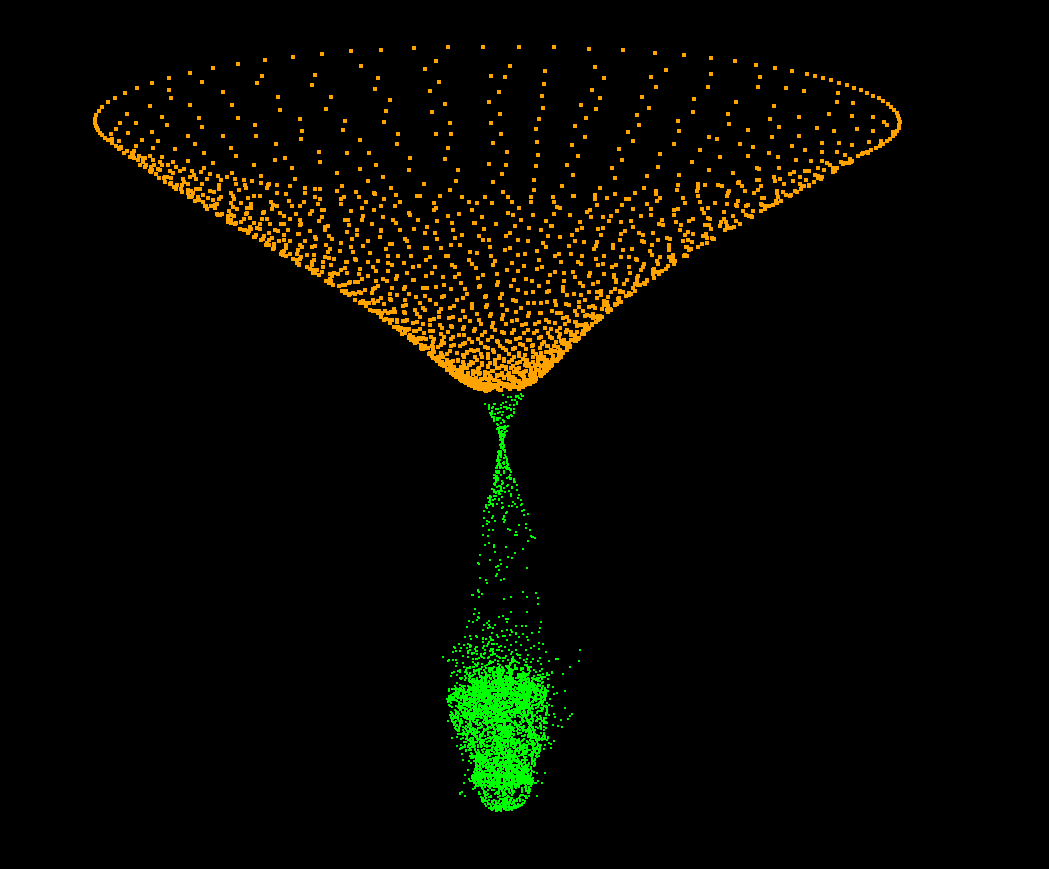}%
\quad
\includegraphics[height=3cm]{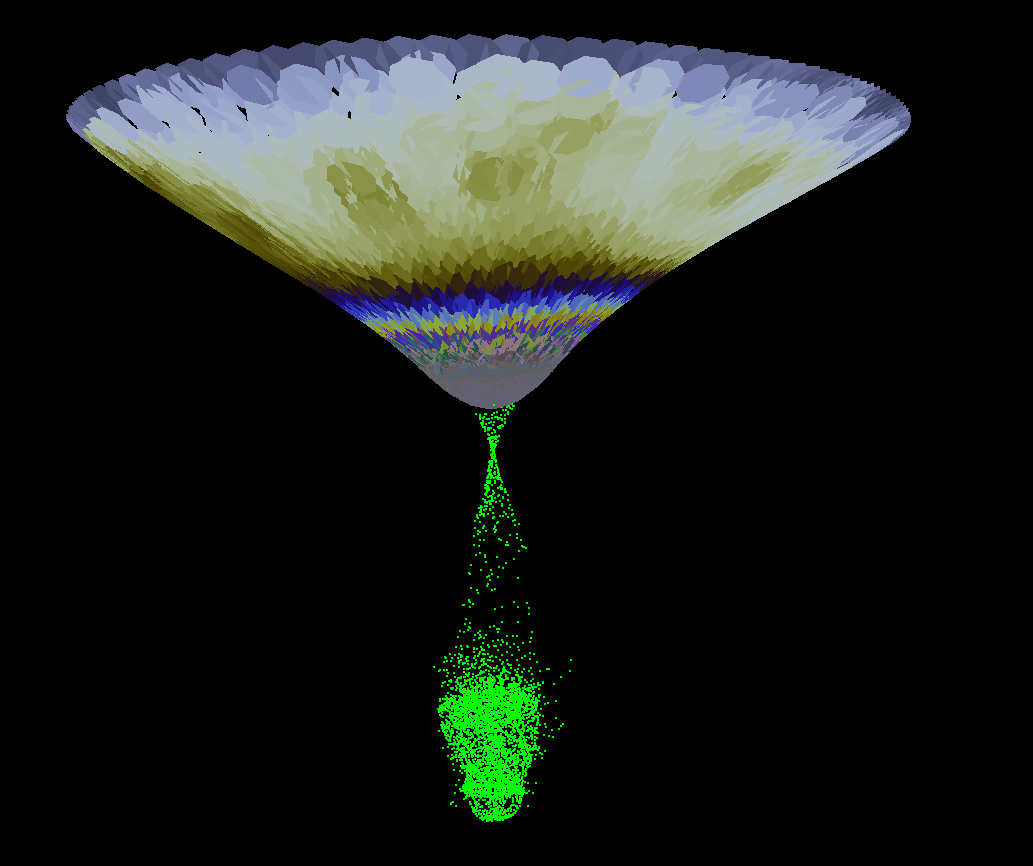}%
}
\end{center}
\caption{
\revisea{An example of codimension transition. In the left subfigure, codimension-1 particles are colored with orange and codimension-0 particles with green. In the right subfigure, the surface consisted of codimension-1 particles is also displayed. One can see that at the tip of the cone formed by the thin film, some particles are transiting to codimension-0 because they passed the threshold.}
}
\label{fig:color-code}
\end{figure}

Under certain circumstances, the codimension-1 thin film will shrink or rupture into volumetric droplets. We capture this phenomenon with the criteria that $\left(\kappa_c\right)_i>\kappa_0$ or $ n_i<2$.
Here $\kappa_0$ is a constant depending on the setup of the scene, and $n_i$ is the number of neighbors within the SPH kernel radius. The first criterion checks if the particle still lives on a relatively even surface, and the second checks if the particle is all alone. When \revisea{both} criteria are met, particle $i$ is deleted from our surface SPH solver and transferred to a volumetric SPH solver \cite{zhu-2010,rim-akinci-2013,rim-yang-2017}. 
%
Conversely, for a volumetric particle $i$, if there is a codimension-1 particle $j$ nearby, and $\left(\bm{r}_j-\bm{r}_i\right)\cdot\bm{n}_j<\textit{threshold}$, particle $j$ is deleted from 3D SPH solver, and transported to our codimension-1 surface SPH solver.
An example of \revisea{codimension-1-to-0 transition is illustrated in Figure \ref{fig:color-code}. We can see that green-colored volumetric particles are detached from the orange-colored codimension-1 thin film, at where the thin film shows a sharp corner. Under surface tension, the bulk fluid tends to form a spherical drop.}

\subsection{Additional Air Pressure Difference}

For a closed bubble, the air pressure differs between the interface, as suggested by the Young-Laplace equation, which tends to contract the bubble into a singularity. Thus an additional force is needed to retain its shape. We calculate the pressure inside the bubble $p_b$ with the ideal gas equation $p_b V_b=\alpha_p$, where $\alpha_p=nRT$ is a constant number. The volume of closed bubble $V_b$ is given by a sum with pyramid volume formula $V_b=\sum_i \frac{1}{3}\frac{V_i}{h_i}\left(\bm{n}_i\cdot \bm{x}_i\right)$, with $\bm{x}_i$ the vector pointing from the gravity center of all particles on the bubble to particle $i$. Afterwards, an external force $\bm{f}_{\textit{ext}}=\frac{p_a-p_b}{2h}\bm{n}$ is added to all particles.

\subsection{Particle Reseeding}

During the evolution of soap films, the total area of films may be enlarged from the boundary. For example, a common way to produce bubbles is to blow air into a circle soap film on a ring soaked with soap water, and as the bubble forms, new portions of soap film are replenished from the ring.

Our algorithm includes a reseeding step for every iteration to implement this feature. At the end of every time step, every fixed boundary particle $i$ is checked with its nearest non-boundary neighbor particle $j$, and if $\bm{r}_{ij}>\textit{threshold}$, a new particle $k$ is reseeded at $\bm{r}_k=\frac{\bm{r}_i+\bm{r}_j}{2}+\bm{\sigma}$, with a random perturbation $\bm\sigma$. \revisea{The perturbation term alleviates artifacts emerging from regular patterns in the distribution of newly-generated particles.} The mass of \revisea{a} new particle $k$ inherits from $m_k=m_j$, and $\bm{u}_k=\frac{\bm{u}_j}{2}$. \revisea{While adding particles to SPH solvers can undermine the physical consistency and cause visible artifacts if not treated carefully, we note that in our case, this reseeding step only takes place near the boundary and is viewed as the physical process of fluid flowing from the boundary, so no mass or momentum redistribution is performed.}


\setlength{\textfloatsep}{20pt}
\begin{figure}[t]
 \centering
 \includegraphics[width=.45\textwidth]{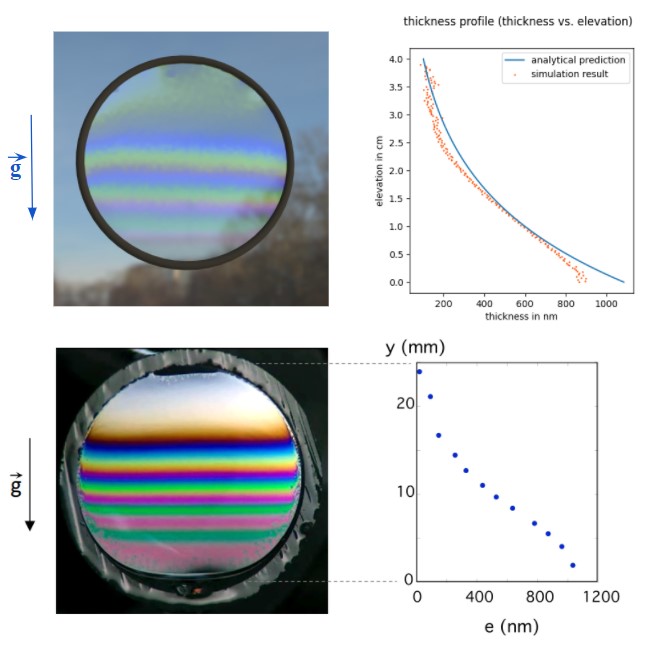}
 \caption{The Newton Interference pattern. The upper two pictures are from our simulation and the lower two pictures shows a real-world experiment \cite{gaulon2017sound}. \revisea{In the right column, elevation(horizontal) vs.\ thickness (vertical) graphs are displayed.} Compared \revisea{with the experimental result}, our simulation can generate visually realistic color fringes. As shown in the \revisea{right} figures, the thickness profile \revisea{generated} by our simulation matches the one measured in \revise{the} real \revisea{experiment}. Overlaid in the top-right figure is the analytical profile \revisea{(blue curve)} predicted according to Couder et al. \shortcite{physics-couder-1989}; the proximity of it with our simulation results \revisea{(yellow dots)} further confirms our simulation result.}
 \label{fig:newton-ring}
\end{figure}

\subsection{Boundary Treatment}
\begin{wrapfigure}{r}{0.26\textwidth}
  \begin{center}
    \includegraphics[width=0.25\textwidth]{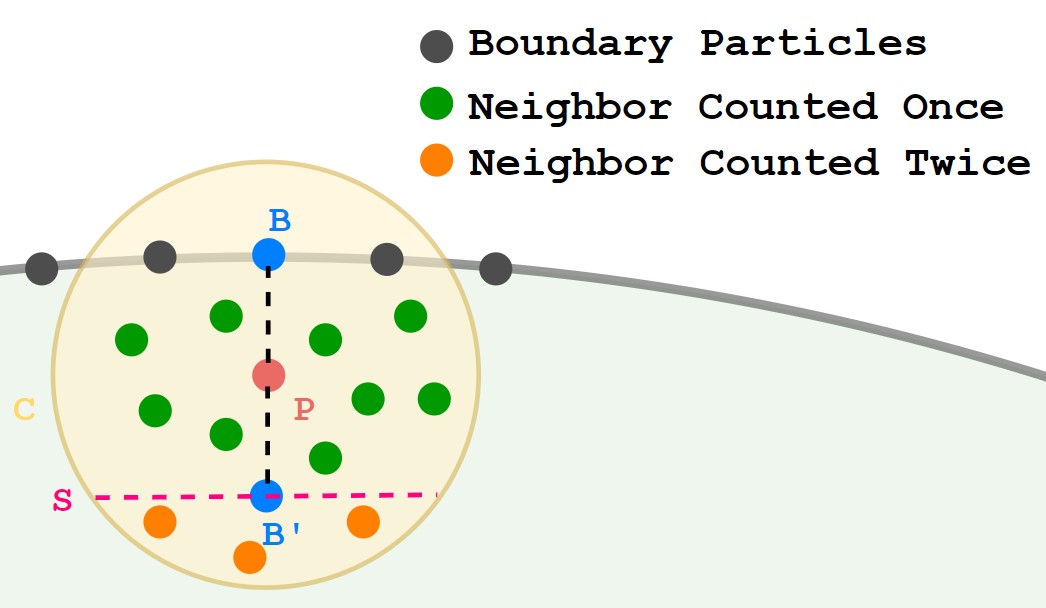}
  \end{center}
  \caption{\revisea{Illustration of the compensation method to treat boundary.}}
  \label{fig:mirrorcomp}
\end{wrapfigure}

For our various examples that involve interaction with circular or square solid rims, it is necessary to devise appropriate boundary handling methods. The task is twofold: First, the particles should not penetrate the solid boundary. Second, a particle should behave consistently when its neighborhood is under-sampled near the boundary.
%
%
For the first task, we correct the velocities of particles \revisea{that} are going to penetrate \revisea{the solid boundary defined by an implicit boundary (e.g., the solid ring in the Newton ring example)}. For the second task, we adopt a geometric compensation method illustrated in Figure \ref{fig:mirrorcomp}.
For a particle $P$ with its neighborhood $C$ truncated by the boundary, we find the nearest neighboring particle $B$ that is on the boundary, take its mirrored point with respect to particle $P$ as $B^\prime$, and draw the secant $S$ across $B^\prime$ perpendicular to $B^\prime-P$.
All neighboring points sampled within the circular segment defined by $C$ and $S$ are counted twice. If the boundary is a straight line, this method would precisely “clone-stamp” a part of the sampled region to fill up the empty region. Otherwise, the cloned region would not fill the empty region perfectly, then we estimate the unsampled area and adjust the compensation accordingly.

\revisea{We adopt this geometric compensation method in place of typical boundary particle method for two reasons: first, the relative positions of the boundary particles to the fluid particles must change as the surface deforms, which adds complication to the traditional ways with multiple layers of boundary particles.}
\revisea{Secondly, we encourage the spatial variance of particle density with low stiffness, so having boundary particles with fixed volume or mass would not do justice. For example, if the particles congregate along the border, we would want the area near the border to rise in height/density, in which case this adaptive clone-stamp strategy is useful as it preserves the state of the fluid. This mirroring approach to enable large density variation is inspired by Keiser et al. \shortcite{keiser2006multiresolution}; the dynamic and adaptive nature is also similar to the Ghost SPH approach \citep{schechter2012ghost}.}

\begin{figure*}[t]
\centering
    \begin{subfigure}[b]{0.245\linewidth}
    \centering
        \includegraphics[width=\linewidth]{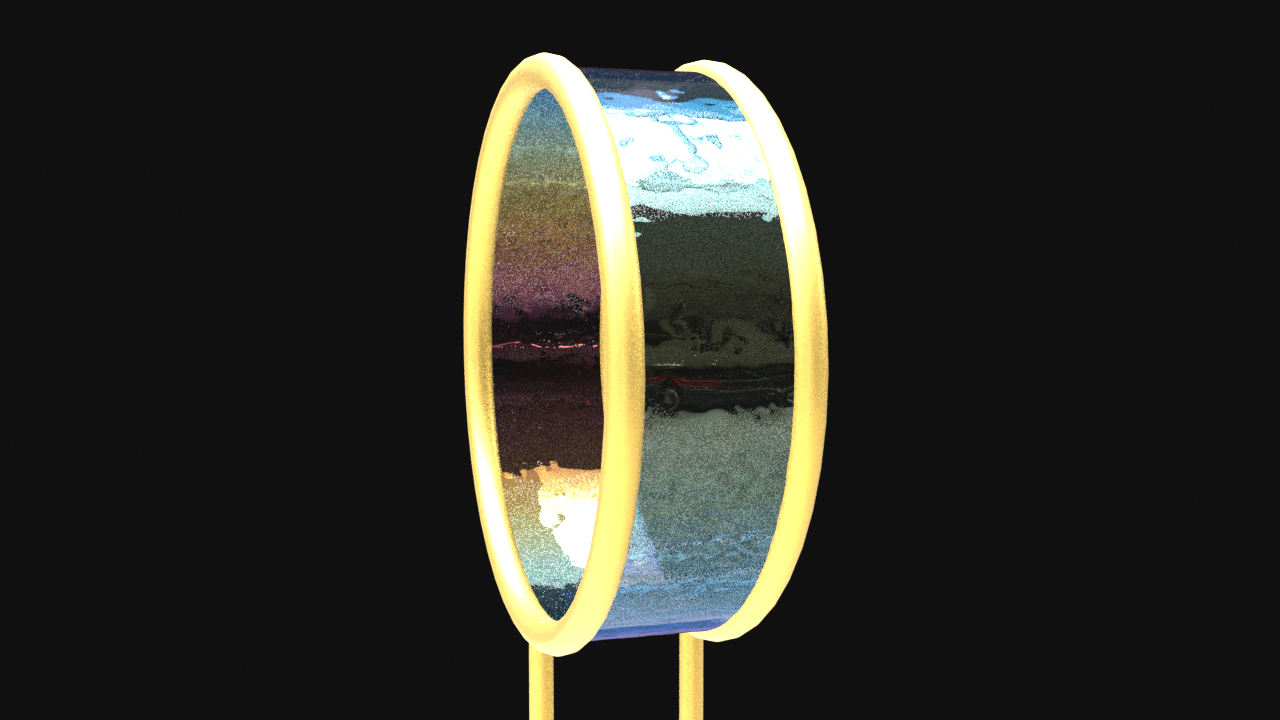}
    \end{subfigure}
    \hfill
    \begin{subfigure}[b]{0.245\linewidth}
    \centering
        \includegraphics[width=\linewidth]{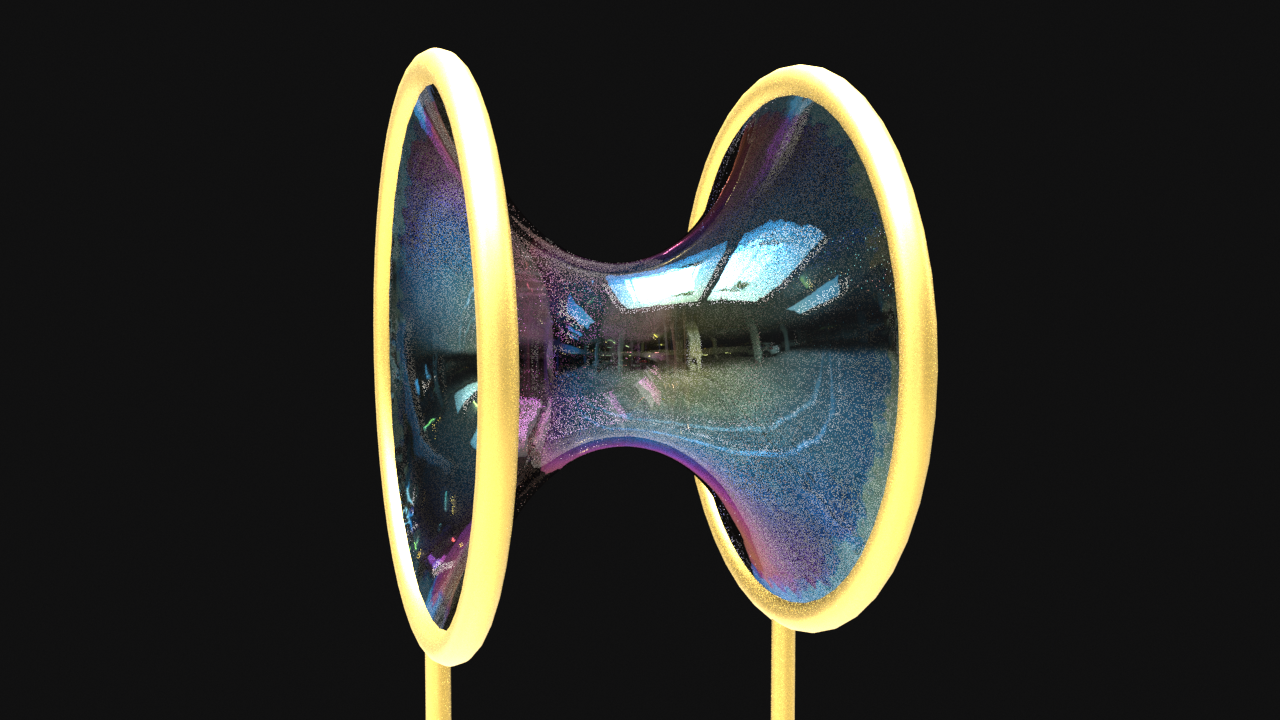}
    \end{subfigure}
    \hfill
    \begin{subfigure}[b]{0.245\linewidth}
    \centering
        \includegraphics[width=\linewidth]{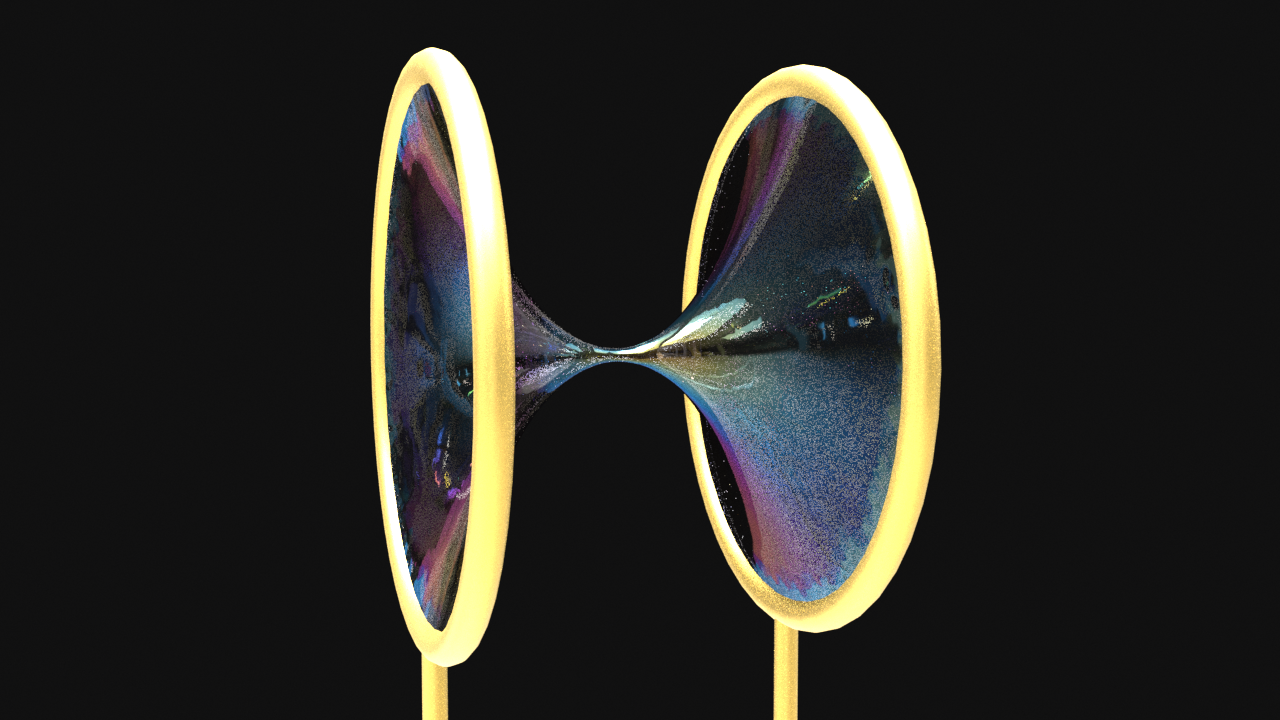}
    \end{subfigure}
    \hfill
    \begin{subfigure}[b]{0.245\linewidth}
    \centering
        \includegraphics[width=\linewidth]{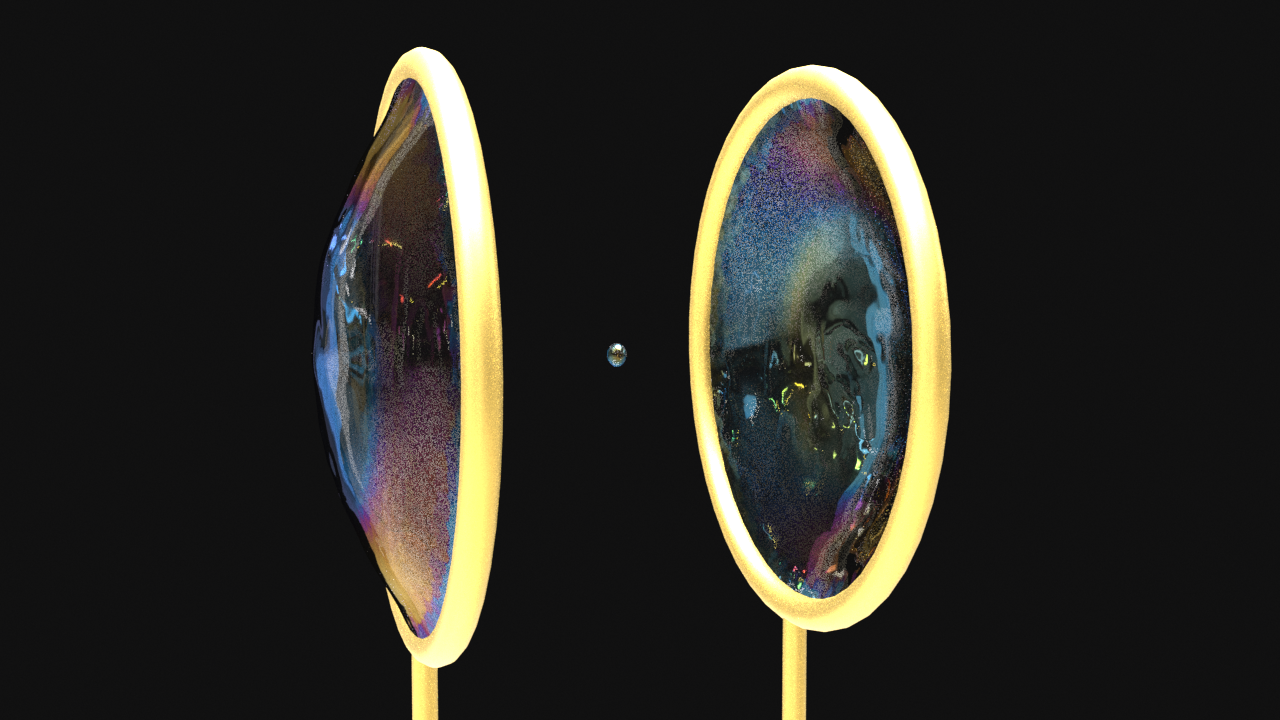}
    \end{subfigure}
\caption{Catenoid: Two parallel, circular rims are connected via thin film, under surface tension, the thin film would contract in the center and converge to the minimal surface of the catenoid. After the rim separation \revise{surpasses} the Laplace limit, the catenoid can no longer be sustained by the thin film, which will eventually be separated and pinch off a small droplet.}
\label{fig:catenoid-bubble}
\end{figure*}

\begin{figure}[t]
\hspace{-0.1in}
\centering
    \supermagicfigure{.155\textwidth}{.4\textwidth}{0in}{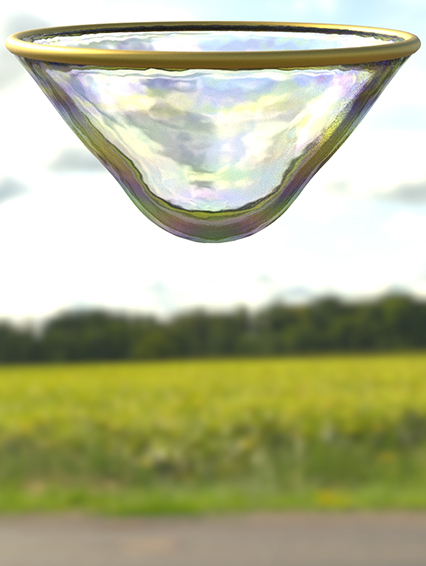}{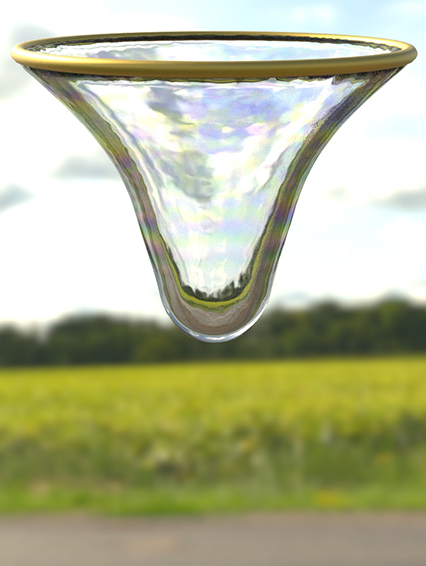}{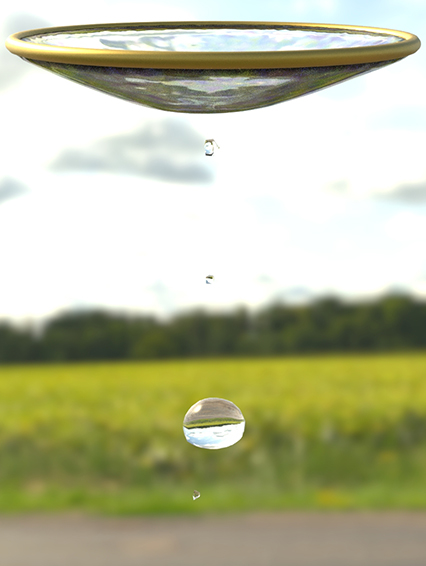}{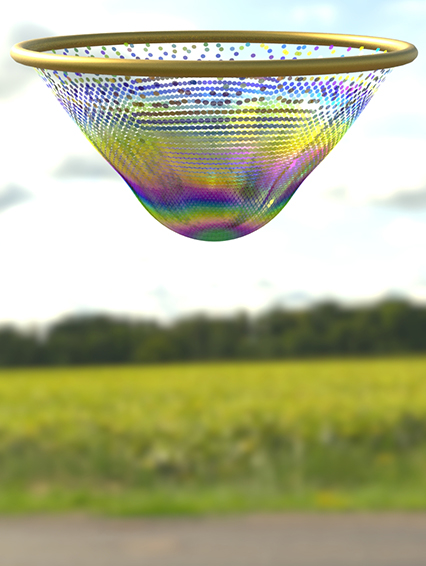}{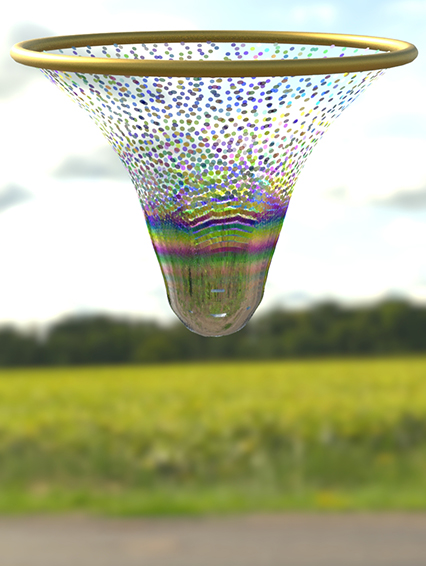}{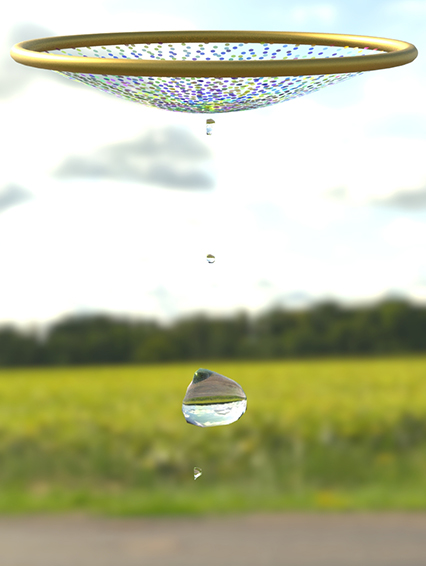}
    \caption{Thin-film dripping, surface particles at the tip transform into volumetric particles, detaching from the thin film and falls down as spherical drips. The upper and lower subfigures depict the same simulation sequence rendered as mesh and a collection of circles, respectively.}
    \label{fig:dripping}
\end{figure}

\section{Thin-Film Visualization}
The color in all of our simulations is computed with the \revise{CIE Standard} Illuminant D65, a standard illuminant that represents the average, open-air illumination under daylight. Given the discretized spectral power distribution, we compute the \revisea{reflection, refraction}, and interference for the individual wavelengths as a function of the thin film thickness along with the refractive indices of the external medium and the fluid. 
As light illuminates a thin film, there exists a difference in distance \revise{traveled} between the reflected lightwaves and the refracted lightwaves. The shift of phase can cause them to interfere constructively or destructively travel, with the exactitude dependent on the exact thickness of the film. The resultant light intensities for all wavelengths will be converted to a single RGB color value by integrating with the CIE matching functions. There exists various open-source software that performs such tasks; we rely on ColorPy \cite{kness-2008} in our implementation.
We utilize OpenGL for generating experimental visualizations and Houdini for rendering photorealistic images and videos. In Houdini, we render the particles both as circles and as mesh. Circles can highlight the intricate flow patterns, and the mesh is used for better realism. The physically-based spectral color calculated in the abovementioned way will serve as the base color of the particles, and additional environmental lighting will be on top to conform it with the rest of the environment.

\section{Results}

\paragraph{Thickness Profile}

Here a piece of \revisea{a planar thin film} is confined in a vertically positioned rim, and the gravitational force pulls the fluid volume to amass at the bottom, resulting in a height field that decreases with the elevation. The vertically varying thickness would result in \revisea{a set of} horizontal color stripes known as the Newton interference pattern. The thickness profile $\eta=\eta(z)$ can be approximated by an exponential function
$
    \eta(z) = \eta^*exp\left((\rho g \eta^*z)/(2\left(\sigma_0-\sigma^*\right)\right))
$
derived by Couder et al. \shortcite{physics-couder-1989}, where $\eta^*$ and $\sigma^*$ represents the surface's equilibrium thickness and surface tension coefficient when horizontally placed. Our experiments \revisea{use} $\eta^* = 387.875 nm$ and $\sigma^* = 71.996 \cdot 10^{-6} N/m$. Since our surface simulation is allowed to be compressible in nature, it handles the volume aggregation at the bottom automatically. As displayed in Figure \ref{fig:newton-ring}, both the visual result of Newton interference patterns and the thickness profile matches the real-world data in trend.

\paragraph{Surface Flow on Oscillating Bubbles}
As shown in the upper part of Figure \ref{fig:oscillation}, we initialize the simulation bubble as a perfect sphere. In the first $1.0$ second, we apply an impulse towards the center at the north and south poles. Meanwhile, \revisea{one hundred} in-plane vortices are seeded randomly. The induced capillary wave by the initial perturbation propagates through the bubble, causing it to deform and oscillate.
The height field is evolved by the surface deformation and vortical flows jointly, forming vivid color patterns. Our method is independent of the initialization method and can be easily extended to bubbles with custom shapes. The lower part of the same figure shows a simulation initialized on an irregular mesh and without the initial impulse. Furthermore, as shown in Figure \ref{fig:half-bubble}, the simulation can also be extended to a half bubble standing on a plane. The initial impulse in Figure \ref{fig:half-bubble} is applied at the top right region at the bubble instead of the north-pole.

\begin{figure*}[t]
\centering
    \begin{subfigure}[b]{0.245\linewidth}
    \centering
        \includegraphics[width=\linewidth]{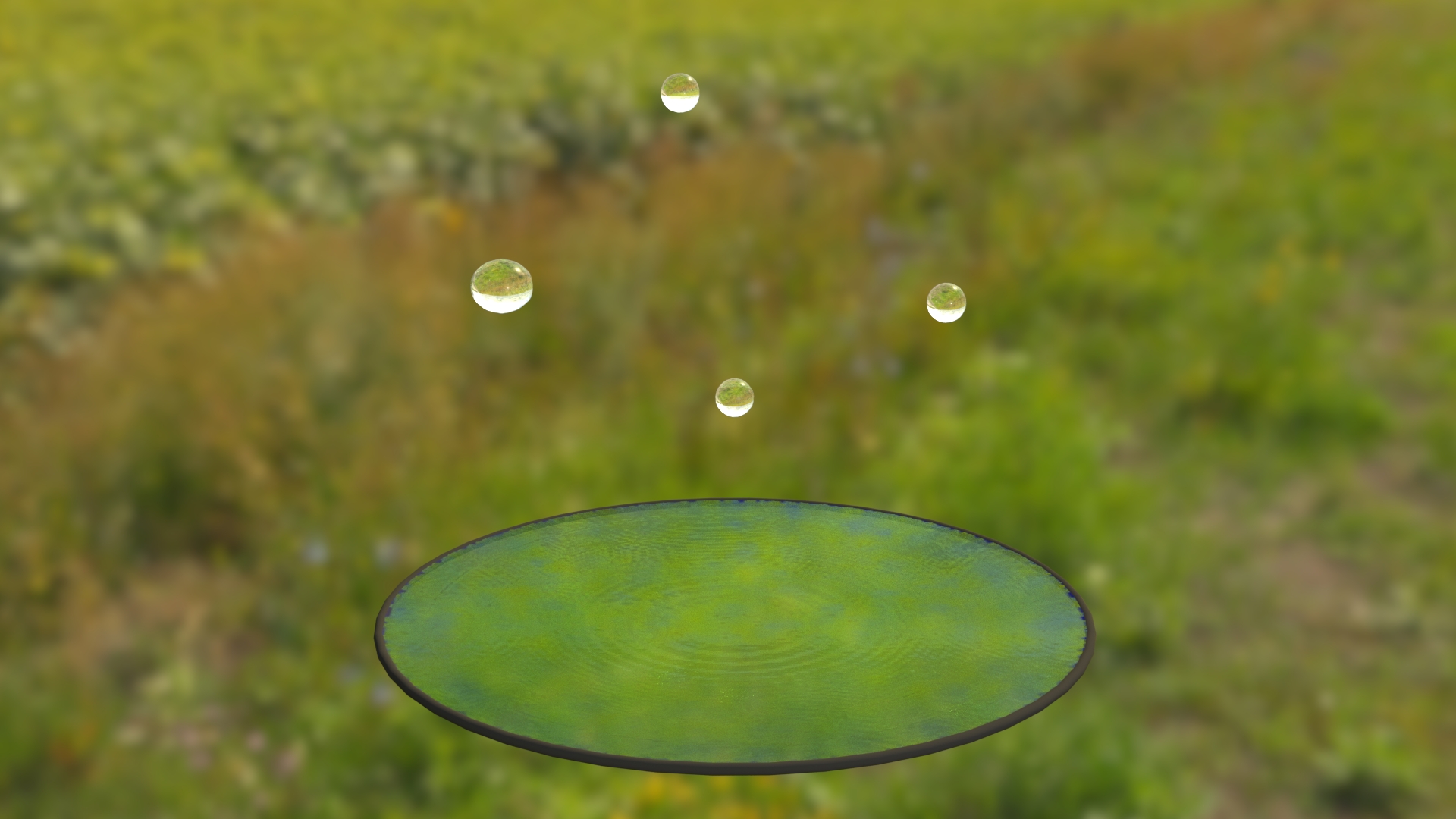}
    \end{subfigure}
    \hfill
    \begin{subfigure}[b]{0.245\linewidth}
    \centering
        \includegraphics[width=\linewidth]{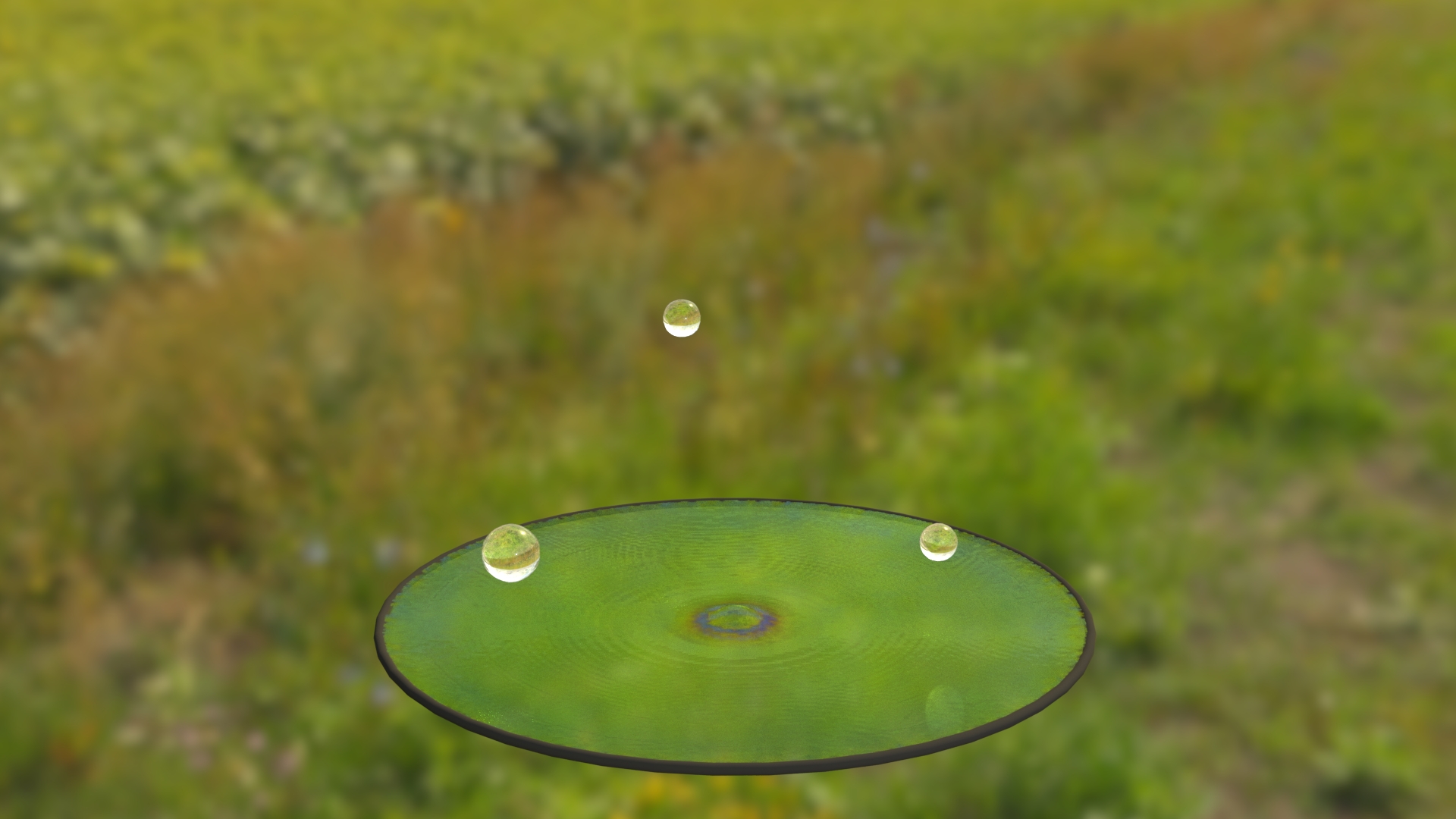}
    \end{subfigure}
    \hfill
    \begin{subfigure}[b]{0.245\linewidth}
    \centering
        \includegraphics[width=\linewidth]{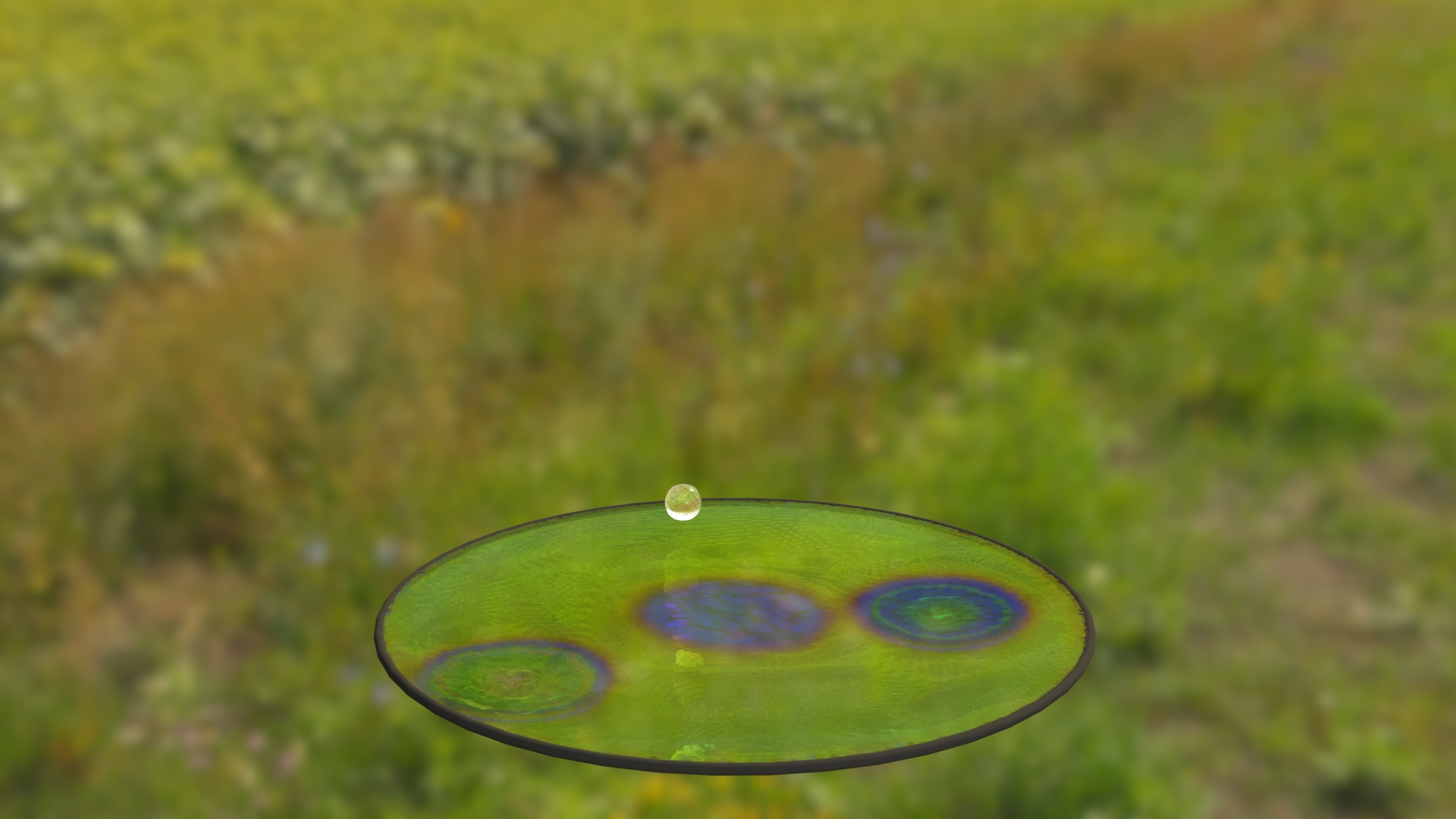}
    \end{subfigure}
    \hfill
    \begin{subfigure}[b]{0.245\linewidth}
    \centering
        \includegraphics[width=\linewidth]{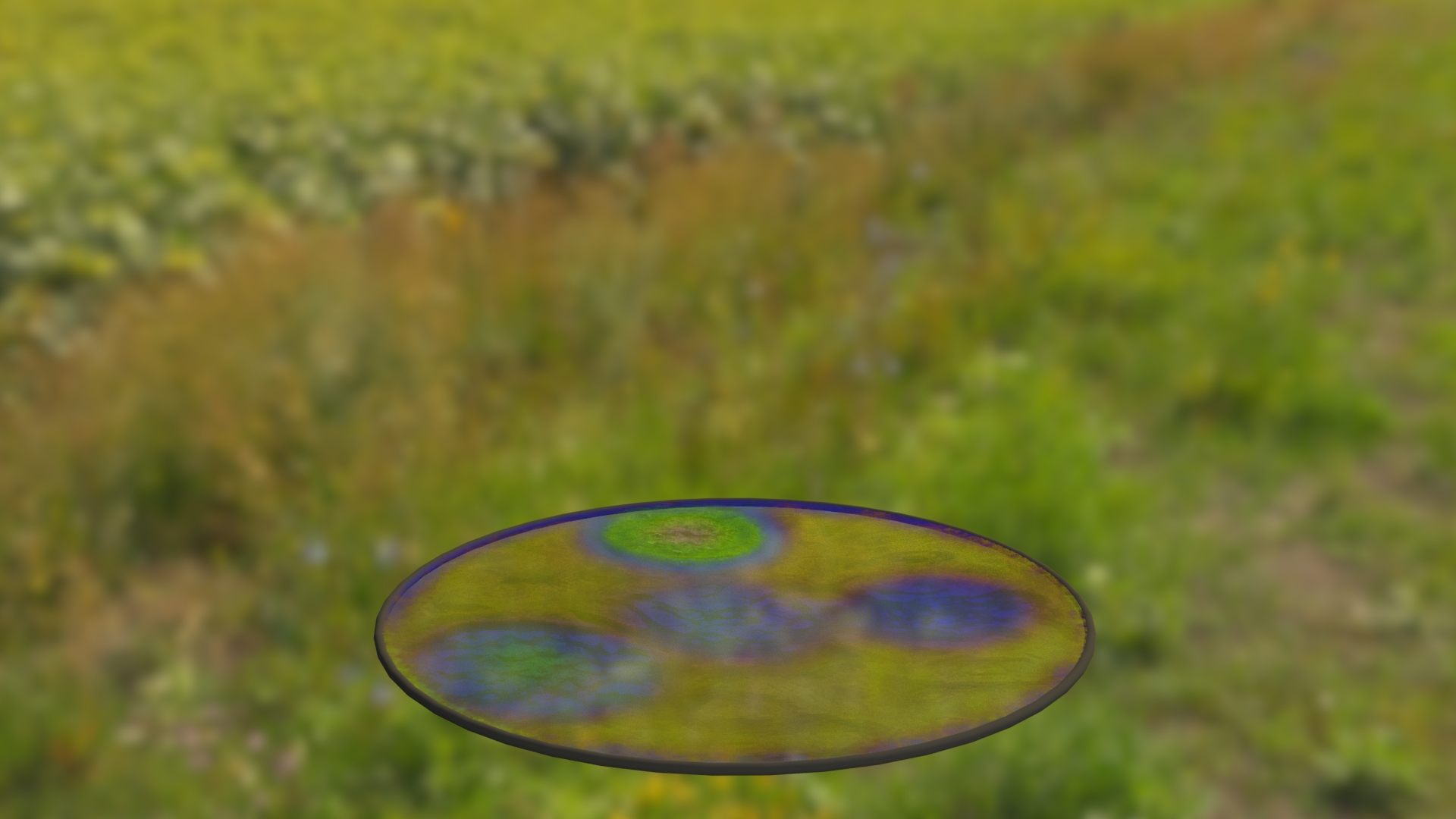}
    \end{subfigure}
     \vskip 0.3\baselineskip
     \begin{subfigure}[b]{0.245\linewidth}
    \centering
        \includegraphics[width=\linewidth]{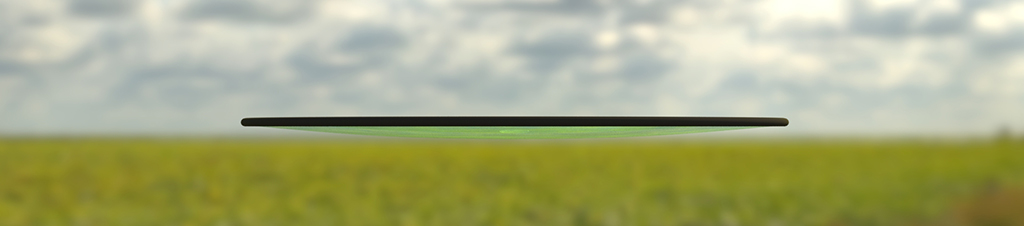}
    \end{subfigure}
    \hfill
    \begin{subfigure}[b]{0.245\linewidth}
    \centering
        \includegraphics[width=\linewidth]{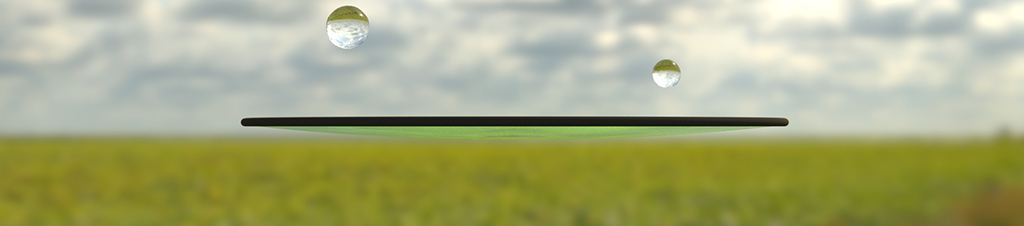}
    \end{subfigure}
    \hfill
    \begin{subfigure}[b]{0.245\linewidth}
    \centering
        \includegraphics[width=\linewidth]{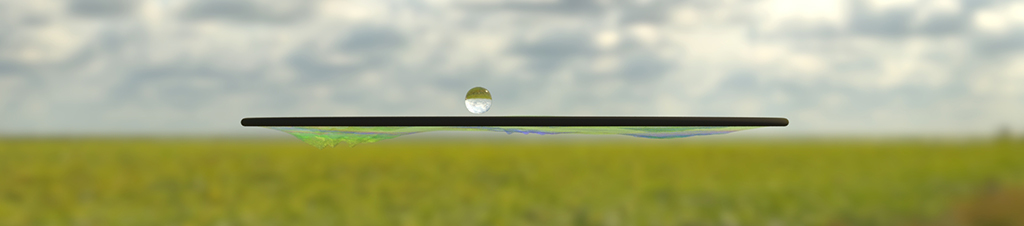}
    \end{subfigure}
    \hfill
    \begin{subfigure}[b]{0.245\linewidth}
    \centering
        \includegraphics[width=\linewidth]{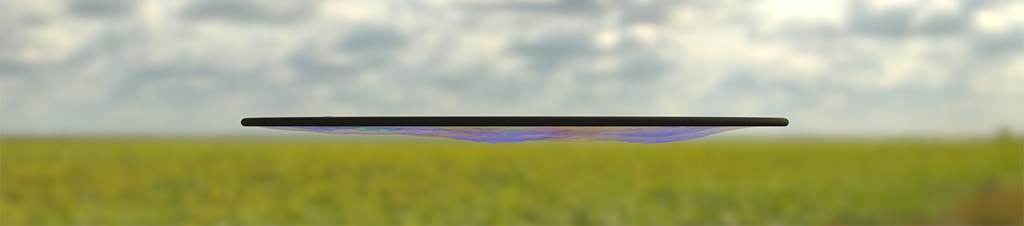}
    \end{subfigure}
\caption{Droplet Marangoni Effect: Four drops of soap solution is \revise{deposited onto} the thin film. After the soap solutions is absorbed by the surface, the surface tension of the the impacted area would be lowered, which enables more high-frequency deformation to be observed. Moreover, the droplets induce local non-zero gradients of surface tension, causing the particles to move towards the high surface tension region, which is known as the Marangoni effect. The color change in the surface would reflect such movements.}
\label{fig:marangoni-bubble}
\end{figure*}

\paragraph{Bubble Rupture}
In the example shown in Figure \ref{fig:bubble-rupture}, we burst a bubble into filaments and droplets. We first manually mark every particle in a small region as "punctured", and give it an inward push, imitating the bubble to be poked by, for example, a human finger. Then, the surface near that region is bent inward by the momentum of the initial impulse, causing more and more codimension-1 particles \revisea{to meet the criteria of codimension transition and hence rupture} into volumetric fluid particles. These particles are then taken over by the 3D SPH solver, morphing them into numerous thin filaments and tiny droplets.
The mechanism for bubble \revisea{bursting} is general to the bubble's shape and dynamics. The shown example directly follows the bubble oscillation example depicted in \revise{the} upper part of Figure \ref{fig:oscillation}. The rupture animation is displayed at $0.1\times$ speed of the oscillation animation.
A feature that separates our animation from the existing ones is that our method models the change in the bubble's hue due to surface contraction. As one can see in Figure \ref{fig:bubble-rupture}, the bubble's color changes from green to purple, providing an extra layer of richness than if such phenomenon is \revise{modeled} by a volumetric SPH solver only.

\paragraph{Large-Deforming Thin Films with Rims}
In Figure \ref{fig:square-RT}, a thin film is confined in a square rim, and we apply a large gravitational pull whose direction circulates periodically around the surface. On the surface, we seed 20 random vortices and perform a slight perturbation to the volume of particle $V_i$ as well as density $\rho_i$ with \revise{Perlin} noise, which reflects in the initial color pattern. The rotating force causes the particles to perform in-plane and out-of-plane motions alternatively, forming appealing color patterns and surface morphology. In Figure \ref{fig:circle-rt}, we use a circular rim to contain a thin film and recreates on it the R-T instability, \revise{a beautiful, classic flow pattern that can be captured by SPH method \cite{extra-solenthaler-2008}}. Unlike the oscillating bubble example, where we desire the bouncy feeling, here, we set the surface tension coefficient low to give the film a soft and mellow feeling, with plenty of high-frequency motions. In addition to the downward gravity, \revise{we apply a large external force $\bm{F}_{ext} = \beta \left[(\cos\alpha, \sin\alpha, 0) + (0, \cos2\alpha, \sin2\alpha)\right]$,
in which $\beta$ is the scalar of force strength and $\alpha = 2\pi t/T$ where $t$ is the simulation time and $T$ is the period.}
We initialize the film so that the upper half has a density \revise{$4$ times as large}, and the lower half has a \revise{height 1.5 times as large}, reflected in the color difference. Under gravity, the heavier top moves down, and the height difference tends the bottom to move up, countering each other and evolve into various finger-like patterns. Besides the flow on the surface,  the aggressive deformation of the surface itself is also reflected by the color. For example, the stretched parts are visibly darker than the rest of the thin film.

\paragraph{Thin-film Dripping with Circular Rim}
As shown in Figure \ref{fig:dripping}, we initialize the thin film as a \revise{half-sphere}. Particles under gravity tend to fall down, and under surface tension tend to contract. When the tip of the thin film contracts, some particles transform from codimension-1 to codimension-0, separating from others and drops down, forming droplets under the surface tension modeled by our auxiliary volumetric SPH solver.

\paragraph{Catenoid}
Consider two rims connected by a thin soap film are pulled away from each other. The thin film will try to shrink under the area minimizing tendency, forming a catenoid,  as have been validated numerically in Appendix \ref{appendix:validation}. Figure \ref{fig:catenoid-bubble} shows the rendered result our simulation. After the separation between the two rims exceeds the Laplace limit, no more catenoids can sustain. Then, the membrane would further contract and collapse into a droplet in the center.
As in the figure, our system can accurately simulate the \revise{surface-tension-driven} deformation while considering the color changes due to the surface elongation.

\paragraph{Droplet Marangoni Effect}
In the previous examples, we showcase the transition from surface particles to volumetric particles.
As shown in Figure \ref{fig:marangoni-bubble}, we highlight the reverse process by releasing four drops of soap solution with high surfactant concentration $\Gamma$ to the thin film with low surfactant concentration. As the droplets impact the film, they are transformed as codimension-1 particles and blend into the surface, significantly reducing the local surface tension of the thin film. The response is twofold: first, the reduced surface tension can tolerate higher surface curvature; therefore, one can observe more high-frequency vibrations \revise{of} the surface; secondly, \revise{a soapy droplet would significantly increase the soap concentration at the impact location, thereby creating a zone with low surface tension. Thus, a nonzero gradient of surface tension would occur at the perimeter of this zone, pointing outwards, pushing particles out, a phenomenon known as the Marangoni effect. As this happens, more particles will cluster at the outskirt of the low surface tension zone, resulting in an increase in film thickness, which is reflected in the color change}.

\section{Performance}
Our algorithm benefits from the intrinsic parallelizability of the particle method and thus is highly efficient. All parts in Algorithm \revise{\ref{alg:update_status}, \ref{alg:compute_force} and \ref{alg:advance_step}} are implemented in parallel with OpenMP for multi-core processors. However, the codimension transition takes a linear time complexity because it needs to move particles between the surface and volumetric solvers. The performance of all scenes is displayed in Table \ref{table:performance} in Appendix \ref{appendix:validation}.

\section{Discussion and Future Works}
We propose a novel particle-based, thin-film simulation framework that jointly facilitates large surface deformation and lively in-surface flows. Our dynamic model is based on the surface-tension-driven 3D Navier-Stokes equations simplified under the lubrication assumption. We devise a set of differential operators on curved point-set surfaces that are \revisea{discretized} by SPH. Our key insight lies in that the compressible nature of SPH, which tends to create artifacts in its typical usage, can be leveraged to our benefit by defining an evolving height field $h$ that enables the incorporation of the surface tension model \revisea{into the thin-film fluid}. Due to the particle-based nature, our method easily handles codimension transitions and topology changes and conveniently integrates an auxiliary 3D SPH solver to simulate a wide gamut of visually appealing phenomena simultaneously \revise{featuring} surface and volumetric characteristics, such as the pinch-off of catenoids, dripping from a thin film, the merging of droplets, as well as bubble rupture.

The main limitation of the proposed algorithm is its incapability of dealing with surfaces carrying non-manifold intersections. In the future, we seek to extend it to simulate the interaction of complex bubble clusters that form a network of plateau \revise{borders}. \revise{
Then, the dilemma between particle density variation required for interesting color patterns and relatively even distribution of particles needed for the SPH framework limits our system in recreating color fields with a steep gradient}, which may be alleviated by adopting the adaptive kernels. We also look forward to exploring more dynamic interactions with solid boundaries, \textit{e.g.} activated by boundary deformation, intrusion, or even inter-crossing.

\revise{
The codimension transition is another challenging topic, for the criterion must be carefully chosen. If the transition happens too easily, the thin film can easily break because a hole of volumetric particles appears somewhere. If it's too hard, the thin film will form unnatural sharp corners instead of turning to the volumetric SPH solver. Due to the lack of adaptivity, the simulation of pinch-off is affected by subjectively defined parameters more or less, which we think is a possible direction for future improvement.} \revisea{
On the other hand, adding and deleting particles during the reseeding step has been a long-standing and difficult problem in conventional SPH simulation, which might cause physical quantity inconsistency. Our thin-film SPH model alleviated this problem thanks to its compressible nature and low stiffness coupling density and pressure on the particle thin film, although the volumetric particles in the simulation (in particular when the codimension transition occurs) could still suffer from such inconsistencies when transferring mass and momentum between the thin film.
}

On the animation side, the level of sophistication in the color pattern that our simulation generates is not quite at the level of real-world verisimilitude. That is because our displayed particles are exactly the ones used for the simulation, which is inevitably limited in the number. In the future, we may further investigate the possibility of using our SPH system as a background simulation, on top of which a larger number of tracker particles are advected, to make for visualization of improved richness and fidelity. Finally, we hope to customize a set of rendering algorithms that can directly take advantage of the geometric representation and computation tools we have already developed for the simulation to allow more consistent and physically accurate visual results.

\begin{acks}
We thank all the anonymous reviewers for their constructive comments. We acknowledge the funding support from NSF 1919647 and Dartmouth UGAR. We credit the Houdini Education licenses for the video rendering.
\end{acks}

\bibliographystyle{ACM-Reference-Format}
\bibliography{refs.bib,refs_compressible.bib,refs_physics.bib,refs_pinchoff.bib,refs_pressure.bib,refs_rim.bib,refs_sph.bib,refs_viscosity.bib,refs_related.bib,refs_foam.bib,refs_wang.bib,refs_extra.bib}

\appendix

\section{Momentum Equation at Center Surface}
\label{appendix:expansion}
The \revise{normal and tangential} components of \eqref{eq:ns_boundary} are  
\begin{equation}
\begin{aligned}
    \begin{dcases}
    p-p_a+2\kappa^{\pm} \gamma&= 2\mu \bm n^{\pm}\cdot (\bm \nabla u)\cdot\bm n^{\pm}+O(h^2),\\
    \bm t_x^{\pm} \cdot (\bm \nabla_{s^{\pm}}\gamma)&=\mu \bm t_x^{\pm} \cdot [\bm \nabla\bm{u}+(\bm \nabla\bm{u})^T]\cdot\bm n^{\pm}+O(h^2),\\
      \bm t_y^{\pm} \cdot (\bm \nabla_{s^{\pm}}\gamma)&=\mu \bm t_y^{\pm} \cdot [\bm \nabla\bm{u}+(\bm \nabla\bm{u})^T]\cdot\bm n^{\pm}+O(h^2),\\
    \end{dcases}
\end{aligned}
    \label{eq:ns_boundary_c}
\end{equation}
respectively.
Using \eqref{eq:geometry}, we obtain
\begin{equation}
\begin{aligned}
\begin{dcases}
    \bm t_x^{\pm} \cdot (\bm \nabla_{s^{\pm}}\gamma)\\
    \quad= (\bm e_x \mp\bm e_y \times \bm \nabla_sh) \cdot (\bm \nabla_s \gamma \pm  \bm \nabla_s h\bm e_z \cdot \bm \nabla\gamma \pm  \bm e_z\bm \nabla_s h \cdot \bm \nabla \gamma) +O(h^2)\\
    \quad=\bm e_x\cdot \bm \nabla_s \gamma+O(h^2),\\
   \bm t_y^{\pm} \cdot (\bm \nabla_{s^{\pm}}\gamma)\\
   \quad= (\bm e_y \pm\bm e_x \times \bm \nabla_sh) \cdot (\bm \nabla_s \gamma \pm  \bm \nabla_s h\bm e_z \cdot \bm \nabla\gamma \pm  \bm e_z\bm \nabla_s h \cdot \bm \nabla \gamma) +O(h^2)\\
   \quad=\bm e_y\cdot \bm \nabla_s \gamma+O(h^2).
\end{dcases}
\end{aligned}
 \label{eq:tgamma}
\end{equation}

Substituting \eqref{eq:geometry} and \eqref{eq:tgamma} into \eqref{eq:ns_boundary_c} yields
\begin{equation}
\begin{aligned}
    \begin{dcases}
    p-p_a+2( \pm \kappa_c
    +\kappa_h) \gamma=\\
    \quad 2\mu (\pm \bm e_z  - \bm \nabla_s h)\cdot (\bm \nabla u)\cdot(\pm \bm e_z  - \bm \nabla_s h)+O(h^2),\\
    \bm e_x  \cdot \bm \nabla_s \gamma=   \\
    \quad\mu (\bm e_x \mp\bm e_y \times \bm \nabla_sh) \cdot [\bm \nabla\bm{u}+(\bm \nabla\bm{u})^T]\cdot(\pm \bm e_z  - \bm \nabla_s h)+O(h^2),\\
      \bm e_y \cdot\bm \nabla_s \gamma= \\
      \quad\mu  (\bm e_y \pm\bm e_x \times \bm \nabla_sh)  \cdot [\bm \nabla\bm{u}+(\bm \nabla\bm{u})^T]\cdot(\pm \bm e_z  - \bm \nabla_s h)+O(h^2).\\
    \end{dcases}
    \label{eq:ns_boundary_ac1}
\end{aligned}
\end{equation}
Further, we substitute the \revise{component-wise forms} of $\bm \nabla_s h$ and $\bm \nabla \bm u$ into \eqref{eq:ns_boundary_ac1} as
\begin{equation}
    \begin{dcases}
    p-p_a+2( \pm \kappa_c
    +\kappa_h) \gamma=\\
    \quad 2\mu\left(\frac{\partial w}{\partial z}\mp \frac{\partial h}{\partial x}\frac{\partial u}{\partial z}\mp \frac{\partial h}{\partial y}\frac{\partial v}{\partial z}\right)+O(h^2),\\
    \frac{\partial \gamma}{\partial x}=\\
    \quad\mu \left[\pm \frac{\partial w}{\partial x}-2\frac{\partial h}{\partial x}\left(\frac{\partial u}{\partial x}-\frac{\partial w}{\partial z}\right)-\frac{\partial h}{\partial y}\left(\frac{\partial v}{\partial x}+\frac{\partial u }{\partial y}\right) \pm \frac{\partial u}{\partial z}\right]+O(h^2),\\
    \frac{\partial \gamma}{\partial y}=\\
    \quad\mu \left[\pm \frac{\partial w}{\partial y}-2\frac{\partial h}{\partial y}\left(\frac{\partial v}{\partial y}-\frac{\partial w}{\partial z}\right)-\frac{\partial h}{\partial x}\left(\frac{\partial v}{\partial x}+\frac{\partial u }{\partial y}\right) \pm \frac{\partial v}{\partial z}\right]+O(h^2).\\
    \end{dcases}
    \label{eq:ns_boundary_ac2}
\end{equation}

Suppose the tangential velocity is even along the local z-coordinate. Similar to the idea of asymptotic expansion \cite{physics-chomaz-2001}, we take the Taylor expansion of $\bm u_s$ as
\begin{equation}
\begin{dcases}
    u = u(z=0) + \frac{\partial^2 u}{\partial z^2}\bigg|_{z=0}z^2+ O(h^4),\\
    v = v(z=0) + \frac{\partial^2 v}{\partial z^2}\bigg|_{z=0}z^2+ O(h^4),\\
\end{dcases}
\label{eq:uvh}
\end{equation}
and with the solenoidal condition, we obtain
\begin{equation}
    \frac{\partial w}{\partial z} =-\bm \nabla_s \cdot \bm u_s = -\bm \nabla_s \cdot \bm u_s(z = 0) + O(h^2).
    \label{eq:wz}
\end{equation}
Combining \eqref{eq:uvh} and \eqref{eq:wz}, as well as using the boundary conditions \eqref{eq:ns_boundary_ac2}, we obtain
\begin{equation}
    \begin{dcases}
    p = p_a-2[\kappa_h\gamma+\bm \nabla_s \cdot \bm u_s(z = 0)] -\frac{2}{h}\gamma \kappa_c z+O(h^2),\\
    \frac{\partial^2 u}{\partial z^2}\bigg|_{z=0} = \frac{1}{h}\frac{\partial \gamma}{\partial x}+O(h^2),\\
    \frac{\partial^2 v}{\partial z^2}\bigg|_{z=0} = \frac{1}{h}\frac{\partial \gamma}{\partial y}+O(h^2).\\
    \end{dcases}
    \label{eq:puv}
\end{equation}
Then, we substitute \eqref{eq:puv} into \eqref{eq:ns_buck} to obtain the momentum equation at $z=0$:
\begin{equation}
\begin{aligned}
    &\rho \frac{D \bm{u}}{D t}+O(h^2)=\\
    &\quad2\bm \nabla_s (\kappa_h\gamma+\bm \nabla_s \cdot \bm u_s)+\frac{2\gamma}{h}\kappa_c \bm n+\frac{1}{h}\bm \nabla_s \gamma+ \mu\nabla_{s}^2\bm{u}+\bm{f}.
\end{aligned}
\end{equation}

\section{Numerical Validation and Performance Table}
\label{appendix:validation}

\begin{table*}
\begin{threeparttable}
\centering
\caption{Performances of Different Scenes. A in table is a 128-core 3.1GHz AMD Ryzen Threadripper 3990X workstation, B in table is a 4-core 2.8GHz Intel(R) Core(TM) laptop, and C is a 6-core 2.6GHz Intel(R) Core(TM) laptop.}
\begin{tabular}{c|c|c|c|c|c}
\hline
Figure & Description & Number of Particles & Computational Resource &FPS\tnote{*}& Time / Frame (Avg.) \\
Figure \ref{fig:newton-ring}&Thickness Profile & 5,154\textasciitilde5,785\tnote{\textdagger} & Laptop C with 6 Cores & 10 & 1.409s \\
Figure \ref{fig:oscillation} (Upper)&Oscillating Bubble (Sphere Bubble)& 163,842 &Server A with 16 Cores&50&10.65s\\
Figure \ref{fig:oscillation} (Lower)&Oscillating Bubble (Irregular Bubble)&24,578&Server A with 128 Cores&50&0.225s\\
Figure \ref{fig:half-bubble}&Oscillating Bubble (Half Bubble)&86,013&Server A with 32 Cores&50&9.6s\\
Figure \ref{fig:bubble-rupture}&Bubble Rupture&163,842&Server A with 128 Cores&50&10.92s\\
Figure \ref{fig:square-RT}&Large-Deforming Thin Films (Square)&58,081&Server A with 128 Cores&5&15.48\\
Figure \ref{fig:circle-rt}&Large-Deforming Thin Films (Circular)&126,282&Server A with 128 Cores&10&13.575s\\
Figure \ref{fig:dripping}&Thin-Film Dripping&5,185\textasciitilde5501\tnote{\textdagger}&Server A with 16 Cores&50&9.74s\\
Figure \ref{fig:catenoid-bubble},\revise{\ref{fig:minimal_surface}}&Catenoid&4,968\textasciitilde5,542\tnote{\textdagger}&Laptop C with 6 Cores&300&0.883s\\
Figure \ref{fig:marangoni-bubble}&Droplet Marangoni Effect&8,173&Laptop C with 6 Cores&50&2.01s\\
Figure \ref{fig:curvature-test}&Mean Curvature Validation & 10,000 & Laptop C with 6 Cores & 10 & 0.35s\\
Figure \ref{fig:height-validation}&Numerical Height Validation& 5154&Laptop C with 6 Cores & 10 & 0.774s\\
Figure \ref{fig:capillary-wave} & Capillary Wave Validation & 1,000 & Laptop B with 4 Cores&50& 0.02s\\
\hline
\end{tabular}
\begin{tablenotes}
    \footnotesize
    \item[\textdagger] The number of particles increases \revise{throughout} the simulation due to particle reseeding, the two numbers listed in the table are the number of particles at the first and last frame.
    \item[*] \revise{The} number of frames per second for simulation. For example, FPS=$50$ means the time step of a frame $\Delta t=0.02s$. We take CFL condition number $C=0.1$, so a frame may consist of multiple time iterations depending on $v_{\textit{max}}$, and the simulation time of a frame is subject to it.
\end{tablenotes}
\label{table:performance}
\end{threeparttable}
\end{table*}

To evaluate the accuracy of our algorithm, we perform a set of numerical tests concerning the main key elements in our algorithm in both the geometric and dynamic computations. We will compare our simulation results with both the ones derived from analytic equations and real-life experiments.

\paragraph{Mean Curvature}

\begin{figure}
 \centering
 \includegraphics[width=.45\textwidth]{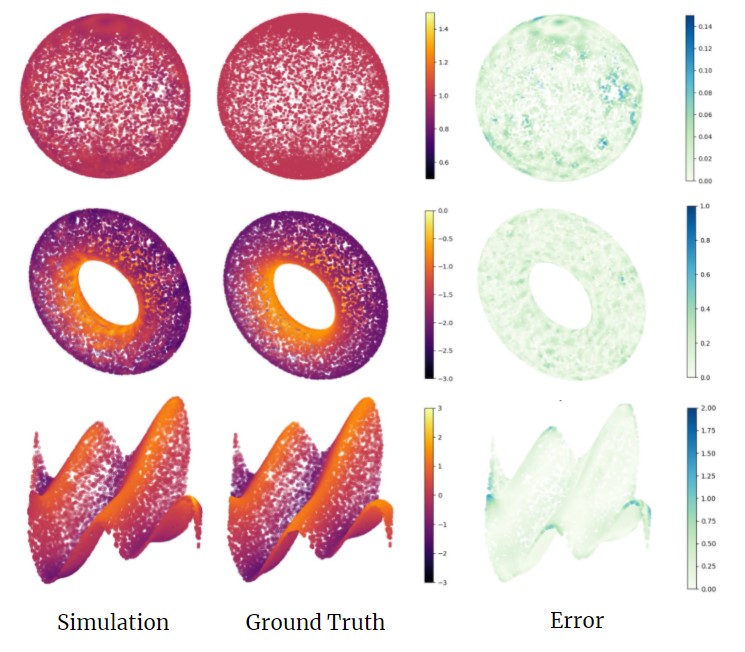}
 \caption{Estimated Mean Curvature using our codimension-1 differential operators vs. the analytical ground truth. From top to bottom are respectively: the sphere, the torus, and the trigonometric \revise{surface}.}
 \label{fig:curvature-test}
\end{figure}

The centerpiece of computing the surface tension behaviors is the estimation of mean curvature $\kappa$, which relates to the pressure jump across \revise{the} air-fluid interface via the Young-Laplace equation $\Delta p = -\gamma \kappa.$

Our method computes the mean curvature $\kappa_c$ of the center surface, which are curved 3D surfaces defined by the particles using codimension-1 planar operators. To verify the correctness and robustness of our method under various surface geometries, we compute the mean curvature of three different shapes: a sphere with $R = 1$; a torus
\revise{
\begin{equation*}
(x,y,z)=R\left((c+\cos^{-1}(v))\cos(u), \sin^{-1}(v), (c+\cos^{-1}(v))sin(u)\right)\\
\end{equation*}
}
with $c = 0.8, a = 0.3, u \in [0, 2\pi), v \in [0, 2\pi)$, and a surface defined by $y=0.1\left(3 \sin(x) + 2 \cos(z) + 4 \sin(2 x + z)\right)$ with $x,z \in [0, 2\pi)$.

To highlight the robustness of our algorithm, we initialize the sample particles from a uniform random distribution to rehearse the scenarios where particles are unevenly distributed in simulation. In Figure \ref{fig:curvature-test}, \revise{
we plot the numerical and analytical estimations of mean curvature on the left two columns and the error between them on the right column.
} We can see that overall, our algorithm does well in estimating the mean curvature for all kinds of surfaces. The error is mostly gathered on the edges due to insufficiently sampled neighborhoods, a shortcoming common to SPH methods \cite{sph-koschier-2019}.

\paragraph{Minimal Surface}

\begin{figure}
 \centering
 \includegraphics[width=.45\textwidth]{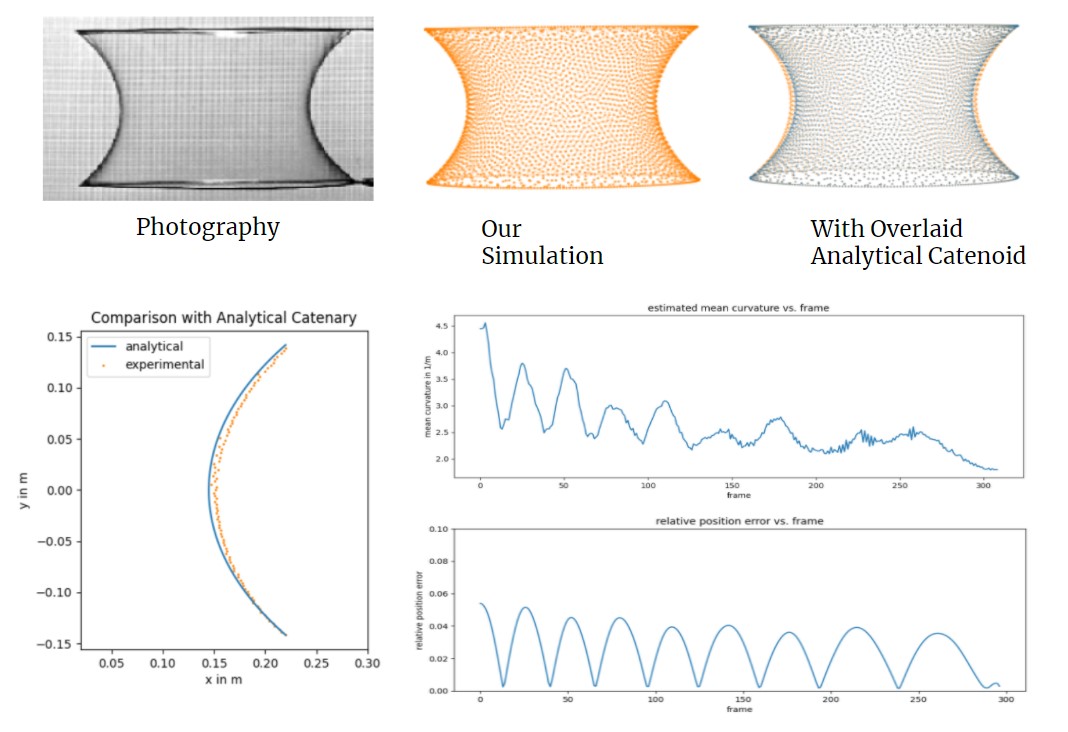}
 \caption{
On the top row, display the thin-film catenoid as photographed, our simulation results, and our results overlaid with a corresponding analytical catenoid. The left figure in the bottom row shows a detailed display of one curve of the cross-section of our simulated thin film with an analytical catenoid. The right two figures plot the average mean curvature and the positional error to the analytical surface.}
 \label{fig:minimal_surface}
\end{figure}

The dynamics of thin-film \revise{are} marked by its tendency to form minimal surfaces under the surface tension, which constantly contracts the surface to minimize the area locally. When two rings are connected by a continuous soap film, and gravity is ignored, the soap film between them will form the catenoid ------ a minimal surface formed by rotating a catenary. The analytical definition of a catenoid is given by
\revise{
\begin{equation*}
      (x,y,z)=\left(c\cosh\left(\frac{v}{c}\right)\cos{u}, v, c\cosh\left({\frac {v}{c}}\right)\sin{u}\right).
  \end{equation*}
}
The way that soap-film catenoids are typically formed is by gradually pulling the parallel rims apart. If we let $D$ denote the diameter of the rims, and let $d$ denote the separation between them, then an analytical catenoid can be solved for when $\frac{d}{D} < 0.66274$.

The constant 0.66274 is the approximation of the Laplace limit, which is the value $\frac{1}{\sinh u}$ for u satisfying $u = \coth u$, $u > 0$. As a result, until this critical condition is met, we can obtain a corresponding analytical catenoid to compare and contrast for every point cloud in our simulation.

In the upper subfigures in Figure \ref{fig:minimal_surface}, we plot the result of our simulation along with the photographed real-world soap catenoid. On the rightmost one, we overlay our simulated results (orange) with the corresponding analytical catenoid(blue). As we can see, our simulation is only marginally different from the analytical solution, and the difference between our simulation and the photography is almost imperceptible.
\revise{The} bottom-left displays a close-up look of a curve from a \revise{cross-section} of our simulated shape (orange) and an analytical catenary. The bottom-right displays two subplots that are time-varying: the upper one records the varying average mean curvature, which is supposed to be $0$ as it is for all minimal surfaces, and the lower one is an average relative positional error, which represents the average distance between a particle and the analytical catenoid relative to the arc length of the catenary. As one can see, both plots illustrate a decreasing trend marked by periodic fluctuation, which corresponds to the oscillating motion of the simulated thin film; until the Laplace limit is violated, the simulated thin film does well in keeping close to the theoretical shape with the relative positional error less than 0.06.\begin{figure}[ht]
\begin{center}
\resizebox{0.45\textwidth}{!}{%
\includegraphics[height=2.8cm]{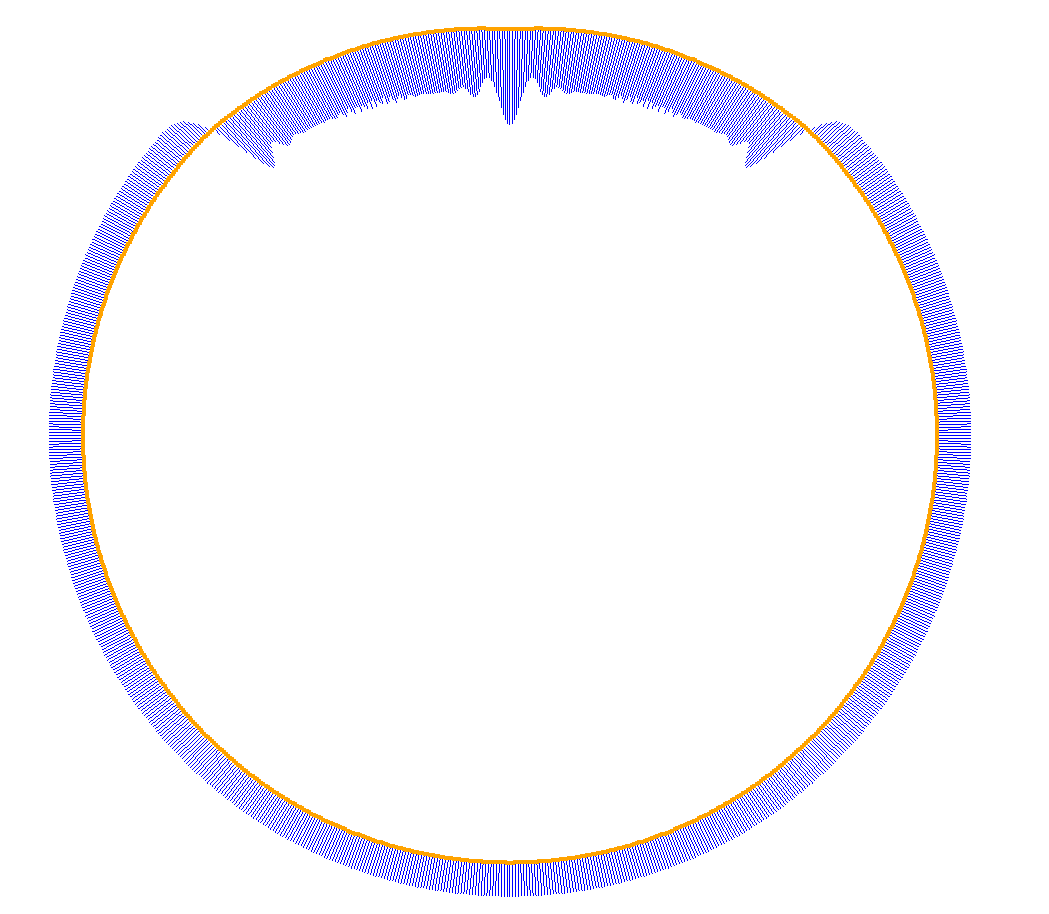}%
\quad
\includegraphics[height=3cm]{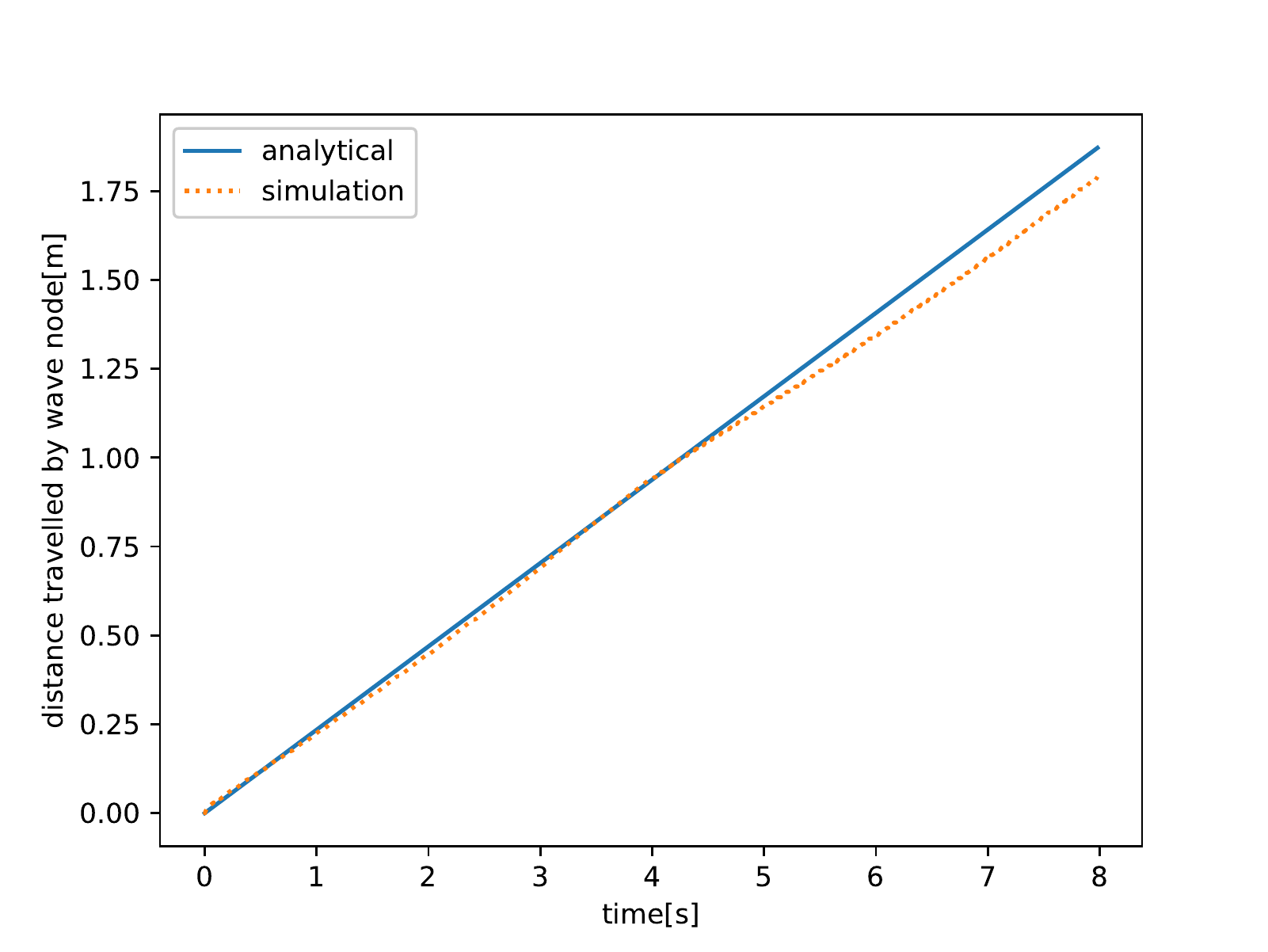}%
}
\end{center}
\caption{The capillary wave on a bubble. We take the surface tension coefficient $\gamma=1.1\times 10^{-5}N/m$, thickness $2h=800nm$, fluid density $\rho=10^3kg/m^3$ and bubble radius $R=0.80m$. The blue lines of the left part are the velocity vectors of particles.}
\label{fig:capillary-wave}
\end{figure}

\begin{figure}
 \centering
 \includegraphics[width=.45\textwidth]{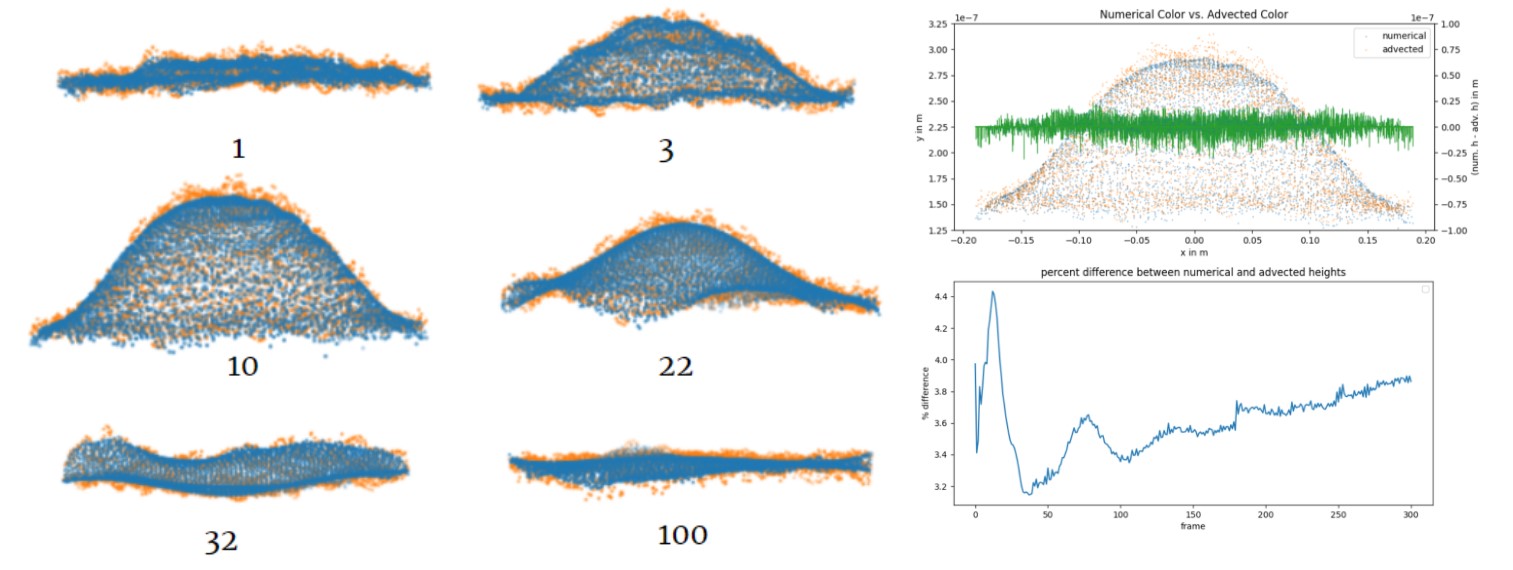}
 \caption{Left: the surface of numerical height (blue), the surface of advected height (orange). Right: The difference between the two surfaces in a single frame (top) and over time (bottom).}
 \label{fig:height-validation}
\end{figure}

\paragraph{Numerical Height vs. Advected Height}
\revise{
We investigate qualitatively and quantitatively the behavioral discrepancy between the two forms of height computation. In the left subfigure of Figure \ref{fig:height-validation}, we exhibit the surface defined by the advected height (orange) and by the numerical height (blue) at frames $[1,3,10,22,32,100]$ of a simulation where a centripetal velocity is initially applied to move particles towards the middle, while the surface tension works to flatten out the curvature. The two surfaces are congruent in trend, and yet the surface formed by the advected height is bumpier than the other. In the top-right subfigure, the difference between the two surfaces is plotted alongside the surfaces to show that their commonality dominates their difference. In the bottom-right subfigure, we show that the change in the average percent difference over time is confined within 5\%.}

\paragraph{Capillary Wave} Figure \ref{fig:capillary-wave} shows the capillary wave on a bubble. The scene setup is a simplification of that in Figure \ref{fig:oscillation}, where initially one downward impulse, instead of two, is applied on the north pole of \revise{the} bubble. It immediately creates a perturbation at the top and then propagated to the bottom. This propagation is dominated by the capillary force $\frac{2\gamma}{h}\kappa_c\bm{n}$, and thus a capillary wave. Note that $\kappa_c$ is calculated by half of the \revise{Laplacian} of surface, then the local dynamics along normal direction system subjects to wave equation

\begin{equation}
    \rho\frac{\mathrm{D}^2 r_z^2}{\mathrm{D}t^2}=\frac{\gamma}{h}\frac{\partial^2 r_z}{\partial x^2},
\end{equation}
with $r_z$ as the local $z$-component of particle position. Thus the phase velocity of \revise{the} wave is analytically given by $v_p=\omega/k=\sqrt{\gamma/h\rho}$.

The right part of Figure \ref{fig:capillary-wave} shows the time-distance relationship traveled by the down-most wave knot, which represents the phase velocity of \revise{the} capillary wave, comparing analytical solution and simulation results. It proves that our algorithm can accurately and convincingly reproduce the effect surface tension.

\end{document}